\documentclass[final,10pt,a5paper]{phdimt}
\pdfoutput=1

\pdfcatalog{
	/PageMode (/UseOutlines)
      /OpenAction (fitbh)
}

\title{Combining Peer-to-Peer and Cloud Computing for Large Scale On-line Games}
\author{Emanuele Carlini}
\mail{emanuele.carlini@imtlucca.it}
\program{Computer Science and Engineering}
\coordinator{Prof. Ugo Montanari}
\coordinatorinst{University of Pisa}
\cycle{XXIV}
\year{2012}

\supervisor{Prof. Laura Ricci}
\supervisorinst{University of Pisa}
\cosupervisor{Prof. Massimo Coppola}
\cosupervisorinst{Istituto di Scienza e Tecnologie della Informazione (ISTI), CNR, Pisa}
\cosupervisortwo{Prof. Alberto Montresor}
\cosupervisorinsttwo{University of Trento}

\tutor{Prof. Marzia Buscemi}
\tutorinst{IMT Institute for Advanced Studies Lucca}

\firstreviewer{Prof. Paolo Costa}
\firstreviewerinst{Imperial College London}
\secondreviewer{Prof. Alexey Vinel}
\secondreviewerinst{Tampere University of Technology}
\thirdreviewer{}
\thirdreviewerinst{}

\newcommand{\timeWait}{\Delta t}

\theoremstyle{definition}
\newtheorem{mydef}{Definition}

\begin{document}

\graphicspath{
{mainmatter/chapter-Related/}
{mainmatter/graphs/}
{mainmatter/./}
}

\def\vec#1{{\bf #1}}

\renewcommand\baselinestretch{1}
\baselineskip=14pt

\hyphenation{Ein-ste-in}
\hyphenation{a-ch-ie-ved}
\hyphenation{ap-pro-ach-es}
\hyphenation{mo-del}

\frontmatter

\pagestyle{empty} 
\maketitle
\makereviewerspage
\pagestyle{plain} 

\tableofcontents
\listoffigures
\listoftables
%

\begin{acknowledgements}
\addcontentsline{toc}{chapter}{Acknowledgements}

Let me be clear from the beginning. This thesis wouldn't have been possible without all the people I have met along the path. And the list is quite long. I would express my infinite gratitude to Prof. Laura Ricci and Dr. Massimo Coppola not only for their scientific guidance but also for continuous support and encouragement. A special thank goes to Prof. Alberto Montresor for all the accurate advices and the fruitful discussions. I would like to thank Prof. Alexey Vinel and Prof. Paolo Costa for their precise feedback and comments. Many thanks to all the people at IMT Lucca, especially to colleagues of the XXIV cycle, with whom I shared most of my time at IMT. I would like to thank Dr. Raffaele Perego, Ranieri Baraglia and all the other friends and colleagues at HPC lab at CNR-ISTI. It has been an immense privilege working with you guys. I wish to thank in particular Dr. Patrizio Dazzi, Stefania Lombardi and Matteo Mordacchini for their support and for an infinite number of fruitful discussions. A particular thank goes also to my friends Beniamino, Daniele, Luca and Iacopo. This thesis wouldn't have been possible without the continuous encouragement of my family, to whom I ex- tend my most deep thanks. Last, but definitely not least, I thank Anna for being on my side all the time and completing my life with her love.

\end{acknowledgements}

%


\begin{center}
\vspace*{0.5cm}
{\Large \bf  Vita}
\addcontentsline{toc}{chapter}{Vita and Publications}
\end{center}
\begin{table}[h!]
\begin{center}
\renewcommand{\arraystretch}{1.25}
\begin{tabular*}{1\textwidth}{l p{8.5cm}}

{\bf August 13, 1981} & Born, La Spezia, Italy \\
& \\
{\bf 2004} & Bachelor of Applied Science Degree \\  
& Final mark: 105/110\\
& University of Pisa, Italy \\
& \\
{\bf 2008} & Master Degree\\  
& Final mark: 108/110\\
& University of Pisa, Italy \\
& \\
{\bf 2009} & Graduate Fellow\\  
& Istituto di Scienza e Tecnologie dellInformazione (ISTI), National Research Council\\
& Pisa, Italy \\
& \\
{\bf 2009} & PhD Student in Computer Science and Engineering \\  
& IMT Institute for Advanced Studies Lucca \\
& Lucca, Italy\\

\end{tabular*}
\end{center}
\end{table}
\clearpage
\begin{center}
\vspace*{0.5cm}
{\Large \bf  Publications}
\end{center}
\vspace*{0.5cm}
{\small
\begin{enumerate}


\item Carlini, E., M. Coppola, P. Dazzi, D. Laforenza, S. Martinelli, and L. Ricci, ``Service and Resource Discovery Supports over P2P Overlays", in \emph{Proceedings of International Conference on Ultra Modern Telecommunications (ICUMT)}, IEEE, pp.1-8, 2009.

\item Carlini, E., M. Coppola, and D. Laforenza, ``XtreemOS, an Open-Source Grid Operating System Targeting the Future Internet", in \emph{III Conferenza Italiana sul Software Libero}, 2009.

\item Carlini, E., M. Coppola, D. Laforenza, and L. Ricci, ``Reducing Traffic in DHT-based Discovery Protocols for Dynamic Resources, in \emph{Grids, P2P and Services Computing}, Springer, pp.73-87, 2010.

\item Carlini, E., M. Coppola, and L. Ricci. ``Integration of P2P and Clouds to Support Massively Multiuser Virtual Environments", in \emph{Proceedings of the 9th Annual Workshop on Network and Systems Support for Games (NetGames)}, ACM/IEEE, pp.1-6, 2010.

\item Carlini, E., M. Coppola, P. Dazzi, L. Ricci, and G. Righetti, ``Cloud Federations in Contrail. in \emph{Euro-Par 2011: Parallel Processing Workshops}, Springer, pp.159-168, 2011.

\item  Carlini, E., M.Coppola, and L.Ricci,`` Evaluating compass routing based AOI-cast by MOGs mobility models", in \emph{Proceedings of the 4th International Conference on Simulation Tools and Techniques}, ICST, pp.328-335, 2011.

\item Carlini, E., M. Coppola, and L. Ricci, ``Probabilistic Dropping in Push and Pull Dissemination over Distributed Hash Tables", in \emph{Proceedings of the 11th International Conference on Computer and Information Technology (CIT)}, IEEE, pp.47-52, 2011.

\item Ricci, L., E. Carlini, L. Genovali, and M. Coppola, ``AOI-cast by Compass Routing in Delaunay Based DVE Overlays", in \emph{Proceedings of International Conference on High Performance Computing and Simulation (HPCS)}, IEEE, pp.135-142, 2011.

\item Ricci, L. and Carlini, ``Distributed Virtual Environments: From Client Server to P2P Architectures", in \emph{In High Performance Computing and Simulation (HPCS), 2012 International Conference on}, pages 817. IEEE, 2012.

\item Carlini, E., L. Ricci, and M.Coppola. ``Flexible load distribution for hybrid distributed virtual environments", in \emph{Future Generation Computer Systems}, Elsevier, \url{http://dx.doi.org/10.1016/j.bbr.2011.03.031}, 2012.

\item Ricci, L., L. Genovali, E. Carlini, and M. Coppola. ``AOI-CastinDistributed Virtual Environments: an Approach Based on Delay Tolerant Reverse Compass Routing, in \emph{Concurrency and Computation: Practice and Experience}, to appear.

\item E. Carlini, L. Ricci, and M. Coppola. ``Reducing Server Load in MMOG via P2P Gossip, \emph{Proceedings of the ACM Workshop on Network and System Support for Games (NetGames 2012)}, Venice, Italy, to appear.

\end{enumerate}
}
%
%
%
%

\begin{abstract} 

This thesis investigates the combination of Peer-to-Peer (P2P) and Cloud Computing to support Massively Multiplayer On-line Games (MMOGs). MMOGs are large-scale distributed applications where a large number of users concurrently share a real-time virtual environment. Commercial MMOG infrastructures are sized to support peak loads, incurring in high economical cost. Cloud Computing represents an attractive solution, as it lifts MMOG operators from the burden of buying and maintaining hardware, while offering the illusion of infinite machines. However, it requires balancing the tradeoff between resource provisioning and operational costs. P2P- based solutions present several advantages, including the inherent scalability, self-repairing, and natural load distribution capabilities. They require additional mechanisms to suit the requirements of a MMOG, such as backup solutions to cope with peer unreliability and heterogeneity. We propose mechanisms that integrate P2P and Cloud Computing combining their advantages. Our techniques allow operators to select the ideal tradeoff between performance and economical costs. Using realistic workloads, we show that hybrid infrastructures can reduce the economical effort of the operator, while offering a level of service comparable with centralized architectures.

\end{abstract}

\mainmatter

\chapter{Introduction}

On-line gaming entertainment has acquired lots of popularity in the last years from both industry and research communities.
This attention is justified by the economic growth of the field, where Massively Multiplayer Online Games (MMOGs, \cite{wow-site,sl-site}) represent a remarkable member.
The market size of MMOG has received a 5 billion \$ evaluation in 2010, while the number of total user have reached around 20 million worldwide\footnote{www.mmodata.net, August 2012}. 

MMOGs are large-scale distributed applications that allowing a huge amount of users worldwide to share a real-time virtual environment.
MMOGs operators provide the necessary hardware infrastructure to support the high requirements of MMOGs. 
The profit of the operators comes from the fee the users pay periodically to participate in the virtual environment\footnote{Other profit sources are gaining attention nowadays. For instance, in the pay-for-win model the access to the MMOG is free, but players pay to be competitive through the game.}.
Regardless of their peculiar business model, operators' profit is directly linked with the number of users that participate to the MMOGs. In fact, the more populated is a MMOG, the higher the probability to attract new users.
For this reason, offering an acceptable level of service is a core necessity for operators. 
Hence, operators size the infrastructure to room the peak number of users thus offering an acceptable level of service.

Currently, most commercial MMOGs rely on a centralized architecture.
This architecture supports a straightforward management of the main functionalities of the MMOG, such as user identification, virtual environment management, synchronization between players, and billing. 
However, with higher and higher amounts of concurrent users, centralized architectures show their scalability limitations, especially in terms of economical return for the operators.

Indeed, server clusters have to be bought and operated to withstand service peaks, also balancing computational and electrical power
constraints. 
A cluster-based centralized architecture concentrates all communication bandwidth at one data centre, requiring the static provisioning of large bandwidth capability.
Further, a large static provisioning exposes the operators to \textit{over-provisioning}, which leaves the MMOG operators with unused resources when the load on the platform is not at peak.

On-demand resources provisioning (where a notable example is Cloud Computing \cite{buyya2009cloud}) may alleviate the aforementioned MMOGS scalability and hardware ownership problems \cite{Nae2008, Nae2011}.
The possibility of renting machines lifts the MMOGs operators from the burden of buying and maintaining hardware, and offers the illusion of infinite machines, with good effects on scalability.
Also, the pay-per-use model adheres with the seasonal access patterns to the MMOGs (e.g. more users in weekends than in the middle of the week).

However, the exploitation of Cloud Computing presents several issues.
The recruiting and releasing of machines must be carefully orchestrated in order to cope with start-up times of on-demand resources and to avoid incurring on unnecessary expenses leaving the server unloaded.
Further, besides server time, bandwidth consumption may represents a significant expense
when operating a MMOG.
Thus, even if an infrastructure based entirely on on-demand resources is feasible, still the profit margin for the DVE operators may be higher. This can be achieved by allowing user-provided resources to share part of the load of a MMOG.

These aspects have been deeply discussed in the research communities in the last decade.
Mechanisms to integrate user-provided resources in a MMOGs infrastructure naturally evolved from peer-to-peer (P2P) classical approaches.
While reducing the load on centralized servers, P2P-based solutions present several attractive advantages.
First, P2P techniques are inherently scalable. When the amount of users grows, more resources are added to the infrastructure. Second, in case of peer failures, P2P networks are able to self-repair and reorganize, hence providing robustness to the infrastructure. 
Third, network traffic is distributed among the users involved, in principle avoiding the creation of bottlenecks.
Furthermore, all these properties pair with little costs for the MMOG operators. 

However, P2P-based infrastructures require additional mechanisms to suit the requirements of a MMOG.
When users leave the system some data must be transferred elsewhere.
In case of abrupt disconnection, a backup mechanism must guarantee for data availability.
The lack of a central authority makes it complex to enforce security and soundness of updates to virtual world at any time. 
Moreover, user machines typically impose strict and heterogeneous constraints on computational power and network capability, which makes them difficult to be exploited.

The high degree of complementarity between on-demand and user-provided resources has been an incentive to combine the two approaches.
The integration of P2P and Cloud computing has recently emerged has an active field of research for multiple applications, such as video streaming \cite{payberah2012clive}, storage \cite{xu2012integration}, replica management \cite{kavalionak2012p2p} and content distribution \cite{montresor2011cloudy}.
This thesis presents a set of mechanisms of a MMOG architecture that seamlessly combines on-demand and user-provided resources. As far as we know, we are the first that propose the combination of P2P and Cloud computing to address MMOGs.

We presented an initial and partial discussion on the topic in \cite{carlini2010integration,pos}. In the same work, we have also presented the initial design of a hybrid and flexible architecture combining the advantages of the two approaches. 

Regardless the particular solution, the mechanisms proposed all share the same core idea, that is to allow operators the possibility to choose the tradeoff between the usage of on-demand and user-provided resources.
In our design, an operator can decide, is some cases even at runtime, to have an infrastructure more reliable and responsive (for example for particularly interactive MMOGs) or to reduce the economical effort and provide a less powerful infrastructure, perhaps suitable for less interactive MMOGs. In other words, the idea is to let the operator to decide how to make profit, by deciding the ideal tradeoff between performance and economical cost.

\section{Massive Multi-player On-line Games}

Massive Multi-player On-line Games (MMOGs) are a synchronous, persistent and interactive virtual environment where players concur and cooperate.
The main distinguishing trait of MMOGs, the one that differentiates them from other on-line games, is the number of players sharing the same virtual environment, which can be in the order of thousands.
The core MMOG player's experience is to take the identity of an \textit{avatar}, an alter-ego in the virtual word. 
The other entities present in multi-player games include characters controlled by the system (usually referred as non-player characters or NPCs) and objects. These  aspects and the terminology are detailed in the Chapter \ref{chap:background} of the thesis.

Avatars are associated with a position in the virtual world, and they are represented by a state (e.g  abilities and health).
Avatars, controlled by the players, travel across the virtual world to solve the so called \textit{quests}, missions which provide rewards upon completion.
Usually this requires to visit and explore the virtual world, interact with other avatars, finding objects, earning money and so on. 
A reward can consist on different prizes, such as increasing the power of avatar abilities, particular items, or virtual currencies. In fact, the rewards largely depends on the genre of the MMOG. 
Similarly to avatars, even NPCs may have a state that usually is less accurate, whose degree of persistence depends on the MMOG genre. 
For instance, vendors NPCs (i.e. characters who sell goods to avatars) may either trace or ignore transactions, according to the game design.
Objects may be classified as \textit{mutable} (e.g. doors) that generally include all the interactive objects and \textit{immutable} such as the natural landscapes, buildings, vegetation \cite{Bharambea}.
In particular, immutable objects correspond to graphic elements that are inserted as static elements in the game client software. The update of these elements occur externally by the game context, or even off-line, usually via patch.
Immutable objects do not require a real-time management, and for this reason in the following we use the term \textit{object} to implicitly indicate mutable objects.

The genre highly characterizes a MMOG.
Massive Multi-player On-line Role Playing Games (MMORPGs), such as Second Life \cite{sl-site} and World of Warcraft \cite{wow-site}, are the category most represented in the today industry. They are characterized by large virtual environments and relatively slow-paced game mechanics. By comparison, Massive Multi-player On-line First Person Shooter (MMOFPS) provide an higher frenetic experience, at the expense of smaller number of concurrent players and less interactive capability of the environment.
In this thesis we mostly consider MMORPGs, as they represent both the most widespread genre and the most challenging from a scientific point of view, considering the high number of concurrent players.

\section{On-demand Platforms}

\begin{figure}[tbh]
\centering
\includegraphics[width=0.9\textwidth]{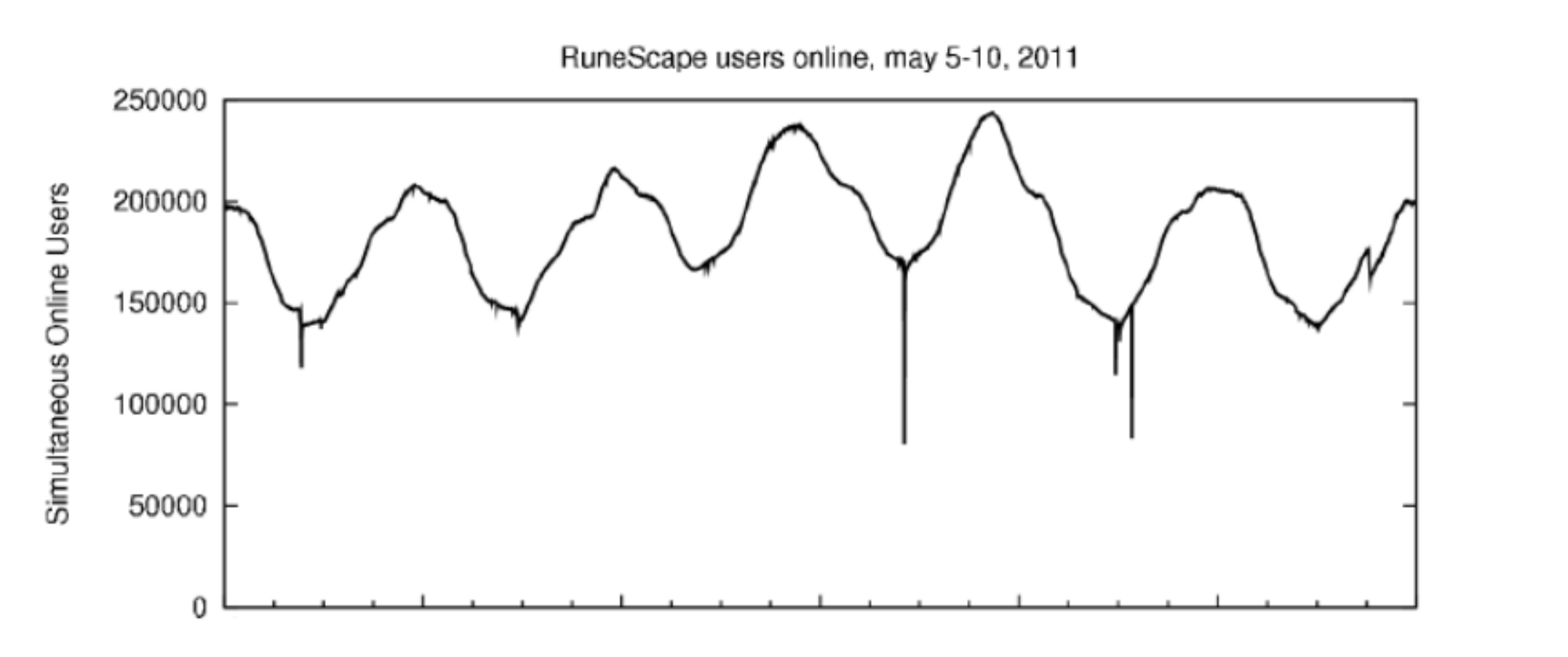}
\caption{Simultaneous Runescape users, source: \cite{Marzolla}}
\label{fig:runescape}
\end{figure}

MMOG operators are forced to over-provision the amount of resources of their architecture to sustain game peak load. 
The work in \cite{Marzolla} collects a set of traces from the {\em RuneScape Fantasy MMOG} and observes that a daily period access pattern is clearly visible (see Figure \ref{fig:runescape}). A daily churn of about 100.000 users is detected between peak and non-peak hours. 
The reservation of a relevant number of resources is mandatory to face a users oscillation of this magnitude.
However, over-provisioning implies the under-utilization of some resources during non-peak periods.

\textit{Cloud computing} is currently examined by the researchers as a suitable computing paradigm able to solve the over-provisioning problem \cite{carlini2010integration, IosupTPDS, Marzolla}.
Cloud computing is a form of \textit{utility computing}, considering computational resources as a service, rather than a commodity. Being a service, customers pay only the computational resources they use.
MMOG operators may exploit clouds by requesting a large set of resources during peak hour and by releasing  them as long as they are no longer needed. However, a deep investigation of several issues is required to fully exploit the potentialities of this novel computational paradigm.  Here we identify and explain two of these issues, namely the overhead of the virtualization and the need to properly orchestrate the provisioning of the resources.

Normally, MMOG operators exploit the Cloud in the form of Infrastructure as a Service (IaaS) \cite{bhardwaj2010cloud}. 
The IaaS offers to the operator the possibility to rent virtualized resources , running their favourite operating systems and services on top of them. In this case, the virtualization introduces overhead mainly for the delays due to the resources instantiation. This delay may depend on multiple factors, such as the particular provider or resource chosen. Mechanisms that dynamically provision cloud resources must consider this delay when exploiting Clouds.

In order to ensure an acceptable level of Quality of Service (QoS), the customer and the Cloud provider sign
a Service Level Agreement (SLA). The SLA defines the guaranteed QoS and the corresponding economical cost for the customer.
In MMOGS, the QoS may be defined in multiple forms. For instance, the response time of the MMOGs service should never drop under a certain threshold. However, the QoS is normally defined against the average behaviour of the resources. Hence, for some time the performance may degrade when the load suddenly increases on the platform.
The infrastructure must define a provisioning mechanism able to recruit the proper  set of resources in such situations. 
On the other hand, a similar mechanism has to release a set of machines in order to reduce the service cost, when the infrastructure is under-loaded.

For these reasons, Cloud-based MMOGs infrastructures should include a set of additional components supporting the dynamic 
provisioning of the resources.
As an example a possible component would be a monitor to collect run time performance metrics.  
It could take into account different performance measures, for instance the
response time of the application, the average system throughput, the amount of bandwidth consumed by the application and the utilization of the rented machines.
Another example would be a provisioner that defines the optimal number of resources and orchestrate their releasing and recruiting. The provisioner could adopt a proactive behaviour by exploiting fast and accurate analytical load models and fast prediction algorithms to foresee load peaks and  under-utilization of resources.

\section{Thesis Contribution}

This thesis covers, and in some cases extends, our work on MMOGs that we have presented in the last few years.
In a recent work \cite{ricci2012tutorial} we have presented an overview on the state of the art and the design issues for MMOG architectures. 
In \cite{carlini2010integration} we have presented an initial design of a virtual environment architecture combining the advantages of the on-demand and user-provided resources, exploiting our prior experience in designing Cloud Computing \cite{carlini2012cloud} and P2P-based architectures \cite{ricci2011aoi,carlini2011evaluating}.

In \cite{pos} we refined the initial design by supporting two core features of the MMOGs with two distinct and independent distributed components. In particular, we separate the management of the \textit{positional actions} (i.e the actions that affects the position of the entities in the MMOG) from the \textit{state actions} (i.e. the actions that modify the state of the entities).
The definition of two distinct components allows for the minimization and the control of the data transfers between the distributed servers due to the movements of the users in the MMOG. 
Furthermore, the definition of two independent components eases the design and the optimization of both of them.
This represents a firm swerve with the current state of the art, where the common approach is to let a single distributed infrastructure manage both positional and state actions.

The \textit{Positional Action Manager} (PAM, in short) is the component devoted to managing the positional actions.
We have fully described PAM in \cite{gossipim, netgames12}.
PAM combines a centralized server and a best-effort mechanism providing support for interest management.
Indeed, PAM employs a combination of a centralized server and epidemic (or gossip) protocols to acquire the position of relevant entities in proximity of the users. PAM exploits two gossip protocols to build an overlay in a completely distributed fashion. The first underlying protocol assures the network to remain connected, and  provides a bootstrap point for the newly arrived users.
The second protocol filters the user according their position in the MMOG, by continuously choosing the best set of users to connect with. This set can be chosen by mean of two different heuristics, one faster and less accurate, and the other one slower but more accurate. Then, following a \textit{wisdom of the crowd} approach, each peer exploits each other knowledge to acquire information about close entities. At fixed intervals, the server provides fresh information to the users.

With the gossip protocols we are able to reduce consistently the load on the central server.
The MMOG operator can tune the PAM by defining the refreshing rate of the server. High rates reduce the cost but also limit the precision of the users' view. On the other hand, low rates increase the cost, and also raise the precision of the results.
PAM has been evaluated with realistic movement traces derived from Second Life \cite{sl-site}. 
The results we have obtained are encouraging. On the average, the accuracy of the players' view in PAM is enough to support a MMOG, and, at the same time, to reduce the load on the server.

The \textit{State Action Manager} (or SAM is short) manages the state actions.
The first version of SAM has been presented in \cite{pos} and in the thesis we present an updated and refined version.
SAM exploits a Distributed Hash Table, equipped with Virtual Servers, to distribute the effort on management of the entities to multiple resources, including user-provided ones.
In the design of SAM, we exploited the knowledge from our prior works on data dissemination in DHTs \cite{carlini2010reducing,carlini2011probabilistic}.
SAM self-adapts to the load of the MMOG, by releasing resources when the load is low and acquiring additional resources when the current infrastructure has not enough capacity to manage the load.
Operators can tune the SAM to decide the maximum amount of entities that can be managed by the user-provided resources. 
SAM recruits on-demand or user-assisted resources accordingly, always trying to minimize the cost and to avoid resources to be overloaded. In order to efficiently provision resources, SAM is designed to adopt prediction mechanisms, so to take provisioning decision well in advance. 

In the last part of the thesis we present a preliminary study on the combination of SAM and PAM in a concrete architecture.
In this context, we focus on the design of a smart client, which allows the players to exploits the advantages of SAM and PAM at the same time. We also propose an enhanced version of the PAM server, which aims to scale up to ten thousands of players, while keeping the economical costs acceptable. 

\paragraph{Thesis structure}
The thesis is structured as follows.
Chapter \ref{chap:background} discusses the challenges and the design issues for modern MMOGs infrastructures. In particular we focus on aspects such as consistency, interactivity, interest management, fault tolerance, load balancing, and security.
Chapter \ref{chap:sam} presents a full description of the State Action Manager, including a in-depth evaluation.
Similarly, the Positional Action Manager and its experimental evaluation are described in Chapter \ref{chap:pam}.
In Chapter \ref{chap:architecture} we provide an insight on the combination of SAM and PAM in a concrete MMOG infrastructure.
An overview and comparison of the approaches in the state of the art is provided in Chapter \ref{chap:related}, which also presents several detailed case studies.
Finally, Chapter \ref{chap:conclusion} concludes the thesis, summarizing the results of the thesis and providing references for future works.

\chapter{Background}
\label{chap:background}

The content of this chapter is divided in two parts.
The first part introduces the basic concepts and terminology that will be used in the rest of the thesis.
The second part provides an overview on the main issues related to distributed architectures for MMOGs\footnote{In literature MMOGs are often referred to as Distributed Virtual Environments (DVEs) or Massively Multi-User Virtual Environments (MMVEs).}.

\section{A model for MMOGs}

A MMOG is a virtual world represented by a collection of entities. 
Entities may be classified as follows.
\textit{Avatars} represent the users in the virtual environment.
Each user commands an avatar, that can travel across the virtual environment and interact with other entities.
\textit{Objects} (e.g. a door) are not directly controlled by the users, but Avatars can interact with them.
A third type of entities are the Non Playing Characters (NPCs).
These entities expose \textit{active} behaviour controlled by the server.
By comparison, objects have a passive behaviour, since they cannot decide to change their state autonomously.

\begin{figure}
\centering
\includegraphics[width=0.8\textwidth]{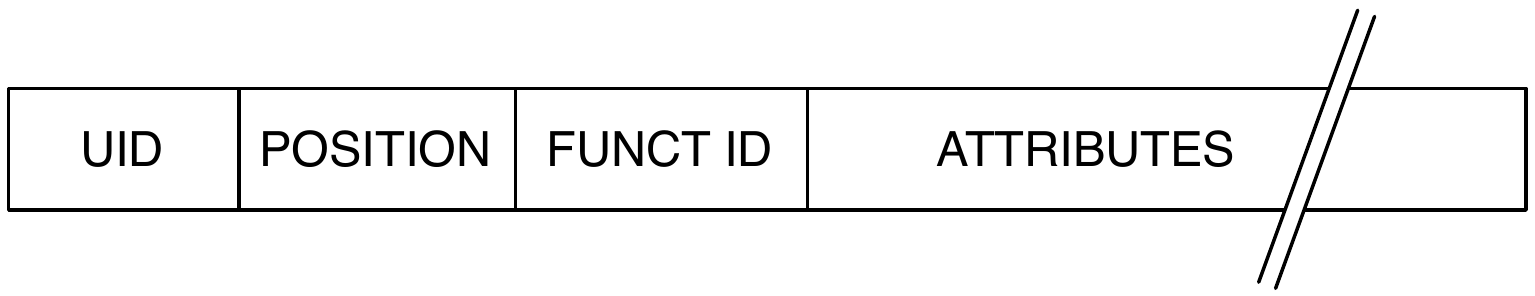}
\caption{The structure of an entity descriptor}\label{fig:descriptor}
\end{figure}

Each entity is represented by an \textit{entity descriptor} (Figure \ref{fig:descriptor}), which contains:
\begin{itemize}
\item an unique identifier (UID);
\item a two-dimensional point, which represents the position of the entity in the virtual environment;
\item where needed, the indication of the procedure to execute for AI-controlled entities (this procedure is sometimes referred to as \textit{think function} \cite{Bharambea});
\item a collection of key-values pairs, which represents the attributes of the entity.
\end{itemize}

%
In order to illustrate some background concepts, let us assume that a central server stores all entity descriptors.
In order to interact with the virtual environment, users connect to the server via a software agent, which we generally refer to as \textit{client}.
The client provides the representation of the virtual environment to the user. It also transforms the action of the user in  messages sent to the server and receives back the modification of the virtual environment and updates. In other words, clients maintain a replica of the descriptors in their local memory that are periodically synchronized with the server.

\begin{figure}[tbh]
\centering
\includegraphics[width=0.8\textwidth]{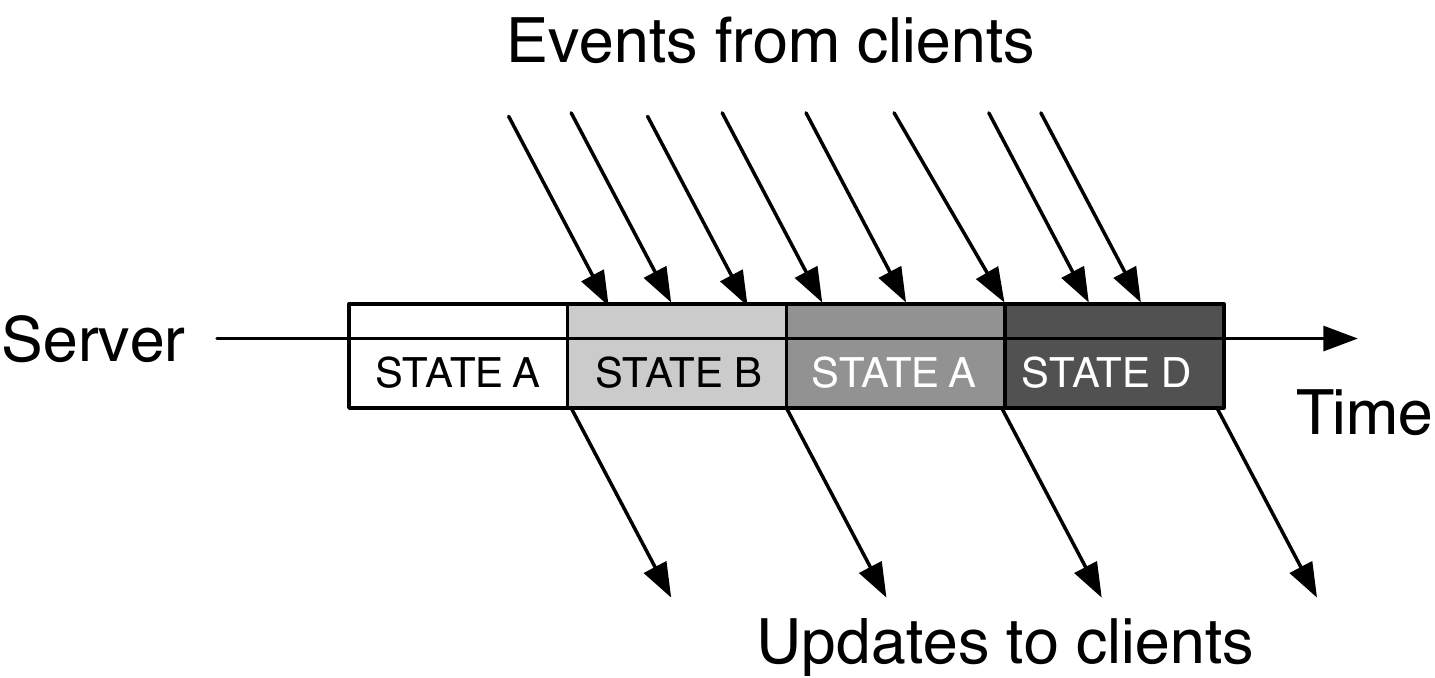}
\caption{Server model}\label{fig:server-model}
\end{figure}

From the point of view of the server, the virtual environment can be modelled as a sequence of states changes over time, in reaction to events issued by clients or NPCs' think functions. This model is depicted in Figure \ref{fig:server-model}.

For the purpose of our work, here we underline a relevant difference in the events (or actions) that server can manage:
\begin{itemize}
\item \textbf{positional actions}. These actions are \textit{single-writer/multiple-reader}, their effect is \textit{volatile} and \textit{error tolerant}.
Since there is a single writer, these actions generate no conflicts to be resolved. However, there is an exception to this, which is the position of the objects. In fact, in this case we have multiple writers and we consider the movement of an object as a state-action.
Positional action are also volatile, meaning that on the exit of the avatar, the actions makes no sense.
Also they are error tolerant, in the sense that a small error in consistency does not compromise the experience for the user.

\item \textbf{state actions}. These actions are \textit{multiple-writers/multiple-readers}, their effect is \textit{persistent} and are \textit{not error tolerant}.
Since these actions are multiple writer, race conditions on entity descriptor may arise. In this case it is a necessity for the server to resolve possible conflicts. The effect of state actions is persistent, i.e. when the last writer leaves the environment, the descriptor must still be available for other possible writers. Since these actions operate on a discrete space (e.g. a door can be opened or closed) there is no room for errors.

\end{itemize}

Normally, the server handles the events in an infinite loop of iteration.
Every iteration has the same finite duration, and at each iteration, the server manages the flow of messages by resolving the possible conflicts on the entities descriptor and by broadcasting the new version of the state.
In this case, it is possible that some clients need to revert the descriptors to a previous state, in order to be synchronized with the server. We discuss more in detail these aspects later in the chapter.

\section{MMOGs: Challenges and Issues}

The definition of a scalable architecture for a MMOG is a complex task.
Due to the high number of users, there is the need to employ multiple servers and to distribute the entities of the MMOGs.
Apart from the strategies for distribution (which we deeply review in the chapter devoted to the related work), a number of issues arises due to the MMOG distribution on more servers.

The communication between clients and servers largely takes place on the Internet. Being a best-effort network, the Internet suffers from unpredictable jitters and delays, that complicates the management of the consistency and interactivity in MMOGs.
In order to have a resource-wise platform for MMOGs, servers send to the user the minimal subset of information necessary, which normally corresponds to the entities users can interact with. However, this apparently easy task may be difficult when entities are spread on multiple servers, as the set of relevant entities may reside on different servers.
If the MMOG platform exploits user-provided resources, backup replicas need to be provided and managed in order to guarantee a certain degree of tolerance to faults.
Further, the load imposed on the platform by users, is not homogeneous. Some servers may be stressed more than others, and in unfortunate situations, an overloaded server may suffer from reduced performance. The exploitation of user-assisted resources amplifies this problem.
Users that participate to MMOGs are often in (real) competition to acquire certain (virtual) privileges in the MMOG. If not properly addressed, this may result in an unauthorized interaction with the system from certain users. This action is typically referred to as \textit{cheating}.
All those issues are described in the following sections.

\subsection{The Consistency-Interactivity Tradeoff}

\textit{Consistency} can be defined as the degree of separation between the descriptors of the users' views in a virtual environment.
In an ideal distributed virtual environments, participants share the same version of the descriptors, resulting in perfect consistency.
This would be possible only by employing a network with a latency close to 0ms, like in case of LANs.
However, most of the MMOGs users are geographically widespread and the messages travel on the Internet.
Due to its best-effort nature, messages have delays which depend on multiple factors and are different from user to user.
%
%
On the other hand, \textit{Interactivity} is a measure of the responsiveness of the virtual consistency, i.e. how fast an action from an avatar produces changes in the virtual environment. As for consistency, interactivity is affected by network latency as well.

\begin{figure}[tbh]
\centering
\includegraphics[width=0.8\textwidth]{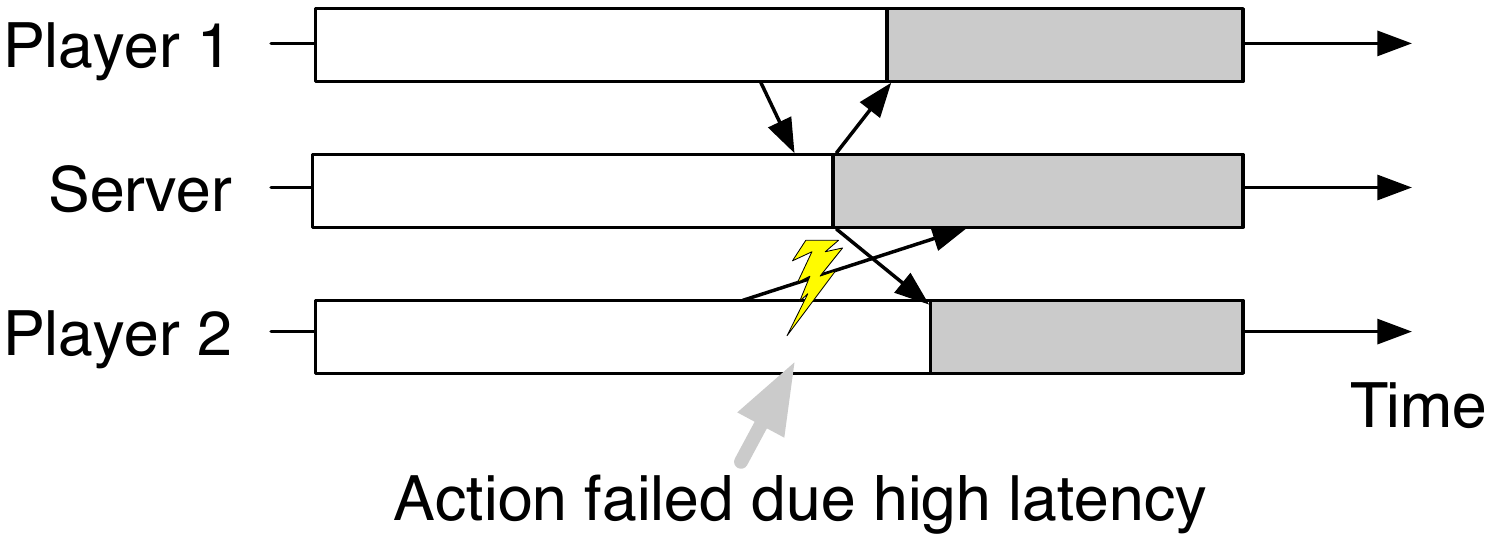}
\caption{Pessimistic consistency management}\label{fig:pessimistic}
\end{figure}

\begin{figure}[tbh]
\centering
\includegraphics[width=0.8\textwidth]{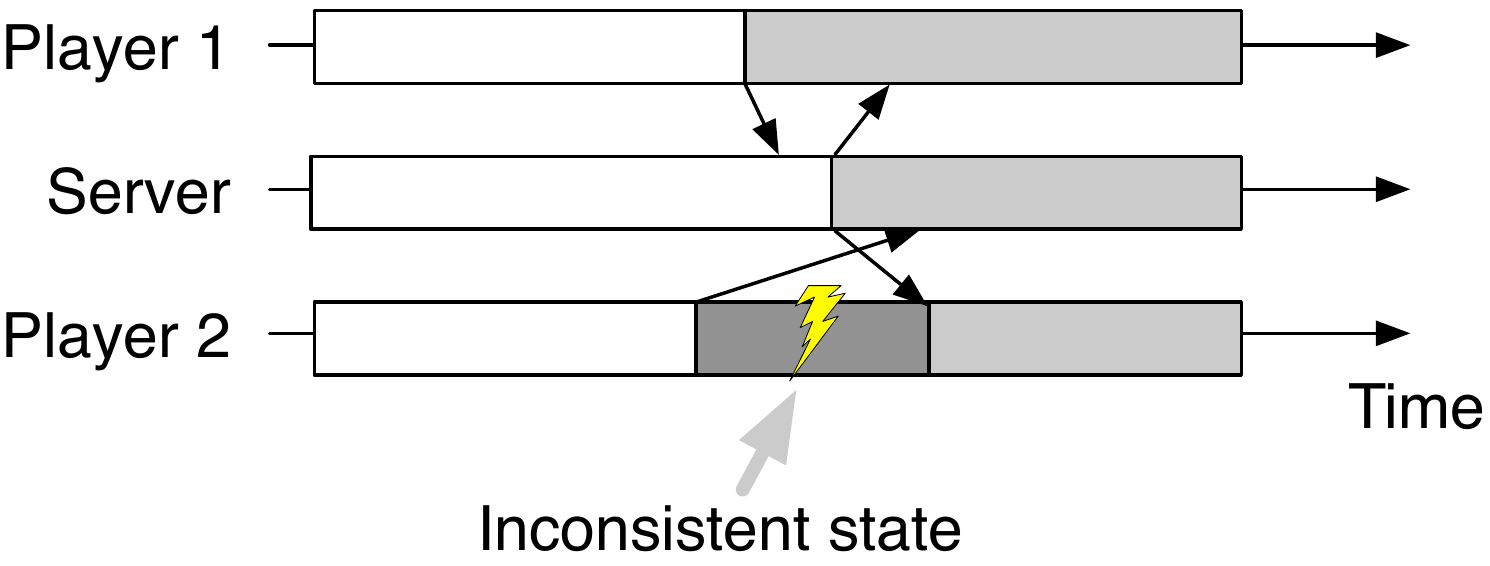}
\caption{Optimistic consistency management}\label{fig:optimistic}
\end{figure}

In a way, consistency and interactivity are two sides of the same coin, as one can be traded to favour the other one.
Approaches that trade interactivity for consistency are referred to as \textit{pessimistic}. 
In pessimistic approaches, an action must be validated by the server before clients perceive the results.
Hence, a non-zero amount of time passes from the issuing of the action until the results are visible to the issuer. 
Normally, to mask this problem an artificial delays are introduced in the MMOG. For instance, for a player to open a door takes several seconds. With those delays there is the time for the message to arrive to the server and for the new state to come back to clients.

It may happen that a state of an entity is the subject of a race condition among multiple players. Let us consider Figure  \ref{fig:pessimistic} as example. In this case, player one and two issue a conflicting action at the same time on their local machines. However, due to the network latency, the request of player one arrives first. Player one then would see the environment changing as expected, whereas player two would have his action neglected.
Hence, a straightforward application of this model disadvantages client with slow connections, as they might regularly lost race conditions.

On the other hand, \textit{optimistic} approaches trade consistency for interactivity. 
When a user issues an action, its client immediately modifies the local state to reflect the actions results.
By comparison with pessimistic approaches, there is no detectable time between the issuing of an actions and its results on the client.
Just after having modified the local state, the client sends the action to the server.
The server computes the new state and broadcast it to the other players.
Note that, due to the latency of the network, there is a non-zero amount of time during which the state of the client is not synchronized with the server.
In case of a race condition, eventually one of the players would suffer from an inconsistent state (see Figure \ref{fig:optimistic}).
In this case, upon the communication of the correct state from the server, she rollsx-back to a consistent state.

LocalLag \cite{Mauve2004} is a representative example of pessimistic consistency management.
LocalLag's goal is to reduce the disadvantages of clients having a slow connection toward the server.
LocalLag delays actions for a certain amount of time before their execution.
This additional delay allows the distribution of events to the clients such that delay due to latency is compensating by the
local-lag. 

Conversely, Dead Reckoning (DR, \cite{pantel2002suitability}) is an optimistic mechanism.
In DR, the server computes the direction and the speed of the entities according to their past movements.
DR allows the server to feed the clients with a stable rate of entities movements, that greatly increasing interactivity.   
DR also reduces the bandwidth consumption of the clients, as they communicate to the server only when a change in speed or direction of their avatars occurs.
However, DR may generate movements artefacts on clients, in case the predicted positions diverge from the actual ones, which triggers a server roll-back.

The most opportune tuning and combination of optimistic and pessimistic approaches depends on the MMOG particular genre. Fast-paced MMOGs would favour an optimistic approach over a pessimistic one, at the expense of occasional rollbacks.
Slow-paced MMOGs would employ pessimistic approaches, masking with game mechanics the delay of issuing actions.
Apart from the genre, we argue that one or the other approach can be used on the same MMOG but applied to different kind of actions.
Optimistic approaches can deal with positional actions, because of their tolerance to errors and they high frequency. On the other hand, pessimistic action can manage state actions, for their lesser frequency and no tolerance to errors.

\subsection{Interest Management}

A common optimization in MMOGs architecture is to communicate to the clients only a minimal, but sufficient, set of relevant information.
This operation is called  \textit{Interest Management} (IM, \cite{morse2000interest}).
Exploiting the concept of IM drastically reduces the size of messages sent by the server to the client. 
From an abstract point of view, IM can be modelled as a publish/subscribe service, in which users have role both of publisher and subscribers. 
In fact, users: (i) publish new entity descriptors by interacting with the virtual environment and (ii) subscribe to the relevant entities and receive updates when their descriptors change.

One of the most effective and straightforward strategies to define the set of a user's relevant entities, is to consider the entities in the spatial proximity of the avatar.
This subset is generally modelled according to the so called \textit{focus-nimbus} model.
In this model, the \textit{focus} refers to the area of visualization of an avatar, while the \textit{nimbus} is the area where an avatar can be seen.
In other words, an avatar $a$ is aware of an avatar $b$ when the nimbus of $a$ intersects with the focus of $b$. 
The shape and the size of the nimbus and focus are MMOG dependent.
In the simplest (yet effective) case, both nimbus and focus are represented by the same circle whose center is the position of avatar in the virtual world.
Generally, this region is called \textit{Area-of-Interest} (AOI, \cite{el2006aoim}).

IM is applied in centralized systems as well, but it acquires even more importance in a case of a distributed MMOG infrastructure. 
Since relevant entities may be spread in different nodes, subscribing to a reduced set of entities also decreases the number of nodes to query when performing IM.
In order to clarify this point, let us consider an example.
Let us assume a virtual environment whose infrastructure is composed by multiple servers, which we refer to as $S_1$ $S_2$ and $S_3$.
Each of the servers manages a contiguous non-overlapping area of the MMOG, called \textit{region}  (see \textit{zoning} in Chapter \ref{chap:related} for a full description of this distribution schema).
Let us also consider a generic client $C_1$, whose Avatar is in the region managed by $S_1$ but its AOI overlaps the additional two servers, $S_2$ $S_3$. 
The first step in order to perform IM, is to find out what are the entities in $C_1$'s AOI.
This step is commonly called as \textit{Neighbour Discovery} (ND).
In order to optimize ND, servers are generally connected through spatial-based overlay (e.g. Voronoi-based) so that servers managing close regions in the MMOGs are connected to each others. 
Once ND is completed, $C_1$ subscribes  (or is subscribed by the servers) to the entities in its AOI.
We refer to this subscription as \textit{State Management} (SM), since from now on, $C_1$ receives updates from the entities.

\begin{figure}[tbh]
\centering
\includegraphics[width=0.8\textwidth]{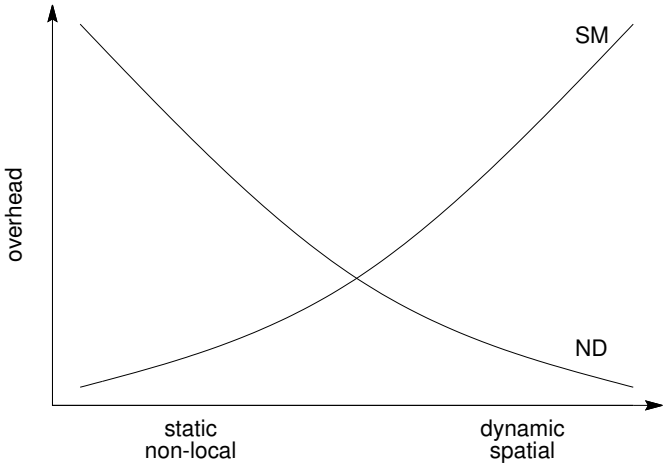}
\caption{The Interest Management tradeoff}\label{fig:im-tradeoff}
\end{figure}

In a centralized architecture, the distinction between ND and SM is blurry, as they are executed at the same time.
When this model is implemented in a classical distributed infrastructure, it presents several issues.
Since each server manages a region, whenever an entity moves from one region to another, its descriptor must be transferred between the servers.
We refer to this as \textit{ownership changing}. 
Ownership changing can create some problems to the users, since a migrating entity cannot be accessed. 
For that reason, it is important to design a MMOG infrastructure so to limit the amount and the duration of ownership changing.

One way to control the amount of ownership changing is to properly tune the size of the regions.
The bigger (smaller) a region is, the less (more) is the probability to trigger an ownership changing. 
However, having the biggest region possible is not the solution. 
Dividing a MMOG into regions helps with the scalability of managing many entities.
Having too large regions may overload some servers, in fact making the situation worse.
Similar issues are in play also when considering a live resizing of the regions.
Indeed, resizing a region would implied entities to move from one server to another.

To deal with the ownership changing, one can be tempted to assign to a sever a bunch of entities not corresponding to a contiguous region of the virtual world.
In this case the problem is completely avoided, since entities are statically assigned to the servers.
However, the immediate drawback would be the complexity of ND: an AOI can, in general, contain entities in any server.
Let us return back to our previous example. 
In the worst case it would be necessary to contact all the servers to discover that $S_1$, $S_2$ and $S_3$ contain the entities in $C_1$'s AOI. 
Clearly, a totally random entities assignment would impose scalability limitations, basically invalidating the benefits from the distribution of the virtual world.

In summary, a tradeoff between region-based and random-based is necessary when dealing with ownership changing.
From an abstract point of view, this tradeoff can be represented as in Figure \ref{fig:im-tradeoff}.
As the locality of the approach increases, the overhead for ND decreases but the overhead for SM increases.
On the contrary, with approaches based on random assignment, the overhead for the ND increases whereas the overhead for SM decreases. 
Typical solutions on the state of the art maintains a locality-based approach while trying to reduce the overhead of the SM (e.g. by using some overlapping among regions to reduce the overhead of ownership changing). However, this does not resolve totally the problem of the ownership changing, and also the problem of determining the size of the regions remains.

\subsection{Fault Tolerance}

One of the main problems of distributed architecture for MMOGs is to cope with unexpected departures of servers.
This problem is of particular relevance when dealing with user-provided resources.
%
%
The unexpected departure of nodes may compromise two important parts of a MMOG infrastructure: (i) the servers overlay (i.e. the network between the servers) and (ii) the management of the state of the objects.

Regarding the overlay, using approaches derived by structured P2P networks is a great advantage with respect to use customs overlays.
Normally, structured P2P-based overlays  employ self-repairing mechanisms that provide robustness even if an high fraction of nodes leave the network. DHT-based overlays have been proved to be resilient to node failures even when a relatively large fraction of the nodes fail at the same time \cite{rowstron2001pastry,kuhn2005self}. In addition, DHT-based approaches often provide stabilization mechanisms that helps in self-repairing the overlay.
For example, the Chord DHT \cite{Stoica2003} employs a solution based on \textit{reverse finger tables} \cite{chen2008towards}. 
A reverser finger table is an additional routing table maintained by each node. It contains a reference to the nodes that contain the table owner in their finger table. This table can be used to inform all the nodes whenever  a change happens in the overlay topology. 
Another example is represented by the Delaunay overlays  built on the bases of Voronoi tessellations \cite{Aurenhammer}. When a peer crashes, the Voronoi tessellation is recomputed and a different Dealunay overlay emerges.
Several works, for example \cite{baragliagodel}, employ approaches that eventually lead to a consistent overlay. 
In these approaches, upon peer failures, affected nodes locally re-compute the overlay.
These approaches are lightweight in terms of the number of messages, at the cost of occasional inconsistencies among nodes view. Other approaches, such as \cite{liebeherr2002application}, employ an higher number of messages to reach a consensus on the tessellation among the nodes involved.
In conclusion, P2P-based failure tolerance mechanisms provide the interesting propriety of the self-adaptation.
In addition, when a peer fails, the modification on the overlay are local to the node involved, causing minimal disruption in the overlay.

Regarding the management of the objects, the problem is more complex, as the mechanisms to assure the availability and the reliability of the information are normally application dependent.
In a MMOG infrastructure without any fault tolerance mechanism, if a node manages an object and such node departs, the descriptor of such object is lost. 
The mechanisms to assure failure tolerance are based on replication of the descriptors in other nodes. These mechanisms can be classified according to the destination of the replicas. In fact we can distinguish two different cases: \textit{locality-based} replicas and \textit{region-based} replicas.

In locality-based replicas, a server maintains replicas in "closer" servers. Here the definition of closeness is the central point.  
Some approaches considers \textit{overlay proximity}, where the replicas are placed in the connected nodes of the overlay. This is the typical case for DHT-based solution like \cite{Knutsson2004,holzapfel2011vorostore}, where the entities assigned to a server are replicated in one or more successors of the server in the DHT ring. Other approaches consider the \textit{virtual environment proximity}, i.e. a server replicates its entities in one or more servers which manage adjacent regions. This approach is typical of solutions that divide the virtual environment in regions, such as in \cite{Frey2008}.
Locality-based replication schemas offer a good degree fault tolerance, however the overhead of synchronize an high number of replicas may be relevant in some situations.

Conversely, region-based replication schemas assign the replicas to a single or multiple backup nodes. These backup nodes share no concept of proximity with the primary server. Whenever a server fails, one of the backup nodes is instructed to take the place of the failed server.  This approach is largely used, such as in \cite{hu2008voronoi,Chen,Kim2004}, for essentially two reasons.
First, it is possible to decide a priori the number of replicas to provide, in fact imposing an overhead for the management of replicas.
Second, it is possible to select cloud machine as backup nodes, and to use them only when really necessary.

In conclusion, for the object management it is important to have a replication system, which must be scalable and whose overhead is controllable. Regarding the overlay, the failure tolerance mechanisms chosen largely depends on the specific P2P system adopted. Without doubts, choosing a well tested and studied P2P-based overlay may turn out to be a great advantage. In any case, fault tolerance mechanisms are an essential component of infrastructures dealing with user-provided resources.

\subsection{Cheating}

In virtual environments, one of the most important aspect of security is called cheating. Cheating is defined in \cite{Neumann2007} as "an unauthorized interaction with the system aimed at offering an advantage to the cheater".
In a MMOG infrastructure, to provide a secure and fair environment is a vital task.
MMOGs gather together users from all over the world that are potentially untrustworthy to each other.
Since MMOGs businesses model is largely related to number of the users, failing to maintain a secure and fair (thus enjoyable) environment is an unfortunate event.
Note that here we refer only to cheating. Principle like confidentiality, integrity and authenticity are general security aspects that are outside of the scope of this section.
Webb et al. \cite{Webb2007} classified cheating as follows:
\begin{itemize}
\item \textit{Game level cheating} the cheater uses bugs or other game mechanics not working as intended to gain an unfair advantage;
\item \textit{Application level cheating} the cheater modifies the client to access the memory and/or sending invalid messages;
\item \textit{Protocol level cheating} the cheater interferes with message packets and/or the network protocols;
\item \textit{Infrastructure level cheating} the cheater interferes with local software library (e.g. display drivers) or the network infrastructure (e.g. spoofing).
\end{itemize}

\begin{figure}
\centering
\includegraphics[width=0.8\textwidth]{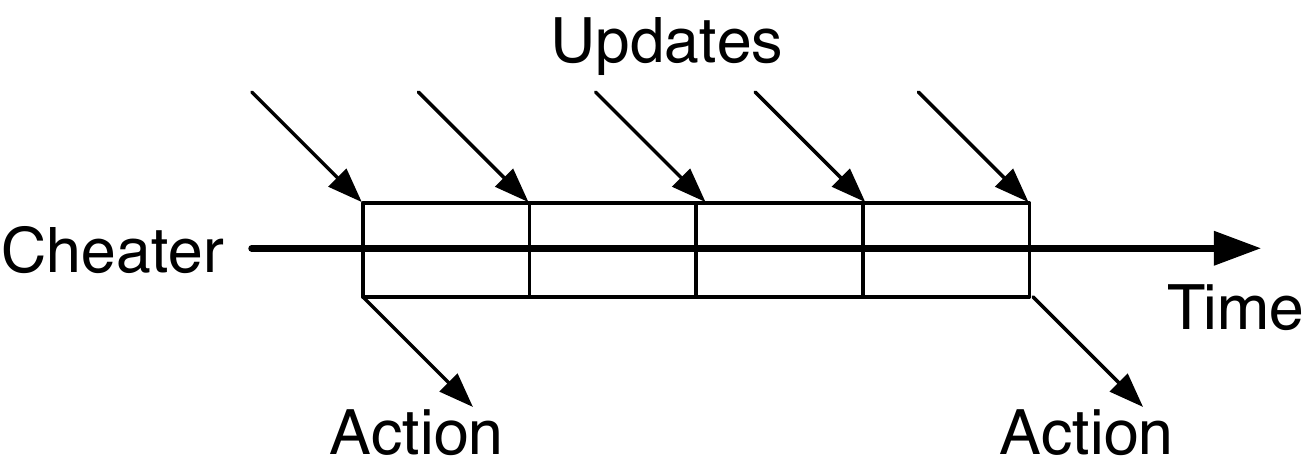}
\caption{Cheating by suppressed update}\label{fig:cheating}
\end{figure}

The most common cheating methods fall into the application and protocol levels.
For example, one of the most common cheating method in the protocol level is the so called \textit{suppressed update} \cite{Webb2007}. Figure \ref{fig:cheating} gives a visual representation of this method from the cheater's point of view.

This cheating mechanism leans on the assumption that most servers exploit Dead Reckoning or similar mechanisms to assure smooth movements to the player, while reducing the load on the server (see previous section for more details).
These mechanisms allow the player to skip up to $m$ messages before considered as disconnected.
To exploit this situation, cheaters avoid to send up to $m-1$ messages in a row, but still listen to the updates. At the $m$th update, cheaters send to the server the best action according to the other player moves. This cheat is resolved in \cite{Cronin} by making mandatory to the client use the state of the server. Similar examples of cheating are to intentionally introducing delays to the packets or to change the timestamps of the actions to fool the server.

The mechanisms to contrast the cheating are called Anti-Cheating (AC).
The related work (Chapter \ref{chap:related}) provides a review of the state of the art for AC mechanisms in distributed and P2P-based MMOG infrastructures.

\subsection{Load Balancing}
\label{balancing}

The management of avatars and objects constitutes the typical computational and bandwidth load of a MMOG.
Avatars move across the virtual environment and interact with each other. This  interaction can be \textit{direct} or \textit{indirect}. In the former case, an avatar directly modifies the state of another one, while the latter case is that of an avatar modifying the state of a passive object so affecting the behaviour or the state of other avatars. Both direct and indirect interactions consume resources on the nodes, in terms of bandwidth and computational power. Such load is assumed to grow exponentially with the number of avatars, with a quadratic or cubic trend according to the game genre \cite{Nae2008}.

Currently, there is no an uniform view on how to measure the load on a MMOG. 
Several approaches \cite{Lee2003,Rieche2008,Ahmed2008b}, including our work \cite{pos}, consider the bandwidth consumption (or equivalently, the message rate) of the server as the measure for the load. 
The works \cite{Lee,Chertov2006} consider the number of connected players as representative for the load.
Other works, such as \cite{Deng1995}, consider the number of entities managed the server. 
Even though the way of measure the load are different, all these approaches strive to avoid overloaded servers.

An overloaded server may deliver with delay the events to the players.
This situation creates visual artefacts and slowdowns on the players, in fact compromising players experience.
An infrastructure where the load is balanced among the servers reduces the possibility of overloaded servers.
However, in a distributed MMOG the load is normally not balanced.
Each server manages a region of the virtual environment, which contains a certain number of entities. 
These entities (avatars and objects) are not uniformly distributed.
Several areas of the MMOGs, in particular the ones corresponding to places of interest, may present a higher density of entities.
These areas, called \textit{hotspots}, are the principal reason of the load unbalancing in virtual environment.

The distribution of the load in a MMOG architecture offers unquestionable advantages and remains an important research issue in these architectures.

\section{Conclusion}
The content of this chapter represents a reference point for the other chapters of the thesis.
We have presented an overview on the basic concepts and issues in the design of distributed infrastructures for MMOGs.
These concepts and issues will be elaborated in the remaining of the thesis according to the particular contexts.

\chapter{State Action Manager}
\label{chap:sam}

This chapter describes in detail the State Action Manager (SAM, in short), a component for MMOG infrastructures that is devoted to the management of the state actions.
This chapter is a revised and extended version of the work we have presented in \cite{pos}.
In the design of SAM, we exploited the knowledge from our prior works on data dissemination in distributed peer-to-peer architectures \cite{carlini2011probabilistic, carlini2010reducing}.
The main goal of SAM is to manage the state actions by orchestrating a seamless combination of user-provided and on-demand resources.
The SAM realizes this task by considering the differences between the two kind of resources. On one hand we have the reliable, powerful and costly on-demand resources, on the other the unreliable, heterogeneous and free user-provided resources. This creates a tradeoff between the cost and the reliability of the platform. The SAM gives the operator the possibility of tuning the platform, in order to control this tradeoff.

From an architectural stand point, in SAM there are four distinct software modules. Each of these modules computes a logical function in the platform.
\begin{itemize}
\item \textit{Client;} This component manages the connections with the servers and visualize them to the player. It is executed by the user-provided resources only;
\item \textit{Server;} This component connects to the SAM DHT. It manages the state actions coming from clients. It resolves possible conflicts and broadcasts the updates to the interested clients. It can be executed both on on-demand and user-provided resources;
\item \textit{Backup server;} The component handles the entities that belong to a failed server. The actions are essentially the same of a normal server. It can be executed only by on-demand resources;
\item \textit{Manager;} This is the central component of the SAM. It orchestrates the combination of user-provided and on-demand resources. In particular, it decides whether and when a particular resource should cover the role of the server. It can be executed by a trusted on-demand resource or can be distributed among multiple on-demand resources.
\end{itemize}

The rest of the chapter is organized as follows.
Section \ref{sec:sam-architecture} presents an overview on the SAM architecture, including the fault tolerance mechanism. 
Section \ref{sec:migration} provides a detailed description and analysis of the entities migration between servers.
We provide a formalization of the entities assignment in Section \ref{sec:problem}.
Section \ref{sec:balancer} focuses on the balancer, which is the component that actively performs the management of the resources.
Section \ref{sec:result} presents an experimental evaluation of the balancer under different scenarios.
Finally, section \ref{sec:conclusion} concludes the chapter.

\section{SAM Architecture}
\label{sec:sam-architecture}

\begin{figure}
\centering
\includegraphics[width=0.8\textwidth]{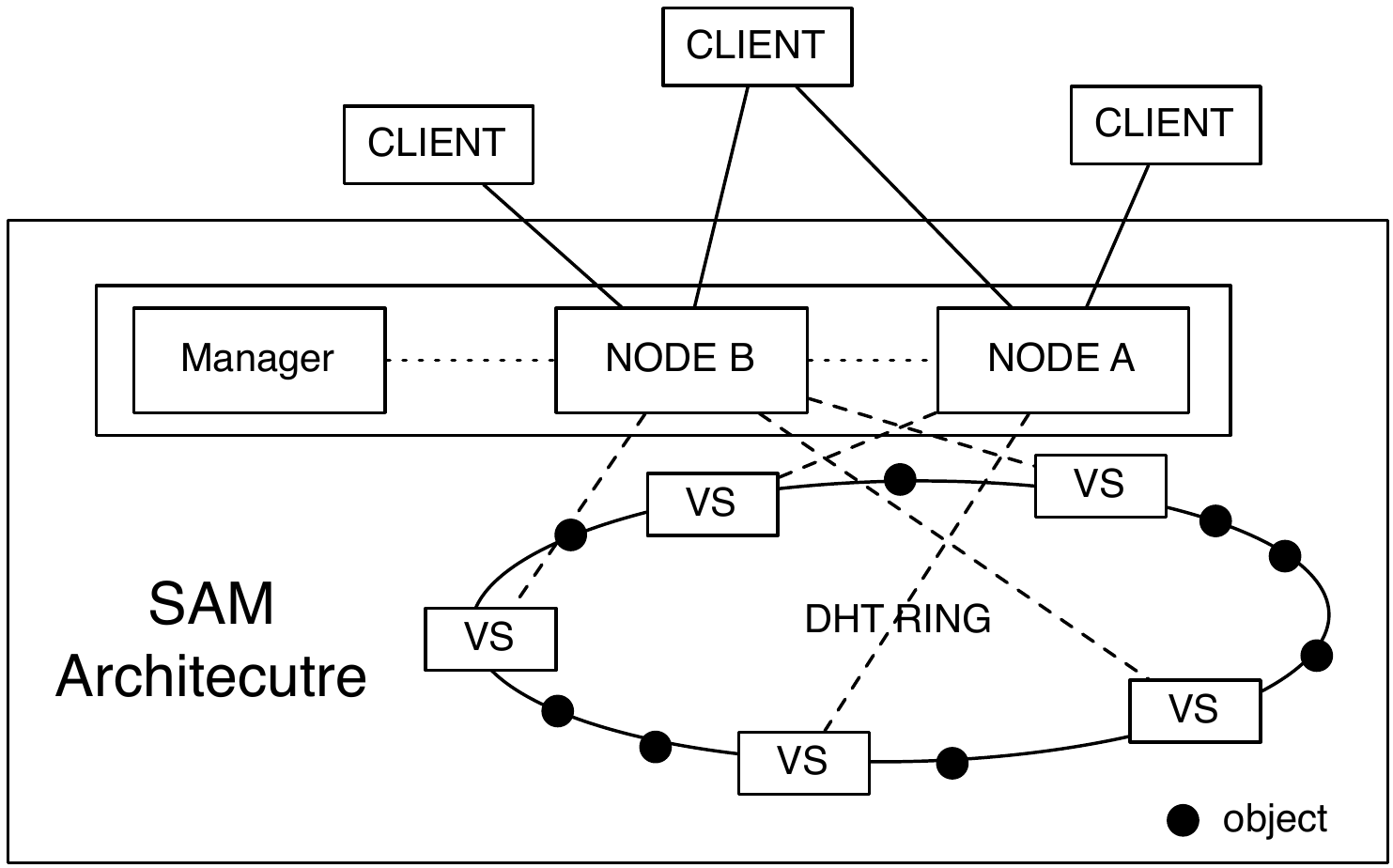}
\caption{Black dots are the objects inside the virtual environment. VS are the Virtual Servers. Node A manages 2 VSs, whereas Node B manages 3 VSs. Client connects to the nodes to modify and read the objects. The manager has a global knowledge of the state of the node and the VSs.}\label{fig:sam-arch}
\end{figure}

Our proposed architecture (see Figure \ref{fig:sam-arch}), exploits the underlying mechanisms of Distributed Hash Tables (DHTs, \cite{Stoica2003, Steinmetz}) in order to build and maintain an overlay for the management of the entities state in the VE.

A typical DHT considers a ring-shaped logical address space (e.g. Chord \cite{Stoica2003} considers a space of $2^{160}$).
Such space is divided among the nodes participating to the DHT.
In fact, each node receives an ID in the logical address space through the application of an hash function.
The nodes are so placed on the DHT, and a generic node manages the address space "close" to itself.
The definition of closeness varies from DHT to DHT. 
For example, in Chord a node manages the address space that goes from its predecessor to itself.
Nodes of the DHT are connected in an overlay, which permits the routing of the messages among the DHT nodes.
The overlay is build to guarantee $O(logN)$ bounds, where $N$ is the number of
nodes of the DHT, both on the routing hops to deliver messages and on the size of the routing tables.

In the proposed architecture, the entity descriptors that compose the virtual environment are placed in the DHT.
Each descriptor receives an ID in the DHT address space through the application of a hash function (i.e. SHA-1) on its initial content.
The application of the SHA1 guarantees IDs to be evenly distributed across the space. 
In addition to the typical DHT mechanisms, we exploit a Virtual Servers (VSs) \cite{Godfrey2004} layer over the DHT.
The Virtual Server approach introduces a clear separation between the logical and the physical nodes.

Each VS is in charge of an address range of the DHT. However, VSs are not permanently paired with the same physical node, and each physical node can host several VSs. Even if this approach presents higher implementation complexity as nodes need to manage multiple VSs, the VSs approach has some evident advantages: (i) more powerful nodes may receive an higher number of VSs than less powerful ones, (ii) heavy loaded nodes may trade VSs with unloaded ones, (iii) in case of a physical node failure, its VSs are possibly transferred/reassigned to different, unloaded, physical nodes, so reducing the risk of overloaded nodes.
Even if the association entities-to-VS remain fixed during the system lifetime, the mapping of VS to physical nodes may change over time. For instance, all VSs may be mapped to cloud nodes at bootstrap time and then transferred to peer nodes afterwards.
In SAM, migrating VS is easy and light. 
Their migration does not affect the organization of the address space at the DHT level.
It only requires the exchange of data managed by the VS as well as the update of the mapping between the logical identifier of the VS and the physical address of the node hosting it.

From a client perspective, virtual servers act as state servers for a set of entities.
Since the relevant entities for a client may be managed in principle by different VSs, each client may have multiple simultaneous connection to different nodes. 
For instance, in Figure \ref{fig:sam-arch}, a client is connected with node A and B at the same time. As a limit situation, each player can connect to a different node per each object, so that the number of connections for each player is bounded by the amount of entities in its AOI.

In our architecture, we consider the load of a VS as the upload bandwidth consumed to broadcast entities state to the clients.
The load depends on the amount of entities that correspond to the VS and the amount of client accessing to them.
The load is changing over time, according on the interaction pattern of the avatars.
The load may be unbalanced due to the presence of more popular entities.
For instance, objects belonging to an hotspot may receive an higher amount of updates.

SAM relies on the \textit{manager} to coordinate the transfer of VSs among the
nodes of the DHT. An important issue is the choice of the nodes actually playing the role of the manager. The simplest
solution is to define a centralized manager executed by a cloud node which may be
either dedicated to this task or share server tasks. 
Nodes of the DHT periodically notifies the central distributor.
The manager periodically computes new assignments node-VSs based on the received information and, if necessary, the enrolment or the disposal of nodes from the DHT.

However, a distributed solution is feasible as well. For instance, it is possible to adopt  
a mechanism similar to the one presented in \cite{Rao2003}. A number $d$ of DHT addresses are chosen  and the nodes handling one of these addresses play the role of \textit{sub-managers}. Each SAM node then chooses randomly, but once for all, one of the $d$ sub-manager. Each sub-manager operates like the centralized manager, but on a reduced number of VSs. Although this mechanism distributes the burden of the manager among multiple nodes, it might also impair the result of the assignments. We leave the analysis of the distributed manager  as a future work.

\subsection{Replication and Fault Tolerance}
\label{replication}

In a distributed system, the need of replication comes from the intrinsic unreliability of nodes. 
Since we target an heterogeneous system including both peer and cloud nodes, a fair orchestration of replication is a relevant issue.
Our approach is based on the reasonable assumption that, in general, cloud nodes can be considered \textit{reliable} whereas peer nodes are \textit{unreliable}, due to the high degree of churn which characterizes P2P systems.
This difference is mainly due to the lack of control over peers, which are prone to unexpected failures, and may leave the system 
abruptly. On the other hand, cloud nodes generally belong to a stable infrastructure based on virtualization, and this greatly increases their robustness and flexibility. 
In order to cope with the unreliability of peers, we propose that every VS assigned to a peer is always specially replicated.
The replica, called \textit{backup Virtual Server} (bVS), is then assigned to a trusted resource, i.e. a cloud node. 
To keep the state of the bVS up-to-date with the original, peers send periodic updates to the cloud nodes.
This periodic updates adds a further bandwidth requirements.
However, the synchronizing is done with long period (e.g. 30 seconds) so to reduce the bandwidth requirement.
The replica schema adopted is \textit{optimistic} \cite{saito2005optimistic}, i.e. players can access to entities without a priori synchronization between the regular VS and the relative bVS.
This schema leads to \textit{eventual consistency}, favoring availability over consistency of the entities.

The presence of bVSs guarantees a certain degree of availability in case of peer failures. 
Let us assume the peer $P$ to manage a single VS and that the respective bVS is managed by the cloud node $C$.
Let us also assume that $P$ departs, either abruptly or gracefully, from the system. 
In this case, $C$ becomes the manager of the primary replica, in place of $P$.
As consequence, users connected to $P$ must then connect to $C$. 
In the case of a gracefully departure of $P$, $P$ itself may inform all the users about the new role of $C$. 
On the other hand, in case of unexpected departure, the involuntary departure of $P$ can be detected either by $C$, since it receives no more updates from $P$, or from the DHT neighbours of $P$, due to the repairing mechanism of DHTs. 
These nodes are able to notify the clients to send their notification to $C$.

\section{Virtual Server}
\label{sec:migration}

One of the main advantages on having a virtual server enabled DHT is the possibility to easily move entities across the nodes of the DHTs.
This ability is a fundamental requisite for enabling pro-active load distribution mechanisms.
To better understand the advantages on exploiting virtual servers, let us spend a few words on the load distribution in classical DHTs (i.e. that does not employ virtual servers).
In classical DHTs there are essentially two ways to dynamically distribute the load:

\begin{enumerate}
\item \textit{Move nodes}. An unloaded node (i.e. $A$) joins a precise address of the DHT, so to unload a heavy loaded node (i.e. $B$). This operation requires $A$ to leave the DHT and rejoin in a position so that part of the load form $B$ is transferred to $A$. Even if this approach may work in a general situation, it is too time consuming and creates too overhead for a live application as a virtual environment. To fully understand the process, let us consider $C$ as the successor of $A$ (i.e. the node that is after $A$ in the ring-shaped space of the DHT)\footnote{We consider Chord in this example, but with small differences the following considerations are valid for all DHTs.}. When $A$ leaves, $C$ becomes responsible of the address space left free by $A$. This information must be spread in the DHT, so that the routing for the former $A$ address space points correctly to $C$. 
In addition, before leaving, $A$ must transmit the descriptors to $C$.
When $A$ joins the DHT and becomes the predecessor of $B$, this information must be spread to the DHT to adjust routing path. $B$ also must send to $A$ the entity descriptors that are in the new address space of $A$. In addition, $A$ must build its routing table, in order to be part of the overlay.
In summary, this process requires two entities transferring (from $A$ to $C$ and from $B$ to $A$), to spread new information about 3 nodes and to build a new routing table.
All these operation take time, and, mostly important, imply a large number of transferring during which the entities are not reachable from clients.

\item \textit{Move descriptors}. This technique requires moving the entity descriptors among node to distribute the load. Practically, a moved entity descriptor changes its ID in the ring-shaped address of the DHT. During the transfer of the descriptor, the entity is not accessible by clients. However, the most relevant drawback of such approach is that any time a client accesses to a new entity, it must query the DHT for its position. This requires to wait up to $\log N$ steps, which may be too long with an high number of nodes.
%

\end{enumerate}

With the virtual server, load distribution is lighter and more flexible with respect to a classical DHT.
Nodes directly exchange virtual servers (we refer to this action as \textit{virtual server migration}), which offers the following advantages:

\begin{itemize}
\item the ID of the entity does not change during time;
\item It is possible to transfer load without nodes leaving the DHT;
\item A moved virtual server does not have to rebuilt its entire routing table. In fact, moving a virtual server requires only to stabilize few routing paths, which is less clumsy than in a classical DHT system;
\item It is possible to partially increase or decrease the load of a node.
\end{itemize}

During a VS migration, entities of the VS cannot be accessed. 
In other words, players cannot interact with the object inside the VS that is migrating.
Also, it can be the case of a player modifying the state of the object locally, just to see it reverted back when the migration of the VS is completed.
To this end, it is important to keep the transition time as short as possible, in order to provide an acceptable level of interactivity for the MMOG clients. 
In the next sections we describe in detail the process of virtual server migration and we empirically evaluate the size of a virtual server and the time for the migration.

\paragraph{Virtual Server Migration}

In order to clearly present the migration procedure, let us consider the following example. 
Let us suppose that a virtual server V migrates from a source node A to a destination node B. 
The actions involved (presented in the sequential diagram in Figure \ref{fig:vs-transfer}) are the following:

\begin{figure}[tbh]
\centering
\includegraphics[width=0.80\textwidth]{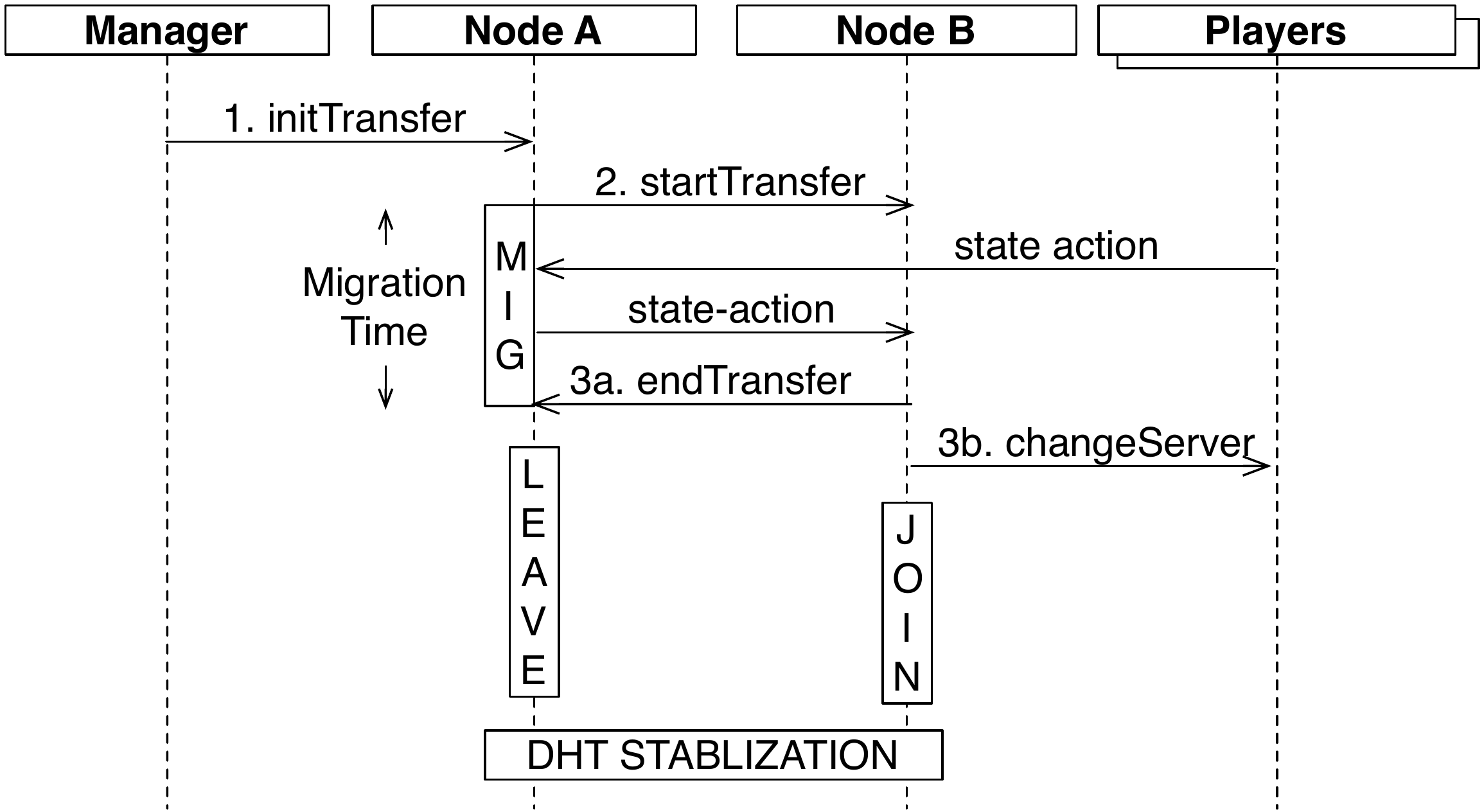}
\caption{Migration of a VS from the node A to node B}\label{fig:vs-transfer}
\end{figure}

\begin{enumerate}
\item The manager notifies to node $A$ a reference to $V$ and the address of recipient node $B$.
\item $A$ sends $V$ to $B$, together with the list of users connected to $V$. 
In the transient time that is needed to complete the transfer, players still send entity update messages to $A$, 
which in turn forwards them to $B$. Note that in this transient period, objects may go 
out-of-sync and, as a consequence, players may perceive some visual inconsistencies.

\item Once received the message, node $B$ notifies the players it has became the manager of $V$.
From this point on, players are able to modify $V$'s objects state.  However, the routing tables of the DHT have to be updated to assure correct routing resolutions.

\item To this end, $V$ executes a \textit{join} operation having $B$ as target in order to update its references in the DHT.
This operation updates the routing table of the node that are in the path from $V$ to $B$, still leaving dangling references to $A$ as the manager of $V$.  To make consistent all references, the stabilization process of the DHT is executed.
\item Finally, a \textit{leave} operation is executed by $V$ on $A$ in order to complete the process.
\end{enumerate}

To understand the impact of the VS migration, we have to consider it from a user's perspective.
From a user point of view, when a VS is migrating the entities in the VS are not accessible.
Longer the time a migration takes, the more the time an user perceive the state of the objects as frozen.
Hence, it is important to measure and evaluate the \textit{migration time} (MT).
The MT goes from the moment in which the server $A$ begins the migration, until the virtual server has been fully copied in $B$.

\paragraph{Migration Time}

\begin{figure}[tbh]
\centering
\includegraphics[width=0.8\textwidth]{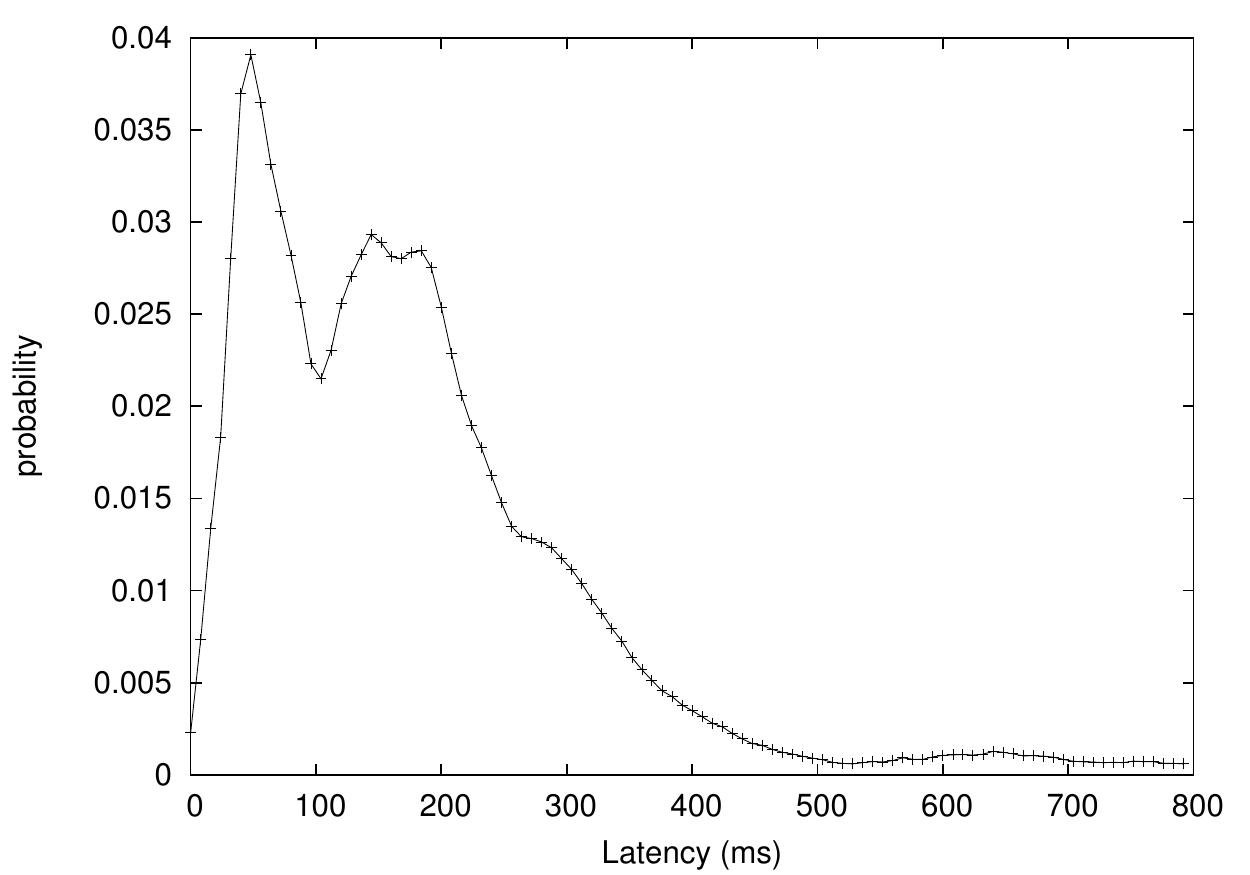}
\caption{Probability density function of RTTs}\label{fig:rtt}
\end{figure}

The \textit{Migration Time} (MT) is the time interval a virtual server takes to migrate from one node to another.
In order to model MT we have exploited (with some minor modifications) the model for TCP latency presented in \cite{cardwell2000modeling}. The MT depends on:

\begin{itemize}
\item \textit{Transport protocol}. We assume nodes to communicate over the Internet using TCP.
Note that this is a worst-case assumption, as a migration between two cloud nodes of the same provider occur in a network that is considerably faster and more reliable than the Internet. In these cases, it would be possible to use UDP to reduce the MT.
Regarding the TCP, we assume that nodes do not maintain active TCP connections with all the other nodes, so the MT should consider the delay necessary for the TCP 3-way handshake. 
Also, we assume that a whole virtual server would be sent during the TCP slow start (i.e. the sender does not trigger the TCP congestion management at the receiver).

\item \textit{RTT}. We model RTT delays according to the traces of the king dataset \cite{gummadi2002king}. 
The probability density function (PDF) of the RTTs is shown in Figure \ref{fig:rtt}.

\item \textit{Message loss probability}. 
As it is in \cite{cardwell2000modeling}, we consider a network loss probability of 0.001.

\item \textit{Virtual Server Size}. The size of the virtual server depends on several factors, such as the size of the routing table, the number of entities managed and the clients accessing the virtual server just before the migration. 

\end{itemize}

\begin{figure}[tbh]
\centering        
\includegraphics[width=0.8\textwidth]{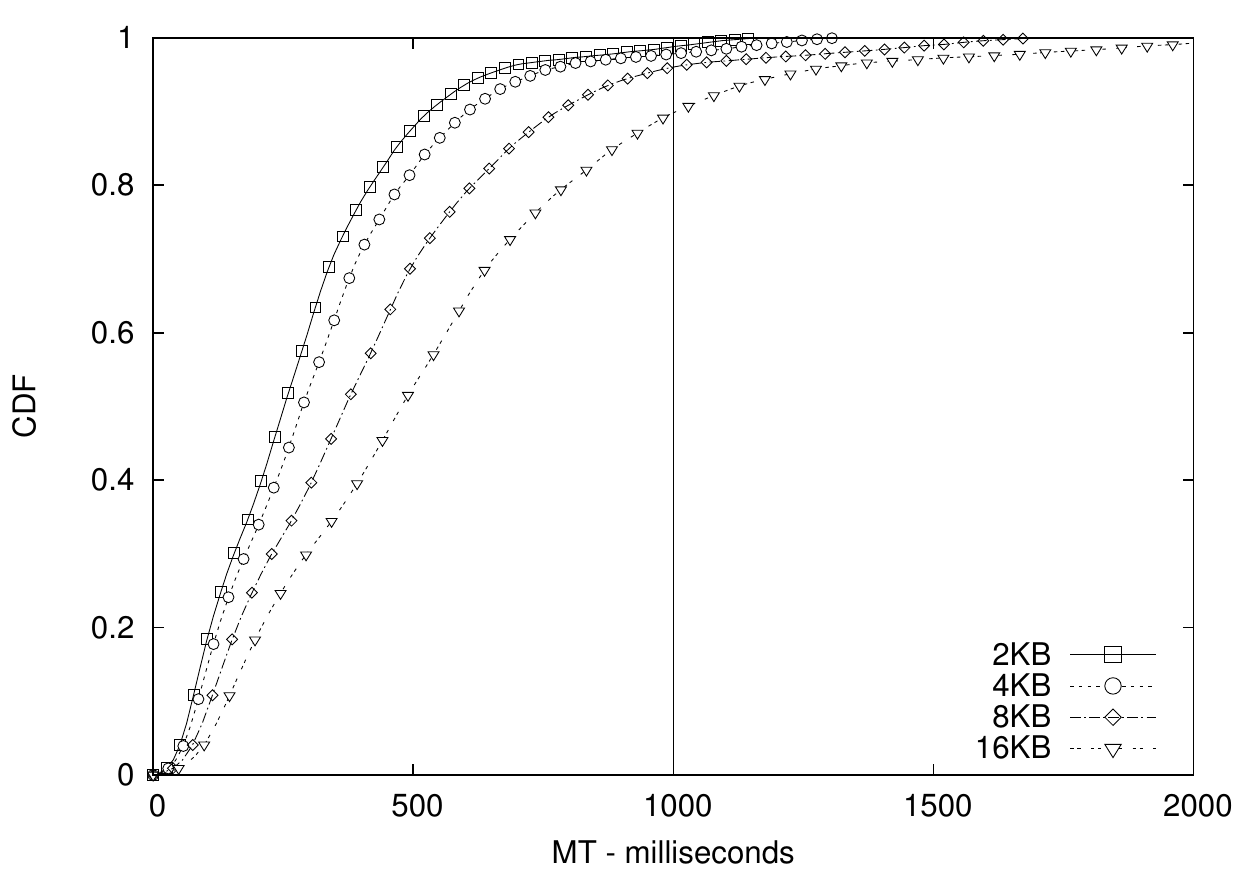}
\caption{Cumulative density function of migration time}\label{fig:nii}
\end{figure}

Figure \ref{fig:nii} shows the cumulative distribution function of the MT considering payloads of different sizes, i.e. 2KB, 4KB, 8KB and 16KB.
For each size, we have conducted 10K migrations.
With a VS size of 2KB the 98\% for migration take less than 1 second, with 4KB the 97\%, with 8KB the 95\% and with 16K the 88\%.

\paragraph{Tuning Virtual Servers Dimension}

In the previous section, we have seen how to compute a migration time of virtual servers with several fixed dimensions.
Instead, in this section we want to compute how large can be a virtual server, so that its MT does not affect the user.

The user is affected when there is a delay between an issued command and the reply from the server. The degree of interactivity expected by the user depends on the kind of the virtual environment she participates. In other words, the amount of latency users may tolerate in response of their action depends on the virtual environment genre.
This latency span from few hundreds of milliseconds in fast-paced MMOGs up to two second in slow-pace MMOGs \cite{claypool2006latency}.
In this analysis we sit in the middle and consider as tolerable latency a delay under 1 second, which fits medium-paced game genre.

\begin{figure}[tbh]
\centering
\includegraphics[width=0.8\textwidth]{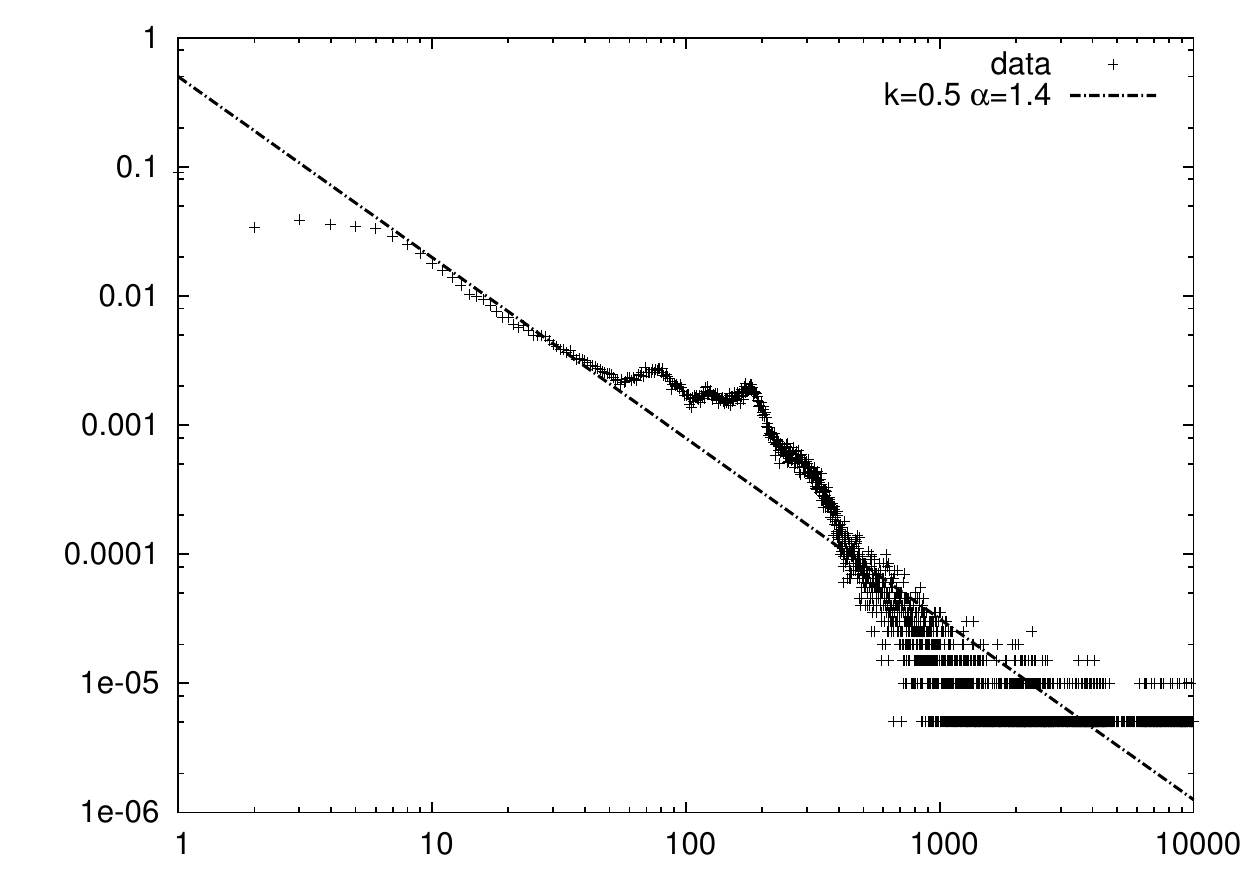}
\caption{Histogram of the client per minutes per entity plotted in log-log}\label{fig:power}
\end{figure}

We are interested in finding how many entities can fit into a virtual server so that on the 95\% of the cases the MT takes less than 1 second. First of all, we need to compute the size of a migrating virtual server, which is composed by:

\begin{itemize}
\item \textit{Entity descriptor}. The content of the entity is composed by: a UID (32 bits), a DHT-ID (160 bits), a point representing the two-dimension position of the entity (32+32 bits), and a list of attributes (integer, 32 bits) with the respective values (double, 64 bits). Let us assume that the dimension of this list is fixed for every entity to 10 elements. We argue that this value is a good average estimate to contain enough information for a general MMOG. Summing up, each entity descriptor has a size of about 140 bytes.

\item \textit{Access list}. The list of clients accessing to the virtual server. The node that receives a migrating virtual server uses this list to communicate with the client. Each entry of this list contains a UID (32 bits), a IP (32 bits) and a port (32 bits).
In order to estimate the number of connected client for an entity, we conducted an empirical analysis.
We counted the clients per entity per minute (hence, we consider a quite large timespan) in a simulation with synthetic generated avatars movements. The movements and the placement of the objects in the virtual environment were generated as described in Section \ref{sec:workload}. 
Figure \ref{fig:power} shows the clients (in percentage) plotted in a log-log scale. The trend of the plot resembles a power law, i.e. a function of the form $y(x) = Kx^{-\alpha}$. By fitting the data, we derived $K=0.5$ and $\alpha =1.4$ (the corresponding function is also plotted in the figure). A number generator based on this function was used to estimate the number of connected clients.

\item \textit{Routing table}. The routing table of the virtual server, which contains the references to other DHT nodes. In a typical DHT this table contains $\log N$ entries.
Each entry of the table is composed by: a DHT-ID (160 bits) and a IP (32 bits). 
By considering a DHT with 10K virtual server, this list contains 14 entries, for a total of 336 bytes.
\end{itemize}

\begin{figure}[tbh]
\centering
\includegraphics[width=0.8\textwidth]{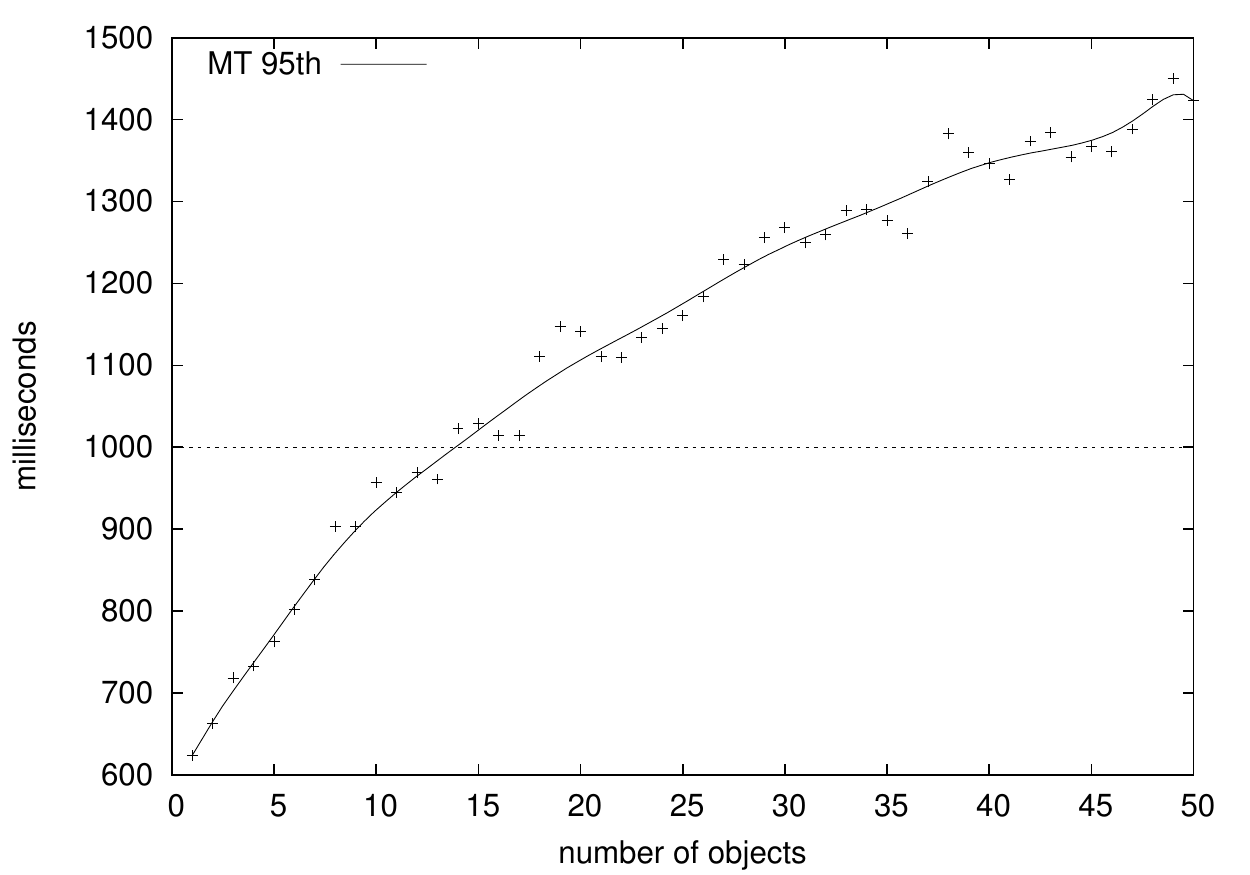}
\caption{95th percentile of MT with different amount of objects}\label{fig:percentile}
\end{figure}

The size of the virtual servers largely depends on the number of entities.
To see the relation between the number of objects and the MT, we conducted experiments considering virtual servers with different amount of entities. 
For each amount, we considered the 95th percentile of 10K runs.
Figure \ref{fig:percentile} shows the result.
Virtual servers managing less than about 15 entities have an MT less than one second.
This result may be used in two ways. Given a virtual environment with a predictable number of entities, it is possible to define the minimum number of virtual servers to employ. On the other side, if it is a necessity to have a specific number of virtual servers, it is possible to know the maximum number of entities the system can support.

\section{Problem Statement and System Model}
\label{sec:problem}

The manager has the task to orchestrate the assignment of the entities to the virtual servers.
The assignment is done according to these three principles: 
(i) the fraction of the entities assigned to user-provided nodes should be below the threshold set by the operator, 
(ii) no overloaded nodes should be kept in the system, and 
(iii) the economical cost should be the lowest that satisfies the last two points.
Note that the point (ii)  would be possible only if the manager had a complete and precise knowledge of the state of the servers.
However, how we will see later, the manager exploits a prediction function to estimate the load of the servers.
Hence, due to the error in the load estimation and the fluctuation of the load, some occasional overloading may happen.
In other world, if the manager would have a perfect load prediction, no overloaded nodes would be kept in the system.

In order to define the risk and the other parameters that the manager uses for the orchestration, here we give a description of the system model. We indicate as $VS$ the set of virtual servers in the system.
Each virtual server has a bandwidth requirement that depends on the amount of requests that are issued to the virtual server. This requirement changes over time, according on the interaction pattern of the avatars. More formally, $\forall v \in VS$ we define $v_{load}(t)$ as the bandwidth requirement of the virtual server $v$ from $t-1$ to $t$, with $v_{load}(0) = 0$. Similarly, $\forall v \in VS$ we define $v_{obj}(t)$ as the number of entities managed by $v$ at time $t$, with  $v_{obj}(0) = 0$.
All the virtual servers in $VS$ must be assigned to a node. The set $N(t)$ contains the nodes that are in the system at the time $t$.
An arbitrary node $n \in N$ is characterized by the following properties: (i) bandwidth capacity $n_{cap}$, (ii) bandwidth cost $n_{bcost}$, (iii) renting cost $n_{rcost}$, and (iv) failure probability $n_{fprob}$.

We consider two different kinds of nodes, cloud on-demand and user-provided nodes. The two kind of nodes have different characteristics.
User-provided resources have no cost for band and renting, while cloud nodes are assigned with the cost from a real price model.
On the other hand, we assume cloud nodes to have no failure probability, whereas user-provided resources have non-zero failure probability. Regarding the nodes we do two relevant assumptions: (i) the value of their properties does not change during time (i.e. the bandwidth cost remains constant) and (ii) we can exploit a price model where cloud nodes are charged per $\Delta{t}$.

The system \textit{risk} $\gamma(t)$ is defined as the number of objects managed by user-provided resources at time $t$, multiplied by nodes failure probability. Risk is computed as follows:

\begin{equation}
\gamma(t) =  \sum_{n \in N, v \in VS} v_{obj}(t) * n_{fprob}
\end{equation}

In order to provide a simpler way to quantify the risk, we mostly consider the \emph{relative risk factor} $\gamma_{R}$ of the system, rather the absolute risk computed above. To compute the relative risk factor, we correlate the absolute risk with the maximum risk, $\gamma_{MAX}$. The maximum risk is computed as if all the objects were assigned to user-provided resource. Therefore, the relative risk factor is simply computed as the ratio between absolute and maximum risk:

\begin{equation}
\gamma_R(t) =  \frac{\gamma(t)}{\gamma_{MAX}}
\end{equation}

One of the principle of our mechanism is to keep the number of overloaded nodes as low as possible, and to intervene when a node is overloaded.
The definition of an overloaded node goes through the concept of \textit{load factor}. The load factor of a node is defined as the ratio between the bandwidth requirement demanded to the node and its capacity.
Given an arbitrary node $n$ and a set $V_n$ that contains all the virtual servers assigned to $n$, the load factor of $n$ is computed as: 
\begin{equation}
lf_n(t) = \frac{\sum_{v \in V_n} v_{load}(t)}{n_{cap}}
\end{equation}
A node $n$ is considered \textit{overloaded} when $lf_n(t) >= 1$.

The last objective of our proposal is to  minimize the cost of the infrastructure. It is possible to compute the cost per interval of time as the sum of the bandwidth cost and the renting cost of the nodes, according on the bandwidth requirements at time $t$. The total system cost $\beta(t)$ is computed as follows:

\begin{equation}
\beta(t) = \sum_{n \in N, v \in V_n} v_{load}(t) * n_{bcost} + \sum_{n \in N} n_{rcost}
\end{equation}

\subsection{Mixed-Integer Programming Modelling}

The problem of assigning virtual server to node can be modelled as a Mixed-Integer Programming (MIP) problem. We define $x_{n,v}$ as binary variable, which value is $1$ if $v \in V_n$, $0$ otherwise. We also define another binary variable $u_n$, whose value is $1$ if $|V_n| \geq 1$, $0$ otherwise. The problem formulation is the following:

\begin{eqnarray}
minimize : & \beta(t) \label{eq:0}\\
\gamma(t) & \leq & max_{risk} \label{eq:1}\\
\sum_{v \in V_n} v_{load}(t) & \leq & n_{cap} * u_n \quad \forall n  \in N \label{eq:2} \\
\sum_{n \in N} x_{n,v} & = & 1 \quad \forall v \in VS \label{eq:3}\\
\sum_{n \in N, v in VS} x_{n,v} & = & card(VS) \label{eq:4}
\end{eqnarray} 

where (\ref{eq:0}) is the objective function, which minimizes the cost.
The constraint (\ref{eq:1}) forces the solution to not overcome the maximum risk defined by the operator.
The constraint (\ref{eq:2}) assures that no nodes are overloaded and, at the same time, force $u_n$ to take the value according the definition.
The constraints (\ref{eq:3}) and (\ref{eq:4}) assure respectively that a virtual server is assigned only once and all virtual server must be assigned.

Due to the quasi real-time constraint of the infrastructure, resolving a MIP every $\Delta_t$ to compute the assignment of virtual servers to node is infeasible. However, the MIP formulation has been revealed useful to compare what would it be an optimal solution with the 
performance of the (faster) heuristics executed by the manager.

\section{The Manager}
\label{sec:balancer}

The \textit{manager} is the component entitled to move virtual servers between nodes, as well as adding or removing nodes from the system.
The \textit{manager}'s goal is to control the system nodes overload and to prevent the risk factor of the system to overtake the maximum $risk_{limit}$ that is predefined by system operator. Moreover, the manager is also responsible for the economical efficiency of the system. In other words, while shuffling virtual servers, the manager takes into account the cost of used resources and chooses the less costly solutions. 

\begin{figure}
  \centering
  \includegraphics[width=0.95\textwidth]{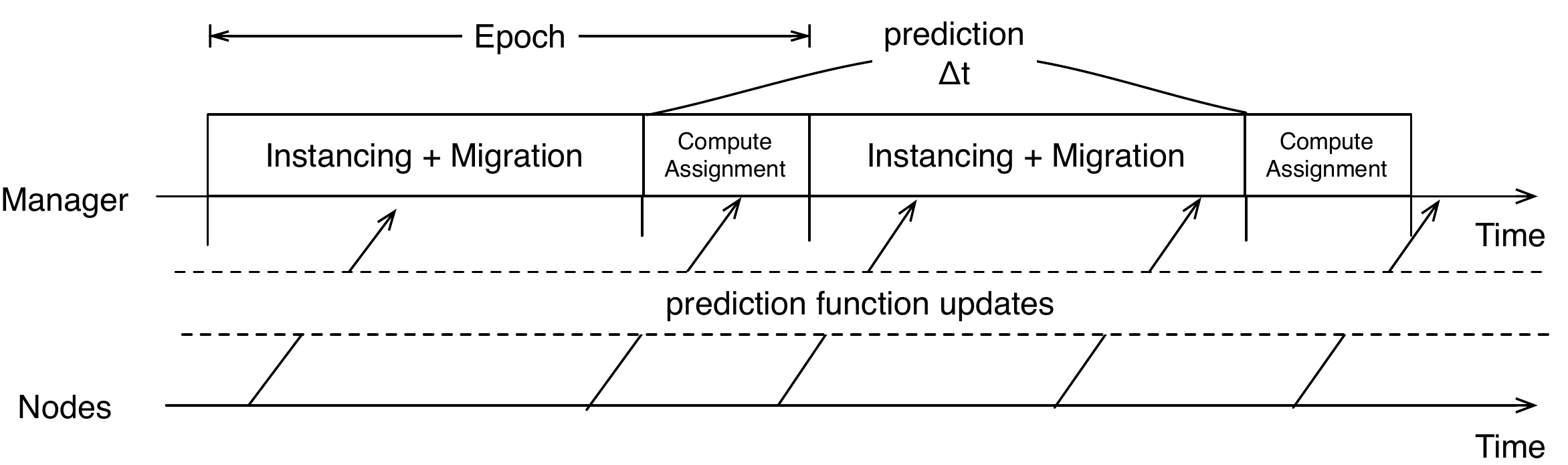}                
  \caption{Manager time management}
  \label{fig:loop}
\end{figure}

The work of the manager is divided into time intervals, which we refer to as \textit{epochs}.
Figure \ref{fig:loop} shows the management of two consecutive epochs.
During an epoch, the manager executes the following: (i) instantiates (or release) any on-demand node, (ii) migrates the virtual servers according to the plan done in the prior epoch, and (iii) computes the new assignment plan for the next epoch. 
The assignment is computed according to load prediction functions, one for each virtual servers and by exploiting an heuristics that considers load predictions in $\Delta t$ time. 
Over time, the manager receives from the nodes the updated coefficient of the prediction functions.
If an update arrives when the new assignment computation is already started, it will be considered in the next epoch.

The duration of an epoch (which we refer to as $\tau_{epoch}$) must be tuned to accommodate the maximum instancing time possible.
This time depends on the particular on-demand platform chosen, and normally it is in the order of few minutes \cite{markatchev2009cloud}. 
 Due to the fact that we use an heuristics to compute the assignments, $\tau_{epoch}$ is largely occupied by the instancing time. As a consequence, we assume $\Delta t \approx \tau_{epoch}$.

\subsection{Load prediction}


%
%
%


The manager computes the load of the virtual servers by using a prediction function for each virtual server.
This means that rather than storing a mere number, the manager stores a set of functions (and their coefficients) defining the load of the virtual servers an arbitrary time. In particular, for each virtual server $v$, the manager stores a pair $(v,L_v)$, where $L_v$ represents the prediction function.

The coefficients of these functions are computed locally by each node, and then sent to the manager.
Each node executes Algorithm~\ref{alg:estimation} in order to compute the load function $L_v$ for each of its VSs $v$.
In case the difference between the actual measured load and the load estimation provided by the function is larger than $\xi_{est}$
the node:
(1) recomputes the load function coefficients for $L_v$ according to the load of the virtual server, and 
(2) sends the renewed function $L_v$ to the manager at the end of the cycle execution. 

\begin{algorithm}[tbh]

\SetKwData{Msg}{msg}
\SetKwData{Load}{load}
\SetKwData{Vspool}{$vs\_pool$}
\SetKwData{Lv}{$L_v$}
\SetKwData{ManagerAddress}{managerAddress}

\KwData{\Lv, the load approximation function for the virtual server $v$}
\KwData{\ManagerAddress, the IP of the manager}

\SetKwFunction{Add}{add}
\SetKwFunction{UpdateFunction}{updateFunction}
\SetKwFunction{Send}{send}
\SetKwFunction{GetLoad}{getLoad}

\Repeat{}
{
\ForEach{VS $v \in$ \Vspool}
{
	\Load $\leftarrow$ \GetLoad()\;
	
	\If{$| \Lv - \Load | \geq \xi_{est}$}{
   		$L_v \leftarrow$ \UpdateFunction{\Load} \;
   		\Msg $\leftarrow$ \Add{$v$, $L_v$} \;
	}
}

\If{\Msg.size $\neq$ 0}
{
	\Send{\Msg, \ManagerAddress} \;
}

wait $\Delta t$ \;

}

\caption{Server's load estimation} 
\label{alg:estimation}
\end{algorithm}

As in Figure \ref{fig:loop}, $\timeWait$ indicates the ahead time of two successive assignment phase predictions.
By waiting this period of time, the servers are \textit{loosely} synchronized with the predictions cycles.
Over time, the \emph{manager} receives renewed load estimation functions from system servers (Algorithm~\ref{alg:passive}). 
Upon reception of a set of pairs ($v,L_v$), the \emph{manager} renews the function $L_v$ for the corresponding $v$ inside its storage. 

\begin{algorithm}[tbh]

\SetKwFor{Receive}{on receive}{}{}

\SetKwData{Lv}{$L_v$}
\SetKwData{Message}{message}
\SetKwFunction{UpdateFunction}{updateFunction}

\Receive{\Message $\leftarrow$ Set$<v, \Lv>$}
{

\ForEach{$(v, \Lv) \in$ \Message}
{
	\UpdateFunction{v, \Lv}\;
}
}
\caption{Load function renewal in manager} 
\label{alg:passive}
\end{algorithm}

%

The described approach in principle allows us to apply a wide range of statistical models for the load estimation, as for example classical methods for data prediction such as ARMA or ARIMA \cite{montgomery2011introduction} models. The choice of the model depends on the expected data fluctuations and the desired accuracy of the prediction $\xi_{est}$, which 
represents the acceptable error of the load estimation of the prediction model. 
High accuracy estimation models allow the manager to predict the load trend for large times interval $\timeWait$ ahead. 

%

In our implementation we use a simple exponential smoothing \cite{gardner2006exponential} as the mean to predict load trends.
As we will see later in the experimental evaluation, this models assures a good prediction power in spite of its simplicity.

\subsection{Virtual Servers Management}

The \emph{manager} employs an heuristics approach to decide what load to move and where to move it.
The \emph{manager} re-distributes the load in the system based on the current and future system states (predicted load and risk factors),  according to the maximum allowed risk and cost balance.
The task of the manager can be considered as the sum of two sub-tasks, \textit{virtual server selection}, and \textit{destination selection}.


\paragraph{Virtual Server Selection}

\begin{algorithm}[tbh]
\SetKwInOut{Input}{input}\SetKwInOut{Output}{output}

\SetKwData{ULimit}{$LF_{up}$}
\SetKwData{BLimit}{$LF_{bot}$}
\SetKwData{V}{v}
\SetKwData{Lv}{$L_v$}
\SetKwData{Vspool}{$vs_{pool}$}
\SetKwData{Npool}{$node_{pool}$}
\SetKwData{Npooltwo}{$node^2_{pool}$}
\SetKwData{Actions}{Actions}
\SetKwData{Psize}{$P_{size}$}

\SetKwFunction{PLF}{predictedLoadFactor}
\SetKwFunction{RF}{riskFactor}
\SetKwFunction{CLF}{currentLoadFactor}
\SetKwFunction{MD}{maxDerivative}
\SetKwFunction{AV}{allVS}
\SetKwFunction{AN}{allNodes}
\SetKwFunction{RC}{recruitCloud}
\SetKwFunction{RVS}{getRandomVS}

\Input{\ULimit, upper load factor threshold}
\Input{\BLimit, bottom load factor threshold}
\Input{\Psize, the min amount of VSs to consider per epoch}
\Output{\Vspool, the list of virtual server to migrate}

\BlankLine

\tcp{1. Take the VSs in nodes whose load is too high}

\ForEach{Node N $\in$ POOL}
{
\While{\PLF{N} $ > $ \ULimit}
{
   \Vspool $\leftarrow$ \MD{\AV{N}} \;
}
}

\tcp{2. Take the virtual server that are backed up}

\Vspool $\leftarrow$ \Vspool $\cup$ backUp()\;

\tcp{3. Take from nodes that are under-loaded}

\If{size(\Vspool) $<$ \Psize}
{
	
\ForEach{Node N $\in$ POOL}
{
	\If{\CLF{N} $<$ \BLimit}
	{
			\Vspool $\leftarrow$ \AV{N}\;
	}
}
}

\tcp{4. Take some random VS}

\If{size(\Vspool) $<$ \Psize}
{
	\Vspool $\leftarrow$ \RVS{}\;
}

\caption{Balancer: Virtual Server Selection} 
\label{alg:first}
\end{algorithm}

In this phase, the manager chooses which virtual servers are good candidates to be moved. The pseudo-code of this phase is in Algorithm \ref{alg:first}. 
Firstly, the manager computes the predicted load factor (PLF) of the nodes in the system, both on-demand and user-provided nodes. The predicted load factor is the ratio between the predicted load at the next time epoch (computed by using the load prediction function) and the node capacity. The balancer removes virtual servers for each node whose PLF overtakes $LF_{up}$ until the PLF drops below the threshold. The removal order of the virtual servers considers the derivative of the load prediction functions; the virtual server with the highest derivative is removed as first.  
The reason behind is that a virtual server with a high derivative would probably have a burst in the load soon, and to reassign it 
may avoid future overloaded nodes in the system.
The removed nodes are added to a data structure (called $vs_{pool}$ in the algorithm), which keeps track of the virtual server to reassign. Note that the removal is \textit{virtual} in the sense that the manager works on a copy of the real system to obtain the list of migrations to execute. 
Further, $vs_{pool}$ receives the nodes that are currently managed by a backup server. These VS have been moved to the back up cloud server due to their regular holder to have crashed or left (see Section \ref{replication}).

If after these first two steps the amount of virtual server in the $vs_{pool}$ is below of the $P_{size}$ threshold, the manager considers virtual servers from \textit{under-loaded} nodes.
A node is under-loaded when its PLF is below the $LF_{bot}$ threshold.
If even after considering under-loaded nodes the $vs_{pool}$ still contains less than $P_{size}$ items, some random virtual servers are inserted in the $vs_{pool}$.

\paragraph{Destination Selection}

\begin{algorithm}[tbh]
\SetKwInOut{Input}{input}\SetKwInOut{Output}{output}

\SetKwData{ULimit}{$LF_{up}$}
\SetKwData{BLimit}{$LF_{bot}$}
\SetKwData{V}{v}
\SetKwData{Lv}{$L_v$}
\SetKwData{Vspool}{$vs_{pool}$}
\SetKwData{Npool}{$node_{pool}$}
\SetKwData{Npooltwo}{$node^2_{pool}$}
\SetKwData{Actions}{Actions}
\SetKwData{Chosen}{Chosen}
\SetKwData{Risklimit}{$risk\_{limit}$}

\SetKwFunction{PLF}{predictedLoadFactor}
\SetKwFunction{RF}{riskFactor}
\SetKwFunction{CLF}{currentLoadFactor}
\SetKwFunction{MD}{maxDerivative}
\SetKwFunction{AV}{allVS}
\SetKwFunction{AN}{allNodes}
\SetKwFunction{RC}{recruitCloud}
\SetKwFunction{CC}{chooseCloud}
\SetKwFunction{REL}{releaseUnusedCloud}

\Input{\Vspool, the list of virtual server to migrate}
\Input{\Risklimit, upper load factor threshold}
\Input{\ULimit, upper load factor threshold}
\Output{\Actions, the list of migrations to execute}

\BlankLine

\ForEach{VS $v \in$ \Vspool}
{
	\Chosen = nil\;

	\Npool $\leftarrow$ (N $\in$ \AN : \PLF{N $\oplus$ v} $<$ \ULimit)\;
	\If{\Npool is $\emptyset$}
	{
		\Chosen $\leftarrow$ \CC{}\;
	}
	\Else
	{
		\Npool $\leftarrow$ (N $\in$ \Npool : \RF{N $\oplus$ v} $<$ \Risklimit)\;
		\If {\Npool is $\emptyset$}
		{
			\Chosen $\leftarrow$ \CC{}\;
		}
		\Else
		{
			Sort \Npool ascending according the cost\;
			\Chosen $\leftarrow$ \Npool.getFirst()\;
		}
	}
	
	\Actions $\leftarrow$ move $v$ to \Chosen\;
}

\caption{Balancer: Destination Selection} 
\label{alg:second}
\end{algorithm}

In this step  the manager executes re-assignment of the virtual servers from the $vs_{pool}$ to available nodes with respect to the system risk and cost. 
The system works with heuristics that considering the predefined maximum risk to select an assignment for virtual servers with minimum cost. 
For each virtual server $v$ in $vs_{pool}$, the manager executes the following. 
At first, the manager selects the node candidates such that, if assigned $v$ their PLF is less than $LF_{up}$.
Note that in the code we use the notation $\oplus$ to indicate that we consider the load as if the node would manage $v$.
If no node can satisfy this requirement, the system recruits a new cloud node and $v$ is assigned to it, and the manager starts with the next virtual server. 
Otherwise, the nodes go through another selection round. In this round, the manager discards all the nodes that would increase the risk of the platform over the threshold $risk_{limit}$. As before, if no candidate remains after this further selection, a new cloud node is recruited.
Otherwise, the candidate that provides the less cost is chosen.

\subsection{Migration}

At the start of the epoch, the manager executes the migrations that comes as output from the assignment computation of the previous epoch. All the migrations are executed in accordance with the virtual server migration algorithm, which is described is Section \ref{sec:migration}.
In this phase the manager also manages the on-demand resources, by actually contacting the on-demand platform for new instances, or by releasing not used instances.


\section{Experimental Results}
\label{sec:result}


\subsection{Workload Definition}
\label{sec:workload}


The management of a MMOG infrastructure generates a certain amount of load on the server, in terms of computational and bandwidth requirements.
A realistic simulation of the load is central to properly evaluate a MMOG infrastructure.
This thesis focuses on one type on load, which is bandwidth.
In particular, the outgoing bandwidth is considered.
This is done for the following reasons.
First, besides machine time, a typical on-demand computing platform charges outgoing bandwidth.
Hence, with the reduction of the outgoing bandwidth it is possible to reduce the operational costs of the infrastructure.
Second, user-provided resources usually have asymmetric connections to the public network.
This implies that the outgoing bandwidth is the resource to optimize, due to the smaller availability compared to the ingoing bandwidth. 
In addition, our workload considers the load related to the management of the users. It does not take into account the bandwidth consumed for other tasks, like backup management, intra-server communications, and other services at application level (e.g. voice over IP).  
In the rest of the chapter the generic term load is used to indicate the outgoing bandwidth load.

Mostly, the load depends on the number of concurrent players connected to the infrastructure.
For instance, \cite{Nae2008} noticed that the load varies from $O(n)$ to $O(n^3)$ according to the pace of interaction provided by the particular MMOG (where $n$ is the number of players). However, a wide analysis is necessary to reproduce the peculiarities of a proper MMOG load. At this regard, we have considered the following aspects when building a synthetic workload for MMOGs: (i) the variation in the number players over time, (ii) the players’ mobility patterns, (iii) the objects distribution, and (iv) the interaction model.

\paragraph{Interaction model}

In our model bandwidth is sampled according to a discrete time step model.
We define each step $t$ to have a duration of $\Delta t$.
During the step duration, for each generic entity $e$ the server gathers all the state actions regarding $e$.
Then, a new state is computed and it is broadcast to the interested players, i.e. the players that have the $e$ in their AOI.
This broadcasting is the action that consumes outgoing bandwidth. Hence, the bandwidth requirement at a generic time $t$ is computed with the formula:

\begin{equation}
\sum_{e \in E}{e_{AOI}(t)} \times M_{len}
\end{equation}

\noindent
where $E$ is the set of all entities. $M_{len}$ is the length of the broadcast message, which we assume to be the same for any entity.
$e_{AOI}(t)$ is the number of AOI's in which the entity $e$ is at time $t$. This value has been defined experimentally, through the simulation of players’ movements and objects placement in the virtual world.

\paragraph{Number of Players}

The number of concurrent players connected to a MMOG infrastructure presents seasonal trends. These trends can regard period of the week (i.e. more players in the weekend rather than in the middle of the week) and period of the day (i.e. more player in the evening rather than in the morning). In any case, it is a central point to evaluate how a MMOG infrastructure adapts itself to these variations. In particular, since the load is in a direct correspondence with the number of players, we are interested in the impact of a variable number of players on the infrastructure.

In order to simulate a seasonal trend in our workload, we vary the number of players over the time. To compute the number of players at a given time $t$ we use the following formula:

\begin{equation}
\sin(\frac{\pi t}{\lambda}) \times P_{max}
\end{equation}

\noindent
where the variation is controlled by two parameters: (i) the maximum number of player $P_{max}$ and (ii) the length $\lambda$, which represents the number of iteration that are needed to perform a season cycle.

\paragraph{Players Mobility Model}

Avatars move on the map according to realistic mobility traces that have been computed according the mobility model presented by Legtchenko et al.  \cite{Legtchenko2010},  which simulates avatars movement in a commercial MMOG, Second Life \cite{sl-site}.
We have presented this implementation in \cite{carlini2011evaluating}, as well as a comparison with other mobility models.
In the model, players gather around a set of {\em hotspots}, which usually correspond to towns, or in general to points of interest of the virtual world. A circular area characterized by a center and by a radius defines each hotspot.
Traces generation goes through two phases: \textit{initialization} and \textit{running}.

In the initialization phases, the area of the of the virtual environment is divided in \textit{hotspot area} and \textit{outland area}. The percentage of the hotspot area is defined by $p_{hot}$ and, consequently $1 - p_{hot}$ represents the outland area.
The hotspots are placed randomly in the virtual environment. The number of hotspot is defined by the parameter $H_{num}$. Their radius is computed such that the total area covered by the hotspots is in accordance to $p_{hot}$.
The parameter $p_{den}$ defines the probability that a player would be initially placed in an hotspot, whereas $1 - p_{den}$ defines the probability for a player to be initially placed in outland. 
If the player is placed in the outland, its position is chosen uniformly at random on the whole map.
Otherwise, an hotspot for the player is randomly selected and the player is positioned inside the hotspot.
The position inside the hotspot is chosen by considering a Zipfian\footnote{The Zipfian distribution was originally studied to show the relation of inverse proportionality between the frequency of words in a text and their rank in the frequency table. Nowadays, it used in many field of computer science, like graph theory \cite{broder2000graph},  to model similar concepts. In our work we use this distribution to model the fact that the frequency of entities is inversely proportional to their distance from the center of the hotspot.} distribution \cite{newman2005power}, so to ensure an higher density of players near the center of the hotspot.

The running phase moves the players across the virtual environment. The movements are driven by a Markov chain, whose transition probabilities are taken from the original paper \cite{Legtchenko2010}. The possible states for the players are the following: 

\begin{itemize}
\item \textit{Halt}(H): the player remains in place.
\item \textit{Exploration}(E): the player explores a specific area. If the player is moving inside an hotspot, the new position is chosen according to a power law distribution. Otherwise, the new position is chosen at random.
\item \textit{Travelling}(T): the player moves straight toward another point in the virtual environment. The new point is chosen in accordance with $p_{dens}$.
\end{itemize}

Initially every player is in state H. At each step $t$, the model decides the new state according to the probability of moving between states.
This mobility model exposes a fair balance between the time spent by avatars in hotspots and outland. 
Furthermore, the path followed by avatars when moving between hotspots is not static, 
i.e. no predefined path connects two hotspots.

\paragraph{Objects distribution}

Like players movement, the distribution of the objects exploits the concept of hotspots.
The idea is to have more objects in the hotspots, which represent a place of interest in the virtual world.
To place objects over the virtual environment, we use the same space characterization in desert and hotspot areas used by the mobility model. 
The total number of objects is defined as $O_{num}$. A  fraction of these objects is placed inside hotspot areas, whereas the rest is placed in the desert area.
The percentage of objects that are placed inside the hotspot area is controlled by the parameter $p_{obj}$. Objects in an hotspot area are placed so that their concentration follows a Zipfian distribution \cite{newman2005power}, with a peak in the hotspot center. Conversely, objects in the desert area are placed randomly.\\


\begin{table}
\centering
\begin{tabular}{|c|c||c|c|}
\hline
$M_{len}$ & 100 bytes & $\lambda$ & 200 \\
\hline
$P_{max}$ & 1000 & $O_{num}$ & 1000 \\
\hline
$P_{obj}$ & 0.7 & $H_{num}$ & 5 \\
\hline
$p_{den}$ & 0.8 & $p_{hot}$ & 0.3 \\
\hline
size & 5000x5000 & $\Delta t$ & 0.2s \\
\hline
\end{tabular}
\caption{Workload's table of parameters}\label{tab:par-work}
\end{table}

\begin{figure}
\centering
        \begin{subfigure}[b]{0.45\textwidth}
                \centering
                \includegraphics[width=\textwidth]{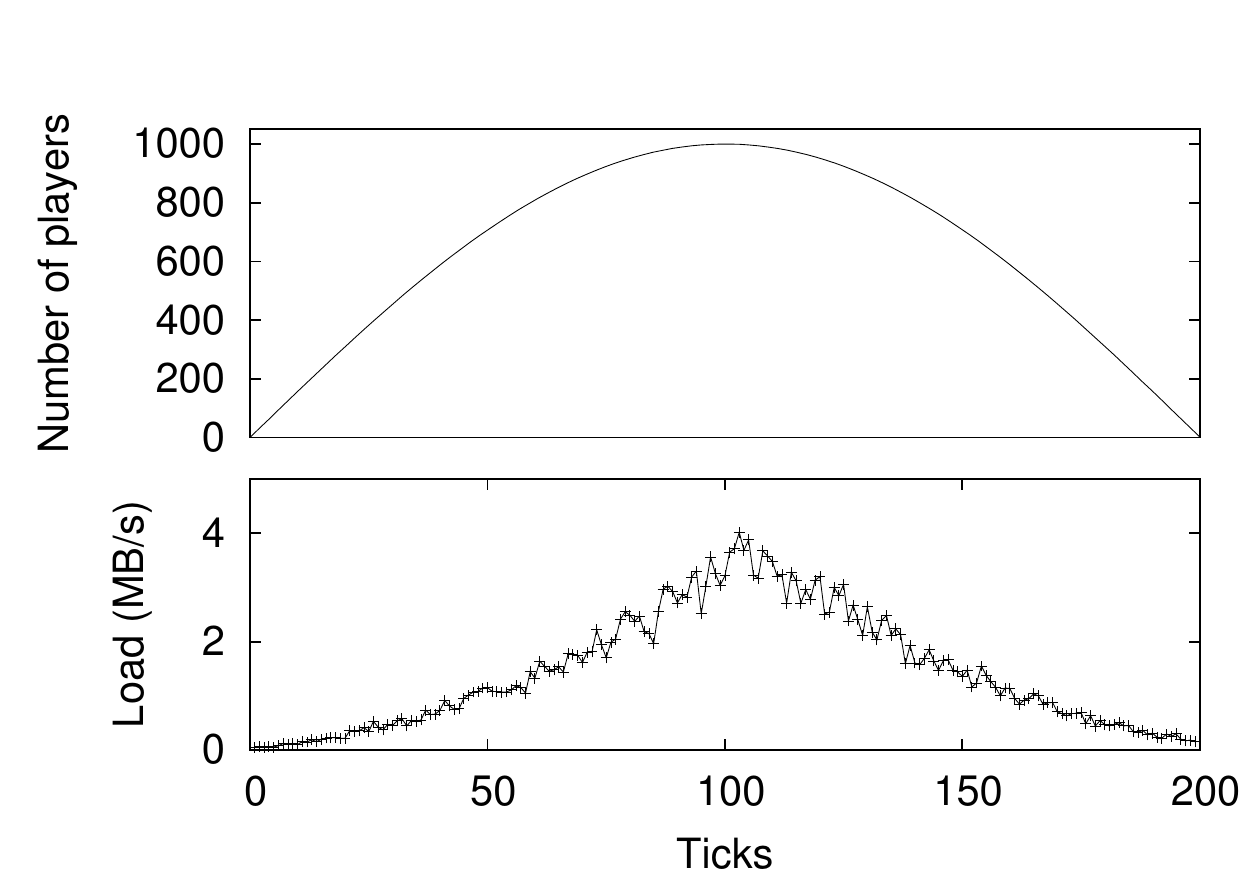}
                \caption{Load variation with number of player over time}
                \label{fig:dataset}
        \end{subfigure}%
        \quad      
        \begin{subfigure}[b]{0.45\textwidth}
                \centering
                \includegraphics[width=\textwidth]{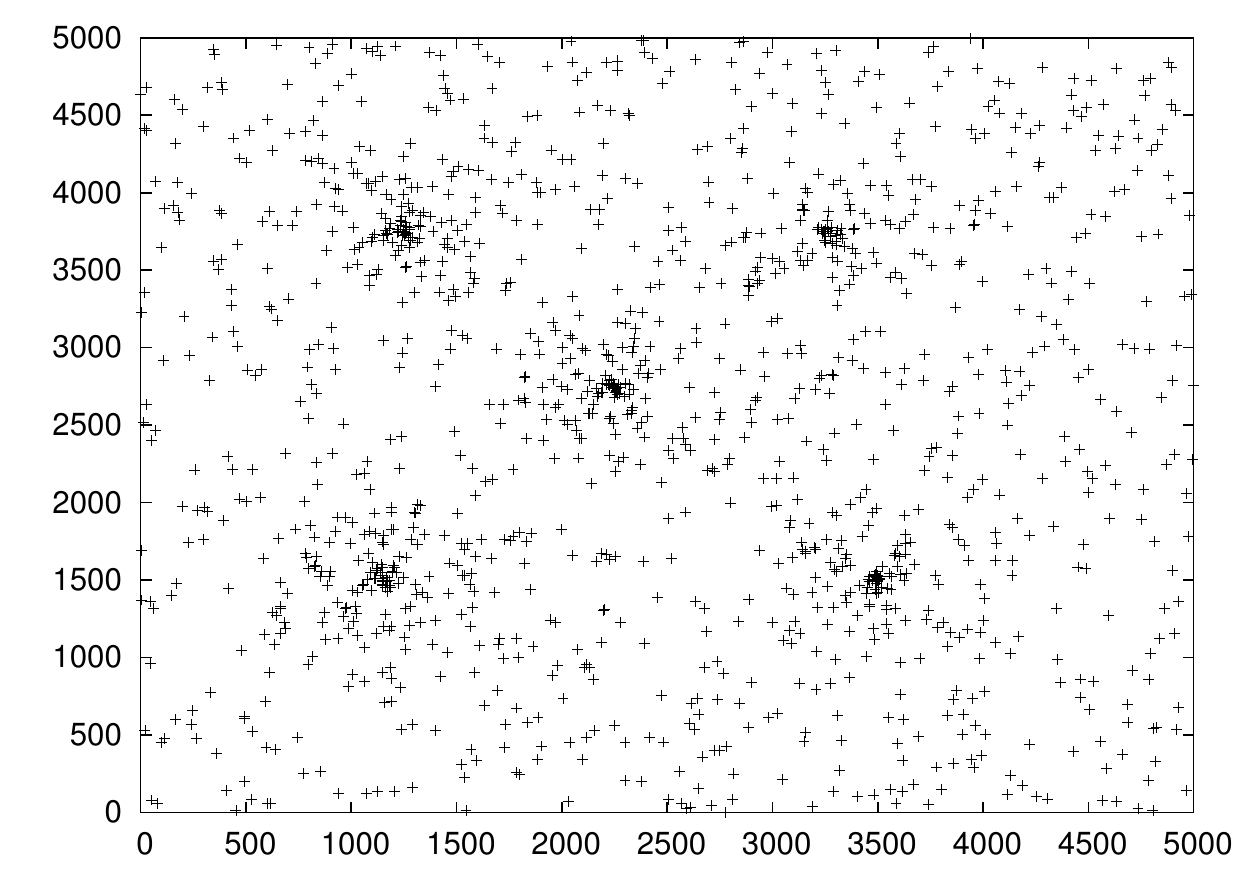}
                \caption{Objects placement in the virtual environment area}
                \label{fig:obj-place}
        \end{subfigure}%
\caption{Workload characterization}\label{fig:dset-char}
\end{figure}

Figure \ref{fig:obj-place} shows the placement of 5000 objects in a virtual environment characterized by the parameters in Table \ref{tab:par-work}.



\subsection{Simulation Environment and Metrics}

The experimental evaluation was carried out by means of simulations exploiting the workload described above.
In the simulation, time is considered as a series of discrete steps.
At each time step, the simulation: 
(i) computes the metrics on the previous iteration and 
(ii) computes the new assignment plan for the virtual servers.
As a reference, here we summarize the parameter that we will use in the rest of the chapter. $P_{max}$ represent the maximum number of players. $risk_{limit}$ represents the maximum fraction of objects that can be assigned to user-provided resources; if not stated differently, its value is $0.1$. 
$\epsilon_{est}$ is the error threshold for the prediction function. If the difference between the predicted value and the measured value is above $\epsilon_{est}$, the coefficient of the function must be recomputed. If not stated differently, its value is $0.05$.

The metrics used to evaluate the proposed approach are the following:
\begin{itemize}
\item \textit{cost per minute}. It represents the amount of US dollars that the platform consume per minute to sustain the MMOG\footnote{Prices are taken from Amazon Elastic Cloud 2, Standard Large instances (http://aws.amazon.com/ec2/pricing/), September 2012.}. In the aggregate form we consider the average.
\item \textit{availability}. It is intended as the fraction of requests from the client that are not replied or that are replied in delay. A server has a delay when the required bandwidth is higher than its capacity (i.e. it is overloaded).
\end{itemize}

\begin{figure}[tbh]
\centering
\includegraphics[width=0.8\textwidth]{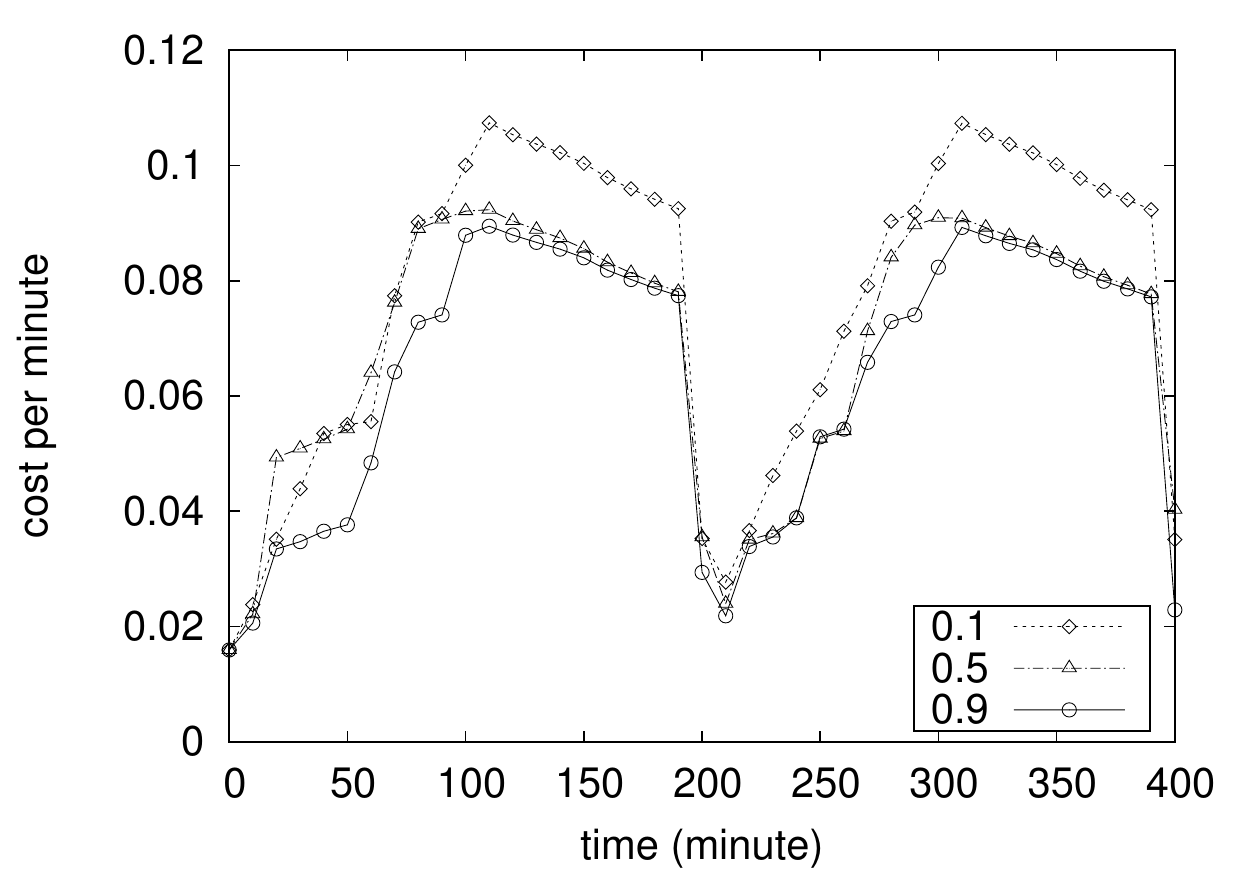}
\caption{Cost per minute with different maximum risk. Prediction error $0.05$, $100$ virtual servers, $5000$ maximum players}
\label{fig:cost-on-risk}
\end{figure}

\subsection{Risk and Cost Trade-off}

One of the strength of our approach is that it leaves to the MMOG operators the ability to set the desired fraction of objects assigned to the user-provided resources. In order to do that, the operator acts on the maximum allowed risk. To show the effect this setting has on the platform, we have experimented with various $risk_{limit}$, from $0.1$ to $0.9$.

\begin{figure}[tbh]
\centering
\includegraphics[width=0.8\textwidth]{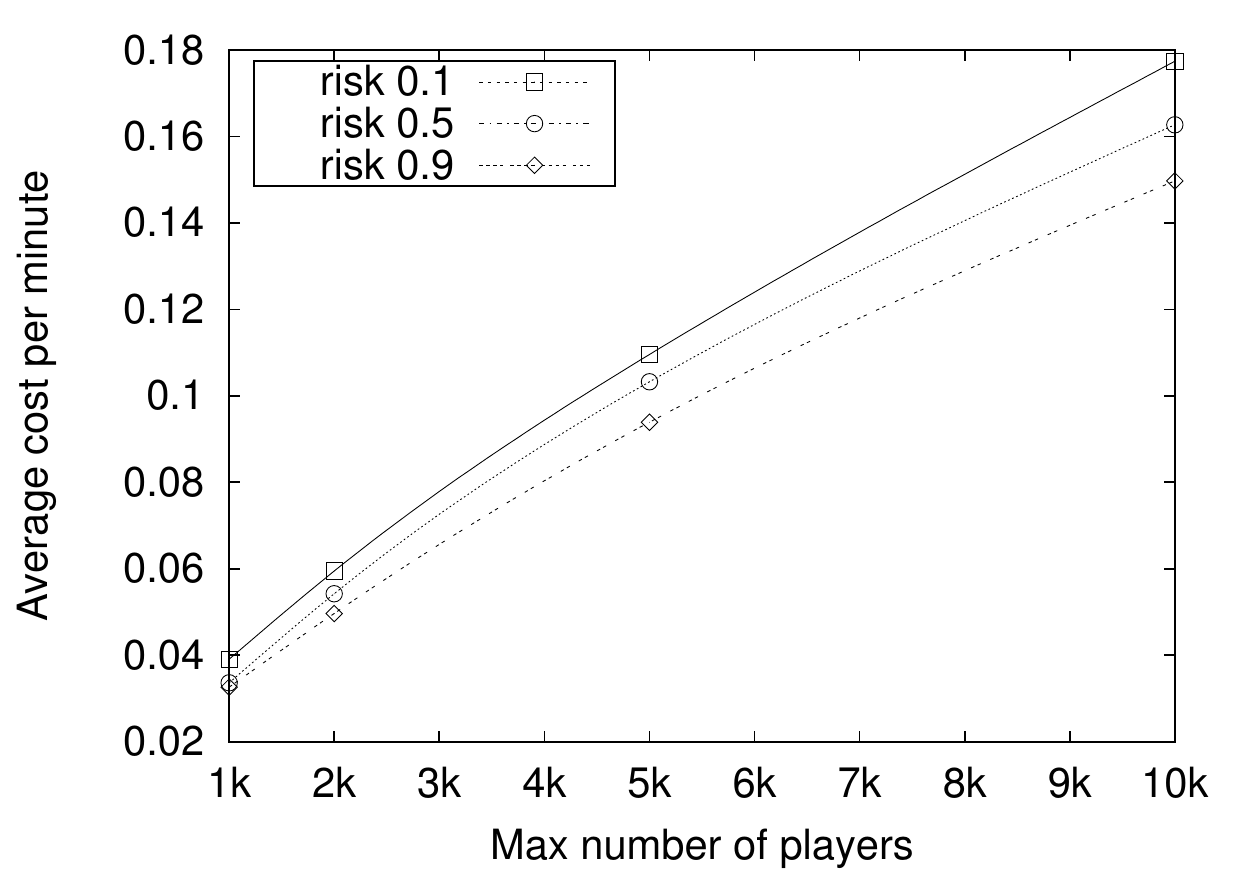}
\caption{Average cost per minute, different number of maximum players and $risk_{limit}$, $100$ virtual servers}
\label{fig:avg-cost}
\end{figure}

Figure \ref{fig:cost-on-risk} shows the cost per minute with three different risk settings, $100$ virtual servers, $\epsilon_{est}=0.05$ and 2000 maximum players.
In general, the data shows that with less risk the cost is higher.
This happens in particular in correspondence of the peak load.
As expected, with $risk_{limit} = 0.5$ and $0.9$ the extra load on the peak is managed by the peers, rather with risk $0.1$ the peak load is managed by the cloud nodes.

\begin{figure}[tbh]
\centering
\includegraphics[width=0.8\textwidth]{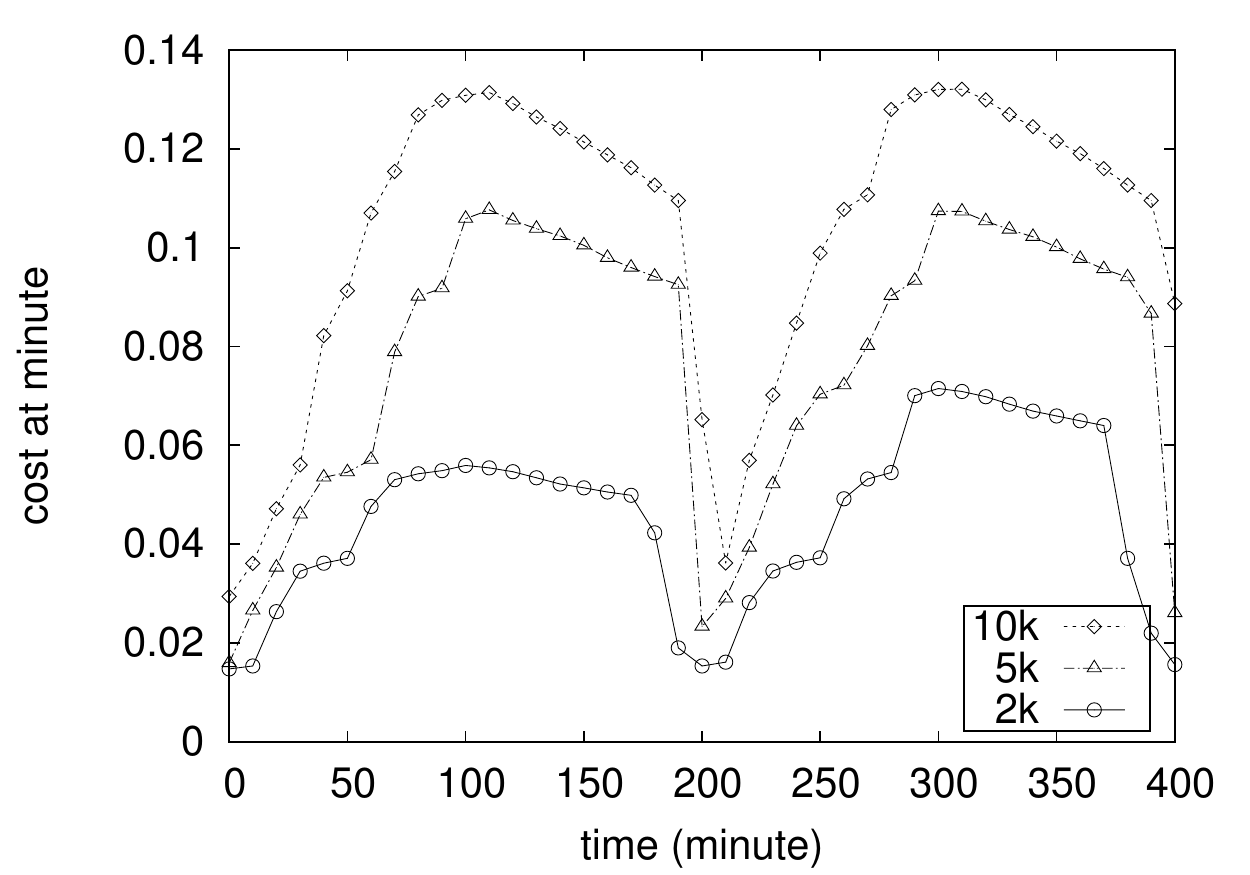}
\caption{Cost trend, different number of maximum players. $risk_{limit}=0.1$, $100$ virtual servers}
\label{fig:cost-on-player}
\end{figure}

Figure \ref{fig:avg-cost} shows the average cost per minute with three different risk settings, i.e. 0.1, 0.5 and 0.9. 
The intermediate values of $risk_{limit}$ offer results in between the extremes. The configuration of the experiment considers from 1000 to 10000 maximum concurrent players, $100$ virtual server and $\epsilon_{est}=0.05$. 
As it can be seen from the figure, the average cost per minute grows accordingly with the number of players. 
Moreover, the gap between average costs for different risk factors increases with the number of players.
For instance, considering 10K players, there is a difference in the cost around 20\% between the maximum allowed risk of 0.1 and 0.9.
Hence, the approach allows the operator to reduce the service cost for large-scale systems.
Nevertheless, even in a case of $risk_{limit} = 0.9$, the system costs remains significant. 
When the load of the virtual servers grows over time, user-provided resources cannot support some of the because of their limited upload bandwidth capabilities.

\subsection{Scalability Cost on the Number of Players}

In order to show how our architecture behaves with different number of players, we have conducted several experiments.
Figure \ref{fig:cost-on-player} shows the cost of the system considering 2000, 5000 and 10000 players. The experiment was conducted with a $risk_{limit} = 0.1$, $\epsilon_{est}=0.05$ and a system with 100 virtual servers.
From the figure we can see that our approach presents good scalability results. The cost is scaled according to the number of players (i.e. VS load) and the system effectively adds cloud resources on demand as well as removing them when the system load is decreasing.


\subsection{Comparison with Optimum}

\begin{figure}[tbh]
\centering
\includegraphics[width=0.8\textwidth]{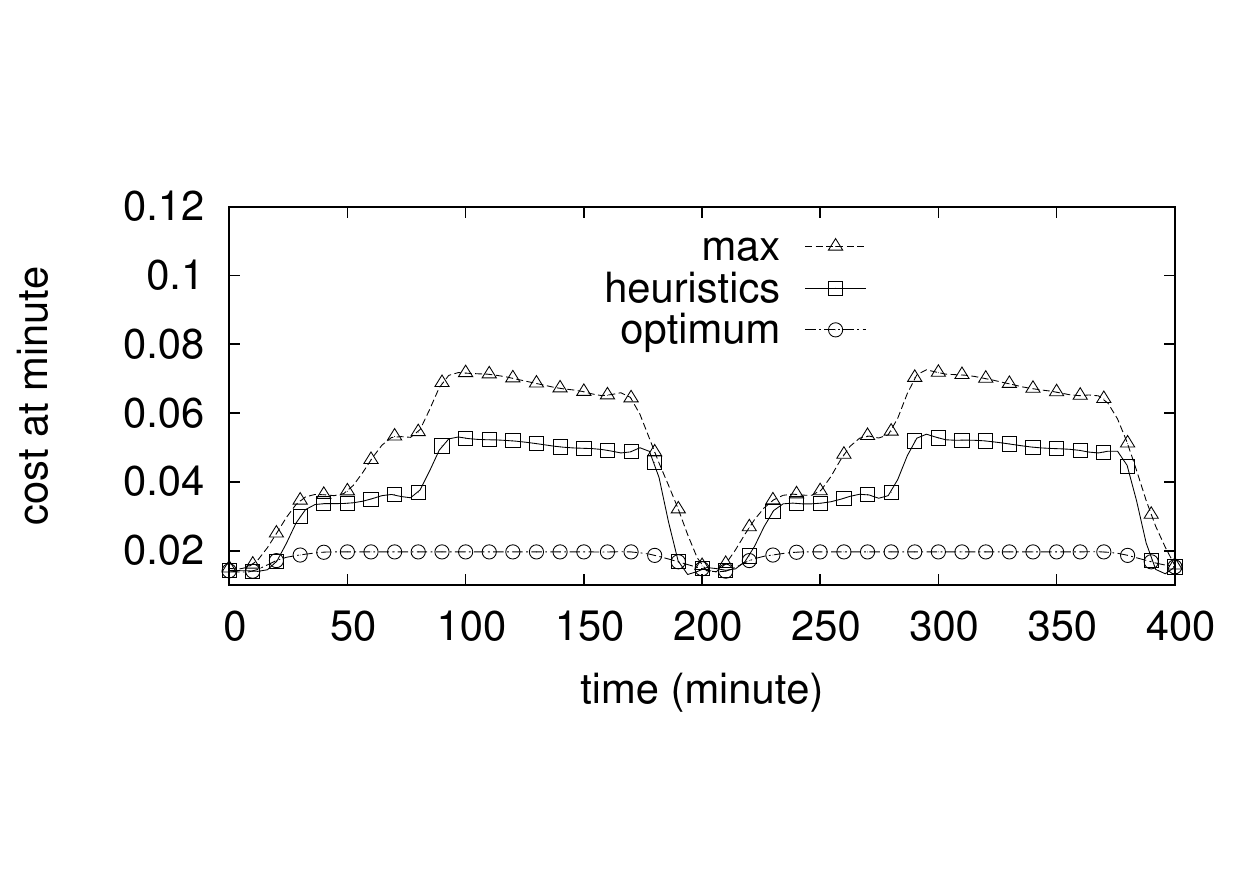}
\caption{Average cost per minute with different strategies in assigning virtual servers. no prediction error, $100$ virtual servers, $2000$ maximum players, $risk_{limit} = 0.9$}
\label{fig:heur-opt-09}
\end{figure}

Figure~\ref{fig:heur-opt-09} shows the cost trend of the heuristic allocation compared with the optimum allocation computed by the MIP-solver\footnote{The MIP problem was solved using the GNU Linear Programming Kit (GLPK) package, www.gnu.org/software/GLPK} and the fully cloud utilization (which we take as the worst case in term of cost).
The maximum number of players is 2000, the error prediction is $\epsilon_{est}=0.05$.
As it can be seen from the figure, the proposed heuristics allocation significantly reduces the service cost. The obtained results are far from the optimum resource allocation since the optimum solver works in much better condition than the heuristics. First, the number of migration per epoch can be as many as the number of virtual servers, which is not the case in the heuristics. Second, the execution time per epoch of the heuristics is around few hundreds of milliseconds, compared with the minutes of the MIP solver.

\subsection{Prediction Error}

The maximum allowed prediction error ($\epsilon_{est}$) regulates the frequency of the recalculation of the prediction function from the servers. In other words, if the difference between the predicted load and the actual measured load is higher than $\epsilon_{est}$, the server must compute the new coefficients for the prediction function.
This operation requires computational time, hence a high value for $\epsilon_{est}$ would imply a lower recalculation frequency, saving precious CPU time that can be used for other tasks. On the other hand, a lower prediction error would improve the prediction precision, which in turn impact positively on the overall performances. In particular, we expect the level of availability to be affected by proportional the value of $\epsilon_{est}$.

\begin{figure}[tbh]
\centering
\includegraphics[width=0.8\textwidth]{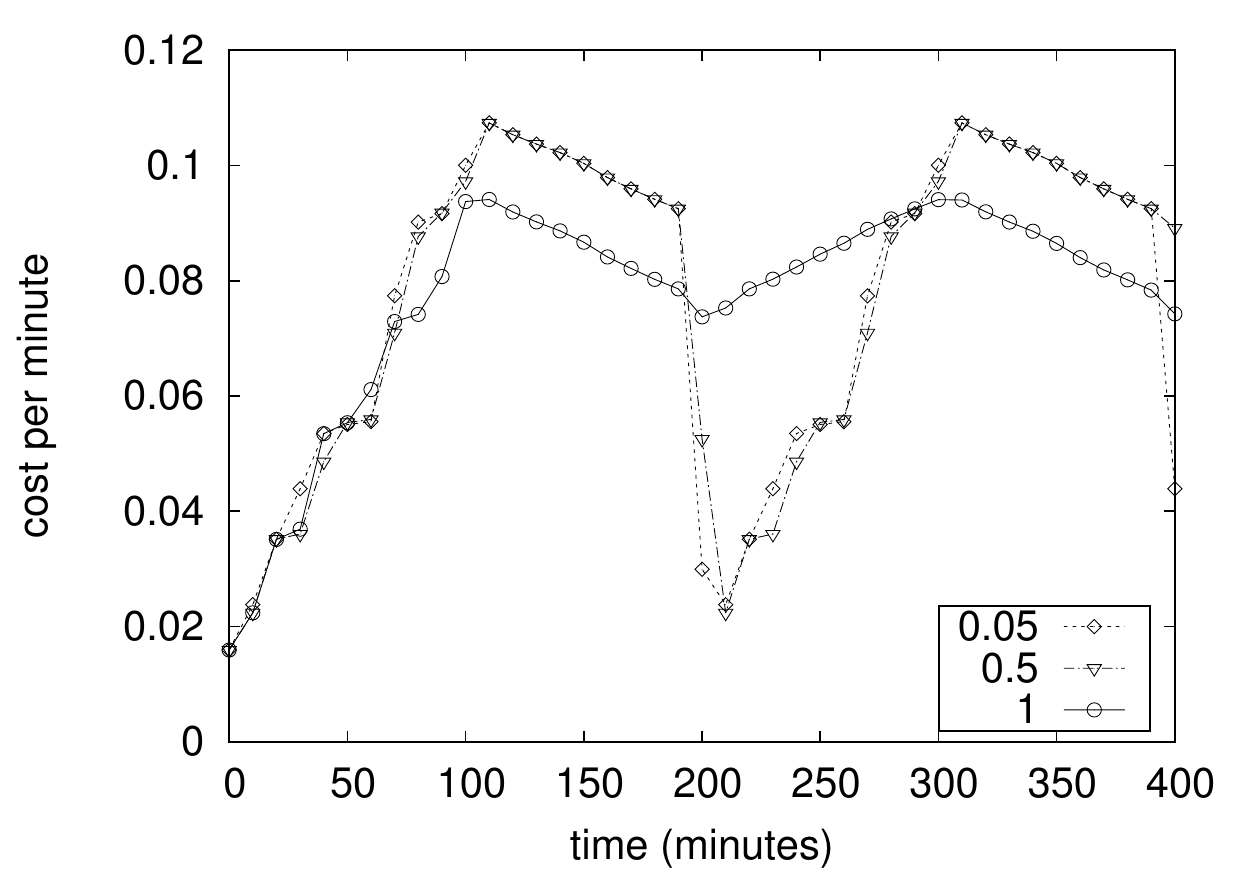}
\caption{Cost per minute, different values for $\epsilon_{est}$. 2000 maximum players, $100$ virtual servers}
\label{fig:cost-on-epsilon}
\end{figure}

Figure \ref{fig:cost-on-epsilon} shows the cost per minute with $\epsilon_{est}$ values of $0.05$, $0.5$ and $1.0$ (2000 maximum players and $100$ virtual servers).
As it can be seen from the figure, $\epsilon_{est} = 1.0$ greatly affects the algorithm behaviour.
Let us analyse this case more in detail. On the peak loads (around the 100th and 300th iterations) the prediction underestimates the load, and the manager assigns load to the peers. However, the actual load is higher, and it should results in a decreasing of availability (we will see this in the next picture). During low load phases (around the 200th iteration) the prediction overestimates the load, assigning more resources to the cloud, which results in a higher cost than necessary.
On the other hand, the picture shows a less significant impact of $\epsilon_{est} = 0.5$. 
This is due to the fact that the upper load factor threshold value ($LF_{up} = 0.8$) is enough to prevent nodes to reach an overloaded state even with a consistent error.

\begin{figure}[tbh]
\centering
\includegraphics[width=0.8\textwidth]{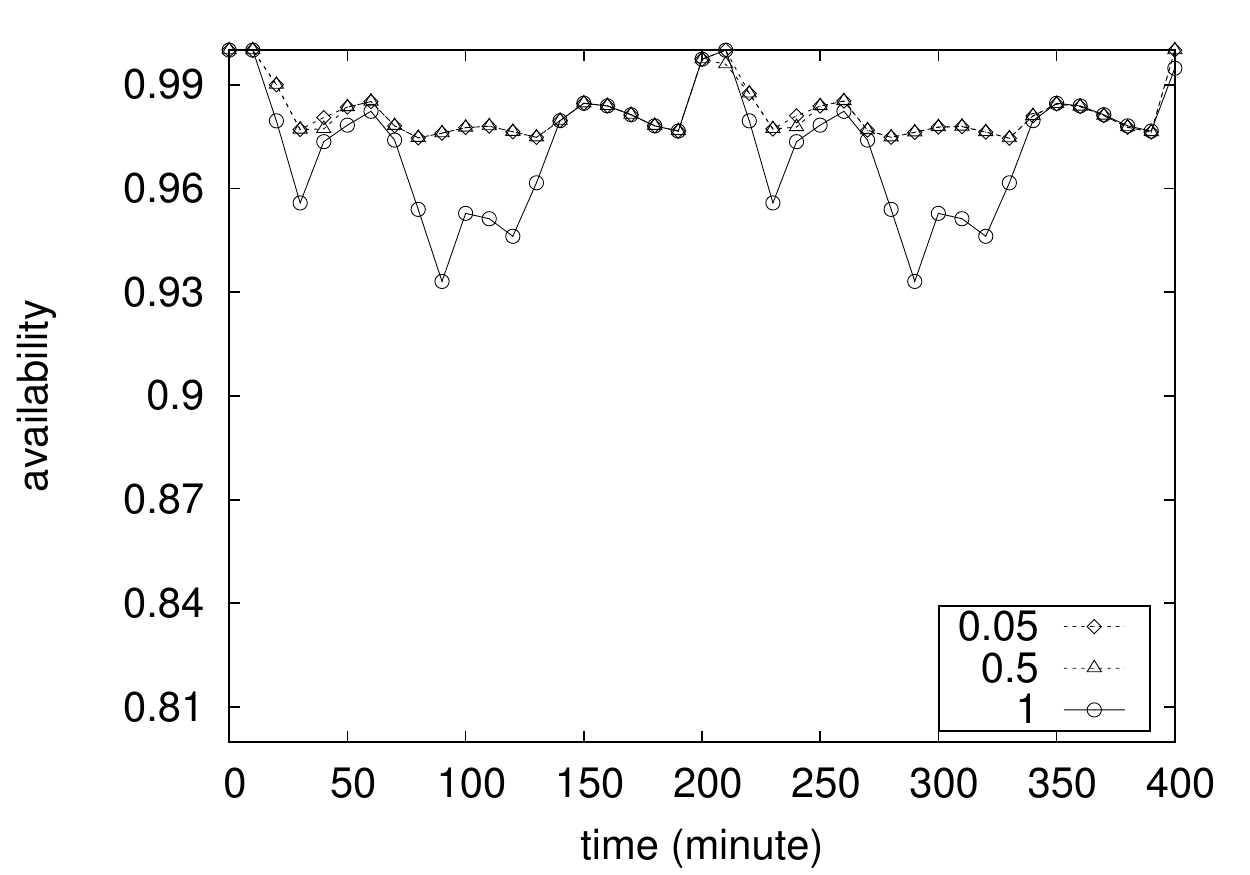}
\caption{Percentage of penalties per minute, different allowed maximum error.  2000 players, $100$ vs}
\label{fig:overload-on-epsilon}
\end{figure}

Figure \ref{fig:overload-on-epsilon} shows the availability with the same values for $\epsilon_{est}$.
As said above, the availability in case of $\epsilon_{est} = 1.0$ is worse than with the other values.
During the peak load the availability drops down of 5 percentage points.
It is interesting to note that $\epsilon_{est} = 1.0$ leads to an availability dropping in case of a load under estimation (around iterations 100 and 300) and to a cost increasing in case of load over estimation (\ref{fig:cost-on-epsilon}, around the 200th iteration).

%

\section{Conclusion}
\label{sec:conclusion}

In this chapter we have proposed an MMOG infrastructure that combines the advantages of P2P computing and on-demand resources.
The embedded flexibility of the proposed architectures is a valuable
characteristic for MMOG operators, which are able to decide which
nodes of the platform to exploit. 
Efficient and effective provision and load distribution algorithms are mandatory to realize MMOGs that may scale
to larger and larger communities of users. 
We have proposed a load distribution and provisioning strategy taking into account a number of relevant issues, such as balancing the infrastructure availability and reducing the economic cost. 
We have designed and a greedy heuristic policy characterized by low computational requirements. 
The experimental results show the effectiveness of our approach.

\chapter{Positional Action Manager}
\label{chap:pam}

As emerges from the background chapter, Neighbours Discovery (ND) is a fundamental issue in MMOGs infrastructures.
In this chapter we present PAM, a stand-alone component to resolve ND.
The main goal of PAM is to assure a cost-effective tunable and up-to-date ND.
To be up-to-date is a strict requirement for ND.
Stale information on neighbours does not value anything in a complex and evolving system like a MMOG. 
On the other hand, cost effectiveness is a requirement necessary to make the component appealing to MMOG operators.
The ability of tuning the trade-off between performance and economical cost can make the difference in a competitive market.
To fulfil both these requirements, PAM is composed by two combined services:
\begin{itemize}
\item a backbone server which we refer to as \textit{PAM server}, or only in this chapter, generically as \textit{server}
\item a fully decentralized network, which we refer to as \textit{PAM overlay}, or only in this chapter with the generic term \textit{overlay}.
\end{itemize}

\begin{figure}
        \begin{subfigure}[b]{0.45\textwidth}
                \centering
                \includegraphics[width=\textwidth]{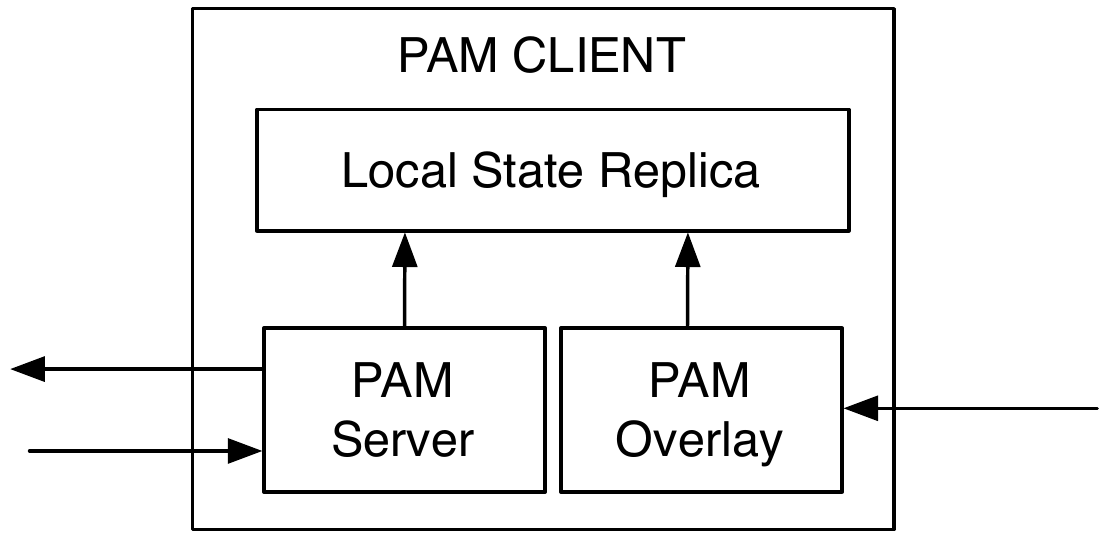}
                \caption{PAM Client}
                \label{fig:pam-client}
        \end{subfigure}
        \qquad
        \begin{subfigure}[b]{0.45\textwidth}
                \centering
                \includegraphics[width=\textwidth]{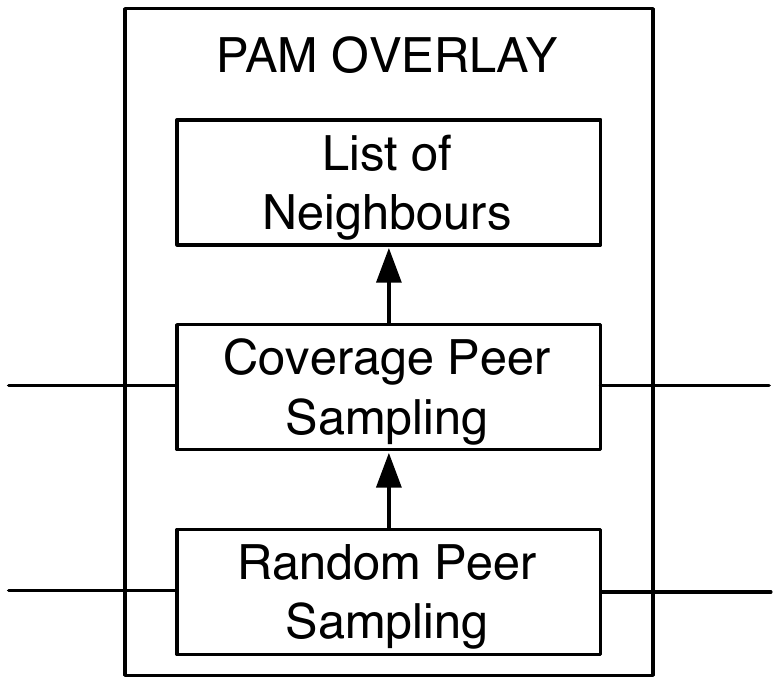}
                \caption{PAM Overlay}
                \label{fig:pam-overlay}
        \end{subfigure}
        \caption{PAM: Client Architecture}\label{fig:PAM-client-architecture}
\end{figure}

Figure \ref{fig:pam-client} shows the logical architecture of a PAM client.
PAM clients maintain connections to both these services in order to receive information about their neighbours. 
Periodically, clients communicate their position to the server.
On the other hand, periodically the server communicates to clients the list of their neighbours.
The rate of the these communications depends on the particular genre of the MMOG.
Fast pace MMOGs require these communication to be less than 100 ms, whereas slow pace MMOGs may employ larger interval on the order of 500 ms \cite{claypool2006latency}.
In this context we are interested in the rate of the communication server-to-client (which we refer to as $T_s$), since it represents the major source of cost of the infrastructure. Indeed, this value can be tuned for more precise ND with higher costs, or, on the other hand, less precise ND but lower costs. 
Besides the server, clients also communicate with a custom overlay.
The overlay is build such that to exploit the "wisdom of the crowd" principle.
Indeed, if a client maintains connections with other clients whose avatars are close in the virtual environment, there is a chance that they know about entities in the client AOI.
The overlay is built such that when a node $n$ has another node $m$ in its view, a connection between $n$
and $m$ exists in the overlay. However, this does not imply that the reverse connection exists. 
Nodes periodically query the overlay to learn about the entities in their AOI.
In this context the overlay plays a fundamental role. If the overlay is effective, the MMOG operator can increase the $T_s$, so to reduce the economical cost, without sacrificing a precise ND.

\section{PAM Server}

Logically, the server is composed by two asynchronous computation flows:
\begin{itemize}
\item a \textit{passive thread} that receives and stores the positions from the clients;
\item an \textit{active thread} that periodically informs clients about the content of their AOI.
\end{itemize}

These operations generate two different kinds of load on the server. First, they generate a computational load, as the server must maintain the connections, store the positions and resolve spatial queries. Second, they generate bandwidth load, as the communications to the clients consume outgoing bandwidth. This second kind of load, besides saturating the bandwidth capabilities, also increases the operational cost of the server, especially if the server is hosted by an on-demand platform.

In order to measure this load, we have conducted an experiment with an average size desktop machine working as a sever.
According to our empirical experience, a medium-sized server can manage around 1000 clients before slowing down. 

\begin{figure}[tbh]
\centering
\includegraphics[width=0.8\textwidth]{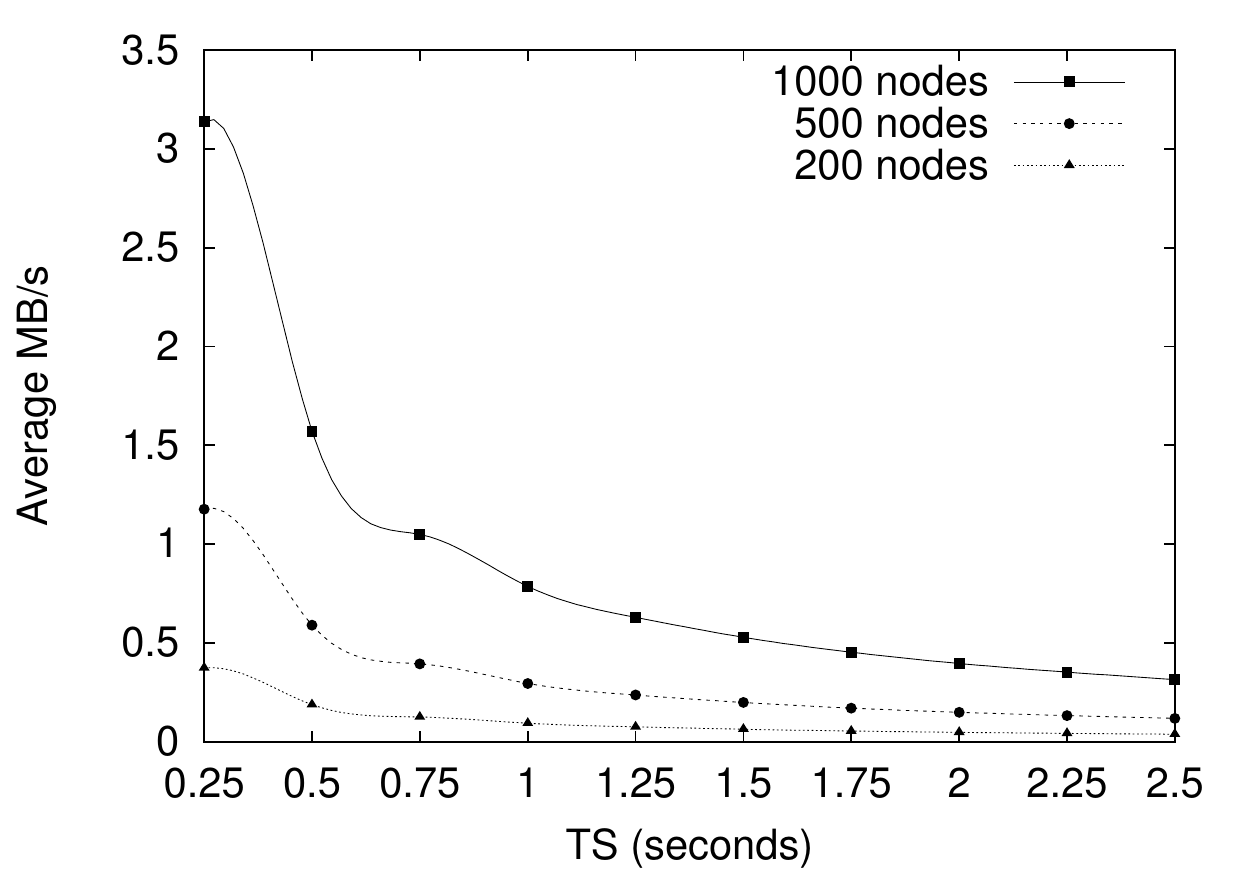}
\caption{Server outgoing bandwidth}\label{fig:band}
\end{figure}

However, reducing the frequency of client updates causes the server to suffer less load.
To measure the reduction of outgoing bandwidth at the server, we have performed several tests by varying $T_S$, which is the distance in time between two consecutive client updates from the server.
We have considered networks with 200, 500, and 1000 peers. 
Results are presented in Figure \ref{fig:band}.

As expected, the amount of outgoing data transfer is greatly reduced by increasing $T_s$.
With this reduction, the virtual environment operator is able to evaluate alternative choices regarding the deployment of the IM server.
For instance, let us consider an operator willing to deploy the IM server on a on-demand platform.
With 1000 nodes, and $T_s = 0.25$ the bandwidth requirement is 3MB/s.
Using the prices of a current commercial on-demand platform \footnote{0.12\$ per GB, Amazon EC2 prices, July 2012}, the deployment would cost 30\$ per day only considering bandwidth. With $T_s = 1$ the cost would be reduced to 10\$ per day.
This simple experiment shows how even a little reduction on the $T_s$ can result in a relevant saving for the virtual environment operator.

\section{PAM Overlay}

The construction of the PAM overlay has been driven by protocols based on epidemic diffusion of information.
These protocols (also known as as \textit{gossip} protocols) are the current reference point to build overlays in a pure distributed fashion. Gossip-based protocols provide seamless techniques for the initial bootstrap of the overlay and recovery node failures, as well as other interesting properties.

Our proposed overlay has been build with a mechanism inspired by T-Man \cite{Jelasity2008}.
T-Man has been one of the first approaches to fully describe the potential of gossip-based protocols in building overlays. 
Here we describe what are the main principles behind the creation of an overlay with gossip protocols, using T-Man as the main reference.

T-Man proposes a gossip-based probabilistic approach whose goal is to build, starting from an arbitrary initial peer configuration, a target overlay characterized by a set of well defined properties. 
These may be inferred by the profiles of the peers or directly characterize the topology of the target overlay. 
In the former case, for instance, a metrics based on the geographical location or on the semantic profile of the peer may be considered to define a  proximity-aware target overlay. An example of the latter scenario is a topology where the nodes are organized in a ring in increasing order with respect to their identifier. 

The definition of a proper {\em ranking function} is a core element to build the target overlay.
Each peer maintains a local view storing the descriptors of its neighbours. At each gossip cycle each peer exchanges a subset of its view with a subset of its neighbours. The ranking function is exploited to select the "best neighbours" according to the properties of the target topology. Hence, using only local gossip messages, the current topology gradually evolves towards the desired target structure with the help of the ranking function.

In large diameter topologies, an underlying random peer
sampling protocol should be exploited in order to speed up the convergence  toward the target topology. The random peer sampling 
serves also in the initial gossip cycles, when the local view of
the peer is empty and a peer needs to know a random sample of
peers to bootstrap on the network. 
Finally, a gossip-based approach for overlay construction is light-weighted, scalable and, when paired with a peer sampling service, it exhibits good convergence speed.

\subsection{Gossip-based Overlay Construction}

The effectiveness of our approach depends on the definition of a proper ranking function. In our case,
it should favour neighbours which may offer a larger number of entities in the interest set of a peer. To this end,
we will consider the spatial coverage of the AOIs of a peer's neighbours.
Unlike most existing T-Man-like approaches, our goal is to build a continuing evolving overlay rather than predefined one. The view of a peer changes continuously in order to reflect the position updates of the peers in the virtual space. In our case, instead of evolving toward a predefined target topology, peer continuously gossip to each other to support the retrieval of new avatars and objects in their AOI.

Our technique to build an overlay supporting IM is based on the following reasoning. 
Let us consider a given peer $P$.
At an arbitrary point in time it has in its local representation of the environment the replicas of the entities that belong to its AOI. 
When $P$ moves, its AOI changes accordingly.
Hence, to maintain its local representation up-to-date, $P$ must discover the new entities belonging to the new AOI.
In order to dynamically acquire this information, $P$ builds an overlay by considering a set relevant neighbours. 

The creation of the overlay poses two issues.
First, $P$ needs to know the identifier of its candidate neighbours; 
second, $P$ needs a mechanism to discriminate among peers, in order to choose the more
promising neighbours from the set of candidates.

The first issue is resolved by continuously refreshing candidate knowledge.
This is obtained with a two-layer gossiping architecture, where each layer runs a gossip protocol (the structure of these layers is shown in Figure \ref{fig:pam-overlay}).
In the underlying layer, called \textit{random peer sampling}, each client runs a random peer sampling protocol, which provide a subset of all the nodes in the system.
This layer enables each peer to maintain a set of long range links that guarantees the connectivity of the overlay.
In the second layer, called \textit{coverage peer sampling}, a gossip protocol connects peers by exploiting a ranking function based on spatial AOI coverage. 
Since the selection of the neighbours is done according the proximity, entities are progressively discarded by the second layer gossip if they disappear from its AOI. 
The two gossip layers are independent, in the sense that layers execute their gossip cycle at their own rate.
The random peer sampling layer communicates newly entered peers to the proximity layer. 
These communications are exploited in situation where a client has few knowledge about its nearby candidate
neighbours and must incrementally acquire new information.
These situations include the bootstrap phase and avatars teleportation, i.e. an avatar "jumping" from one place to another of the virtual environment.

The second issue is related to the AOI coverage offered by the neighbours of a peer. 
Each peer should choose the best configuration of its neighbours in order to optimize the number of entities which may be retrieved from them.
At each iteration the peer adapts its overlay neighbours set by providing a partial order from multiple configurations of neighbour sets. 
Since avatars are continuously moving, a large part of the
IM performances depends on the freshness of peers knowledge. In order to maintain the selection of the neighbours as
fresh as possible, each entry in the view of the peers is marked
with a time-stamp. Time-stamps provide an estimation on the
freshness of the entry. Our mechanism considers the age of the
entries in two situations. First, before to
rank the neighbour candidates, all the candidates whose age is
greater than a certain threshold are not considered. Second,
during the ranking, fresh configurations are favoured with
respects to the stale ones. In principle, the internal clock of the
peers can be used as the source for the time stamp. However,
for simulation purposes, we model the time as a discrete
successions of iterations. The simulation starts at iteration zero
for all the nodes, and for each gossip-cycle the count is
increased by one. When an entry is created, the iteration count
is used as time-stamp for such entry.

\subsection{Ranking Function}

In this section we explore in details the principles behind the ranking functions.
The definition of our ranking function posed two distinct challenges: (i) to measure the amount of area covered by neighbours peers and (ii) to determine the best subset of neighbour peers that maximize the area coverage. In the rest of this section we formalize these two problems and we provide a description of the adopted solutions.

\paragraph{Measuring Coverage}

\begin{mydef}[AOI coverage]
\emph{
Given a set of AOIs $\mathscr{N}=\{N_1 ... N_n\}$ and an AOI $P$ such that $P \notin \mathscr{N}$ we define as the coverage of P given $\mathscr{N}$, $c(P,\mathscr{N})$, as the area of $P$ that is overlapped by the AOIs contained in $\mathscr{N}$.}
\end{mydef}

\begin{figure}[tbh]
\centering
\includegraphics[width=0.45\textwidth]{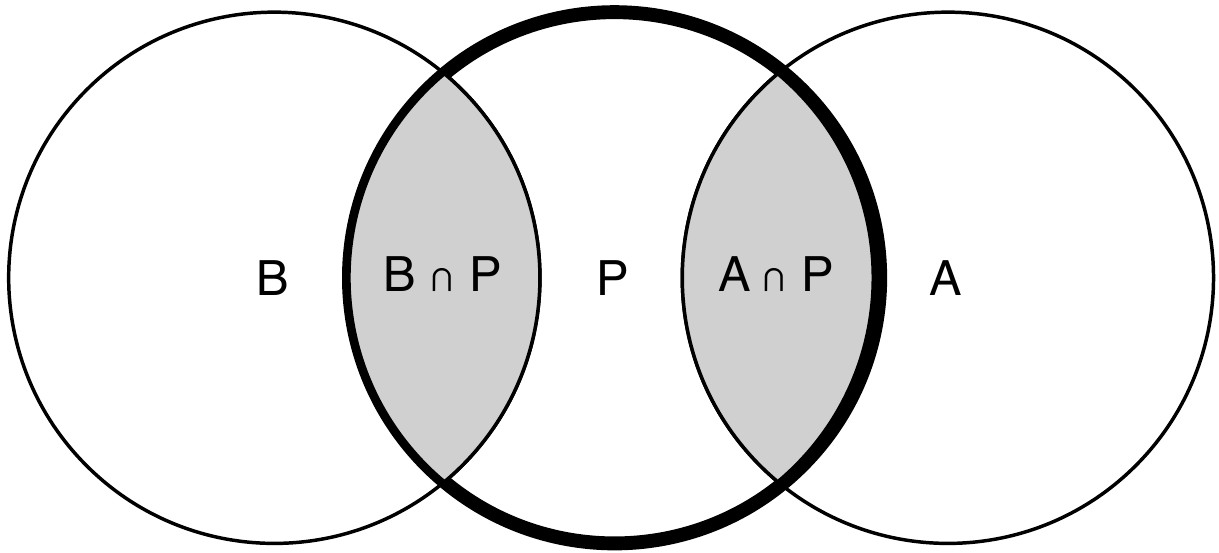}
\caption{Simple continuous $c(P)$ with $\mathscr{N} = \{A,B\}$.}
\label{fig:simple}
\end{figure}

\begin{figure}[tb]
\centering
\includegraphics[width=0.9\textwidth]{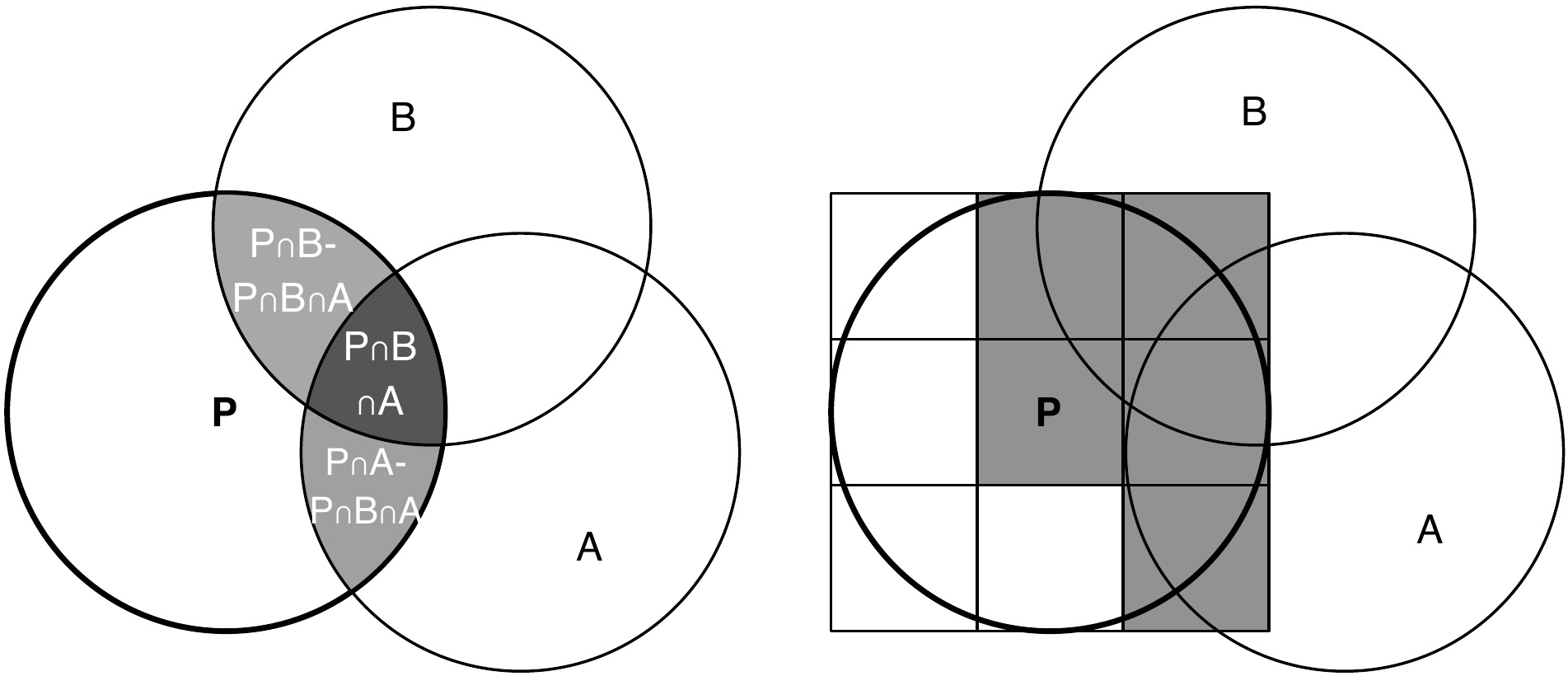}
\caption{Continuous and approximate coverage with $\mathscr{N} = \{A,B\}$. In this case the AOI coverage is approximated to $5/9$}\label{fig:app-cont}
\end{figure}

Computing $c(P,\mathscr{N})$ requires to compute all the unique intersections of AOIs in $\mathscr{N}$ with $P$'s AOI and to evaluate their area.
In trivial situations this is easy to compute. For example, Figure \ref{fig:simple} depicts a simple scenario where $\mathscr{N}=\{A,B\}$.
In this case the coverage is just the sum of the intersections of A and B with P, i.e. $ c(P,\mathscr{N}) = B \cap P + A \cap P$. 
However, in real situations, computing the AOI coverage is far from a trivial problem.
For instance, in the case depicted by Figure \ref{fig:app-cont} we have that $ c(P,\mathscr{N}) = (P \cap B - P \cap B \cap A) + (P \cap B \cap A) + (P \cap A - P \cap B \cap A) $.

When many peers are close to each other, to compute the effective coverage may be prohibitively expensive in terms of computational effort. Practically, this happens for two reasons. First, the number of the intersections grows quadratically with the number of peers. Second, it might be computationally costly to evaluate the area of an intersection resulting from many AOIs. 
For this reasons, we approach this issue considering an approximation.
The idea is to approximate the continuous surface of the AOI as a grid of disjoint tiles. In this way, instead of dealing with custom-shaped areas, we consider the tiles as the units to compute the coverage. This approximation reduces the complexity of the problem, since it makes easy to compute the area of each tile.
Moreover, the amount of tiles is a parametric value and does not depends on the number of peers.
Figure \ref{fig:app-cont} shows an example on how to compute the coverage of a given AOI (P in the figure) considering a 3x3 approximation.
The number of tiles varies proportionally with the degree of the approximation. A high number of tiles leads to higher precision, in principle increasing the performance of our mechanism. Besides, since the AOI to approximate is a circle, tiles at the corners of the approximation square might be out of the actual AOI area. In this case we do not consider such tiles for the coverage area estimation.

The pseudo code of the function $coverage()$ that realizes the tile-based coverage approximation is presented in Algorithm \ref{alg:approximate-coverage}. For each AOI $\in \mathscr{N}$ and for each tile, we check whether the AOI intersects with the tile. If it is, we check the counter associated to the tile. If the tile counter is zero, it means the tile is overlapped for the first time so we increase the \textit{covered\_tiles} counter. If the tile counter is greater than zero, we just increment it. Besides the number of the tiles covered, this function also counts the number of AOIs that cover each tile. It is easy to show that the complexity of the function is $O(n \times t)$, where $n$ is the cardinality of $\mathscr{N}$ and $t$ is the number of tiles.

\begin{algorithm}[h]
\SetKwInOut{Input}{Input}
\SetKw{Return}{return}
\SetKwInOut{Output}{Output}

\Input{P, the considered peer}
\Input{$\mathscr{N}$, the set of neighbours AOIs}
\Output{the approximated coverage given $\mathscr{N}$ and $P$}

\BlankLine

\KwData{covered\_tiles $\leftarrow$ 0}
\ForEach{AOI $\in$ $\mathscr{N}$}
{
  \ForEach{tile $\in$ getTiles(P)}
  {
    \If {intersect(AOI, tile)}
    {
      \If {tile.count = 0}
      {
      covered\_tiles $\leftarrow$ covered\_tiles + 1;
      }
      tile.count $\leftarrow$ tile.count + 1;
    }
  }
}
\Return covered\_tiles\;

\caption{Coverage(P, $\mathscr{N}$)}\label{alg:approximate-coverage}
\end{algorithm}

\paragraph{Maximizing AOI Coverage}

The aim of the network is to discover the larger amount of objects in the AOI of the peer (possibly all of them). The straightforward solution is to to keep links with the neighbours that maximizes the coverage. This indeed requires peers to make a choice, due to the bound imposed by the gossip view size. Hence, very often a peer needs to choose what is the best subset of peers to keep in its view. This subset is defined as follows:

\begin{mydef}[Maximum AOI coverage]
\emph{
Given a set $\mathscr{N}$ of AOIs, $\mathscr{N} = {N_1...N_n}$ and a natural number $d \leq n$ and an AOI $P \notin \mathscr{\mathscr{N}}$, we define $S_d = \{ X \in \mathscr{P}(N): |X| = d \}$, find the set $M \in S_d$ that maximizes the coverage of $P$.}
\end{mydef}

This problem is NP-complete. To prove that, we show how it corresponds to an instance of the \textit{set cover} problem. 
The set cover problem has been proved to be NP-complete by Karp in 1972 \cite{karp1972reducibility} and it is defined as follows.

\begin{mydef} 
[Set cover problem]  
\emph{Given a set $U$ of elements (called the universe) and $n$ sets of elements whose union comprises the universe, identify the smallest number of sets whose union contains all elements in $U$.}
\end{mydef}

The correspondence with the Maximum AOI coverage problem is resolved by considering: (i) $n$ as $\mathscr{N}$, (ii) elements as the tiles, and (iii) $U$ as the tiles covered by the AOIs in the optimal solution.

A naive solution to this problem would be to enumerate the possible combinations of peers and for each of them compute the coverage.
Unfortunately, this is highly impracticable since the combinatorial nature of the problem. 
Hence we propose two heuristics algorithms with different characteristics, a score-based and a greedy one.

\paragraph{Score-based Heuristics}
The rationale behind this heuristics algorithm is to assign a score to each tile. The tiles that intersect with few peers will receive a higher score than tiles intersected by a larger amount of peers. The idea is then to favour such peers that overlap high score tiles. The heuristics works as in the pseudo code in Algorithm \ref{alg:score-heuristic}. 

First, it computes the coverage of the AOI by considering all the peers in $\mathscr{N}$. 
Each tile has a score that is the reciprocal of the number of intersected AOIs.
Second, it computes the score for each AOI as the sum of the scores of each intersected tiles.
Finally, it sorts the AOIs in descending order according to their score, and it chooses the first $d$ entries.

\begin{algorithm}[tbh]
\SetKwInOut{Input}{Input}
\SetKw{Return}{return}
\SetKwInOut{Output}{Output}

\Input{P, the considered peer}
\Input{$\mathscr{N}$, the set of neighbours AOIs}
\Input{d, the size of the returned set}
\Output{a subset of $\mathscr{N}$ with cardinality $d$}
\BlankLine

coverage(P, $\mathscr{N}$)\;

\ForEach{AOI $\in$ $\mathscr{N}$}
{
  \ForEach{tile $\in$ getTiles(P)}
  {
    \If {intersect(AOI, tile)}
    {
      AOI.score $\leftarrow$ AOI.score + $\frac{1}{tile.count}$\;   
    }
  }
}

sort AOIs in descending order according to score\;
\Return the first $d$ AOIs\;

\caption{Score-based Heuristics}\label{alg:score-heuristic}
\end{algorithm}

The complexity analysis of the score-based algorithm heuristics goes as following: 
(i) the coverage procedure, which we have already seen to be $O(nt)$, 
(ii) the computation of the score, that can be considered as $O(n)$, 
and (iii) the sorting, which is $O(n \log n)$. 

\begin{figure}
\centering
\includegraphics[width=0.7\textwidth]{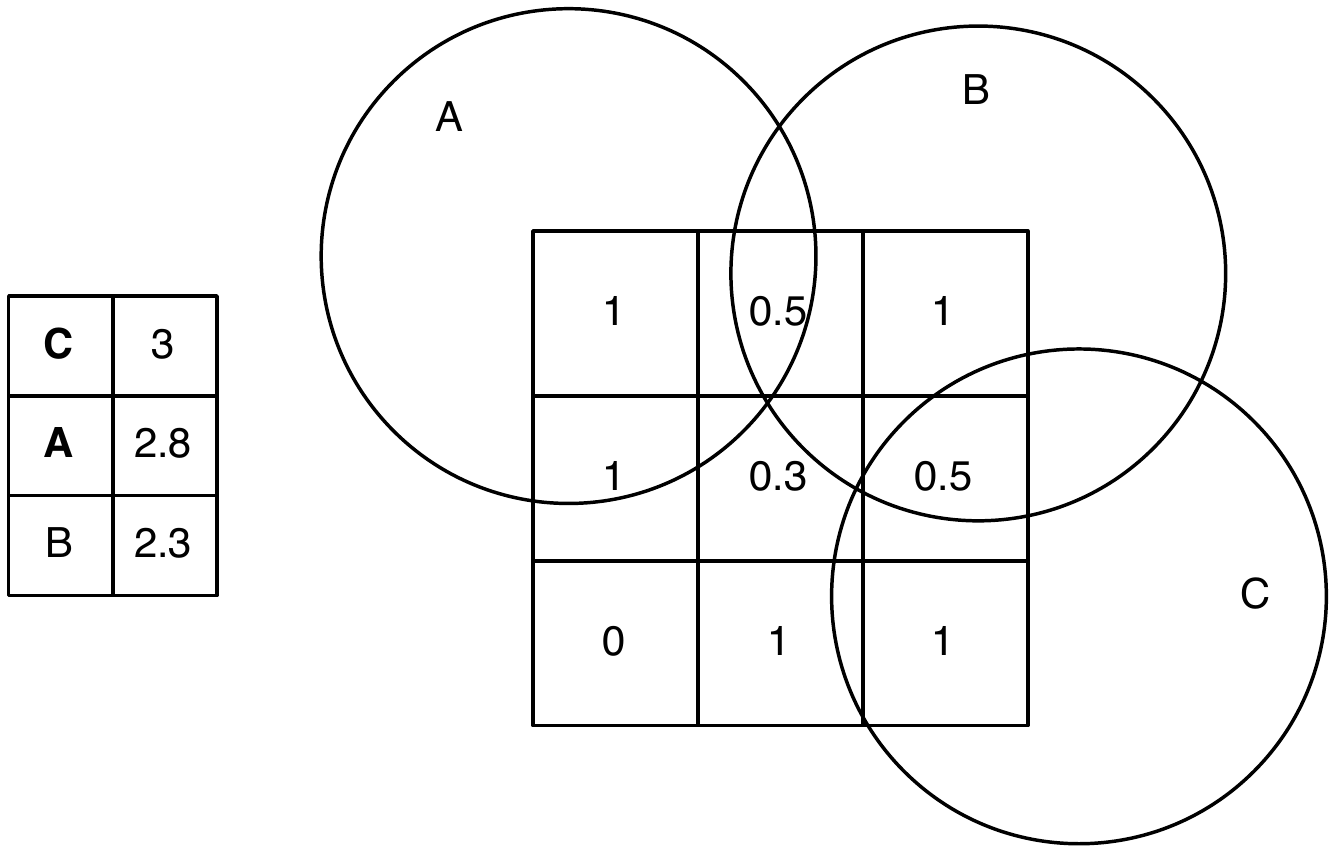}
\caption{Graphical examples of the score-based heuristic}\label{graph:score}
\end{figure}

\begin{figure}[tbh]
\centering
\includegraphics[width=0.7\textwidth]{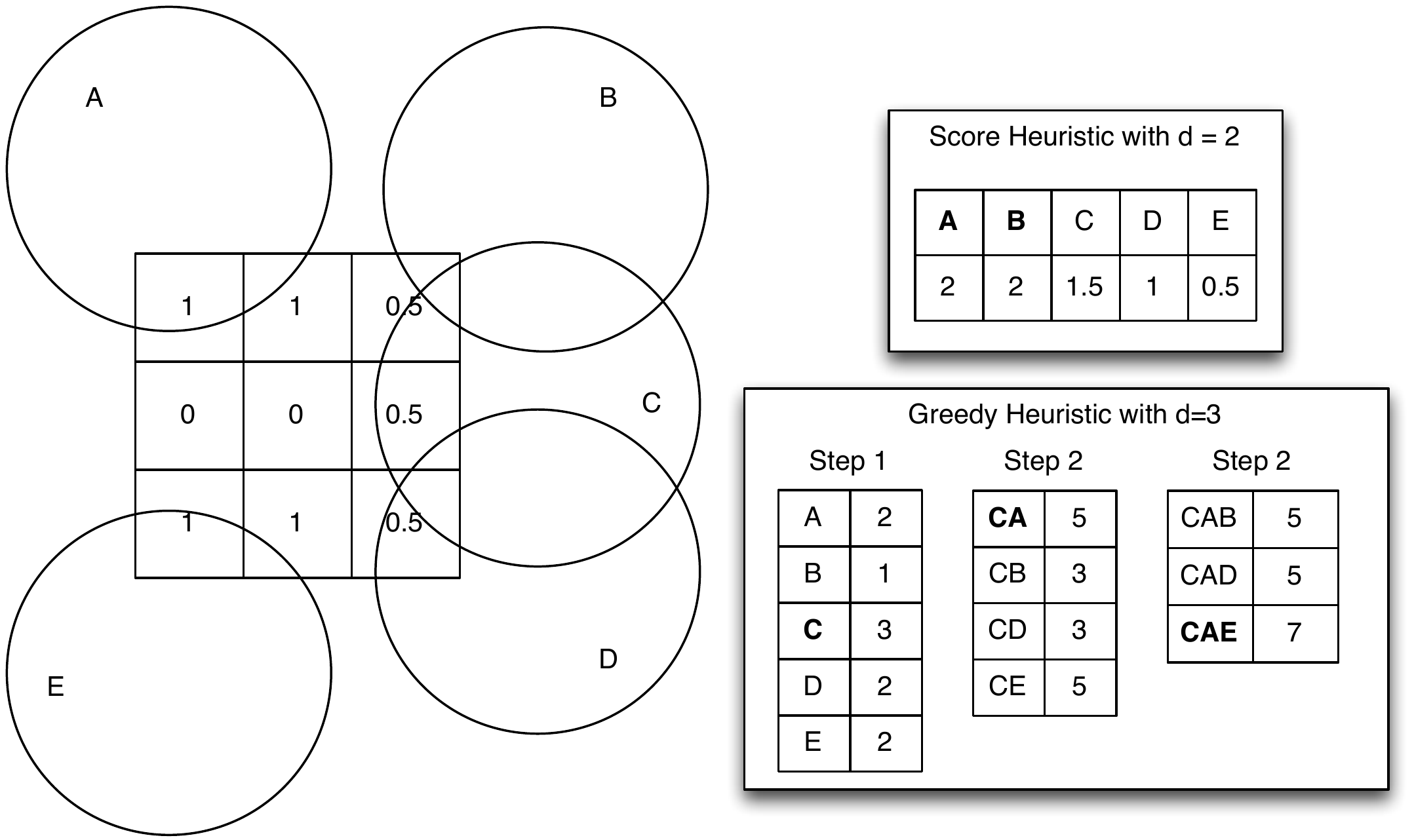}
\caption{Greedy and Score Heuristic with $\mathscr{N}={A,B,C,D,E}$}\label{fig:gheur}
\end{figure}

Figure \ref{graph:score} shows a graphical execution of the score-based heuristic algorithm. 
For example, the central tile has a score of 0.3 since A and B and C intersect with it. If we consider $d=2$, the heuristics chooses the combination $\{A,C\}$, which is also the best combination possible. However, the heuristics not always finds the optimum. Let us consider the example in Figure \ref{fig:gheur}. In this case the score-based heuristic algorithm chooses as the best combination $\{AE\}$ that covers 4 tiles instead of $\{AC\}$ or $\{EC\}$ that cover 5 tiles each.

\paragraph{Greedy Heuristics}
The idea behind the greedy heuristics is simple: at each step to choose the peer that yields the higher increment on the number of unique tiles covered. The pseudo-code of the greedy heuristic algorithm is represented at Algorithm \ref{alg:greedy-heuristic}.
For each peer in the view, it is selected the AOI that maximizes the number of further covered tiles considering the already chosen AOIs. Note that: (i) an AOI can be selected only once as, upon selection, it is removed from the list of candidates, and (ii) to evaluate the number of tiles covered we use the function $coverage()$ described and evaluated in the previous section. 

\begin{algorithm}[tbh]
\SetKwInOut{Input}{Input}
\SetKwInOut{Output}{Output}
\SetKw{Return}{return}

\Input{P, the considered peer}
\Input{$\mathscr{N}$, the set of neighbour peers}
\Input{d, the size of the returned set}
\Output{a subset of $\mathscr{N}$ with cardinality $d$}

\KwData{C $\leftarrow \emptyset$}
\BlankLine

\While{$|C| < d$} 
{\label{greedy:outer}
  chosen $\leftarrow \emptyset$\;
  max\_score $\leftarrow$ 0\;
  \ForEach{AOI $\in$ $\mathscr{N}$}  
  {\label{greedy:inner}
    score $\leftarrow$ coverage(P, C $\cup$ AOI)\;
    \If{score $>$ max\_score}
    {
       max\_score $\leftarrow$ score\;
       chosen $\leftarrow$ AOI\;
    }
  }
  remove chosen from $\mathscr{N}$\;
  add chosen to C\;
}
\Return C\;
\caption{Greedy Heuristics}\label{alg:greedy-heuristic}
\end{algorithm}

The complexity analysis of the greedy heuristic algorithm goes as following.
The outer cycle (line \ref{greedy:outer}) is repeated $d$ times. 
The inner cycle (line \ref{greedy:inner}) is repeated at maximum $|\mathscr{N}| = n$ times. 
The function $coverage()$ has a complexity of $O(nt)$.
Hence, the total complexity in time is $O(td \times n^2)$.

Figure \ref{fig:gheur} shows a graphical execution of the greedy heuristic algorithm. At the first step, the heuristics chooses $C$, as it is the AOI that covers the most tiles. At the second step, $A$ is chosen so that the current combination becomes $\{CA\}$. At the third step $E$ is chosen, and the final combination is $\{CAE\}$. Note that at the second step, the heuristics could have chosen $E$.
In such case the second combination was $\{CE\}$ that would have lead to the same results (i.e $\{CAE\}$).

To prove approximation guarantees of the greedy heuristics, first we have to introduce \textit{submodular} function \cite{nemhauser1978analysis}. 
Consider $\Omega$ to be a finite set and an arbitrary function $f:2^{\Omega} \rightarrow \mathbb{R}$, we can say $f$ is submodular if it satisfies the following property: the marginal gain of adding an element to a set $S$ is at least as high as the marginal gain from adding the same element to a superset of $S$. 
More formally a submodular function must satisfy 

\begin{equation}
f(X \cup {x}) - f(X) \geq f(Y \cup {x}) - f(Y)
\label{eq1}
\end{equation}

for all elements $x \in \Omega$ and for all pairs $X \subseteq Y$.
Now, suppose $f$ to be submodular, \textit{non-negative} (i.e. takes only positive values) and \textit{monotone} (i.e. adding an element to a set cannot cause $f$ to decrease). Let also suppose that our aim is to find a set $S$ of cardinality $k$ such that $f(S)$ is maximized. It has been proved in \cite{nemhauser1978analysis} that a greedy algorithm resolves this problem with a worst-case approximation of $(1-1/e)$, where $e$ is the base of the natural logarithm. In other words, if the optimum value is 100, the greedy algorithm is guaranteed to find a solution with a value of \textit{at least} 63.

In order to apply this result to our greedy algorithm, $c(P, \mathscr{N})$ must be submodular, non-negative and monotone. Non-negativity is immediate, since we measure an (approximation of). Monotonicity is also immediate, since adding an AOI to a set cannot change the number of tiles already counted. 
To prove submodularity, we show how it satisfies (\ref{eq1}). Let us consider what happens when we add an arbitrary AOI $x$ to a set $X$ whose $Y$ is a superset of: (i) $x$ neither intersects with AOIs in $X$ or AOIs in $Y$. In this case the equality holds since the marginal gain for both sides of the equation is zero; (ii) $x$ intersects only with AOI's in $X$. In this case we possibly have an increment on the left side, so the equality holds; (iii) $x$ intersects only with AOI's in $Y$. In this case the left part of the equation is greater, since it considers all the area covered by $x$, whereas the right part is incremented only of the part that is non overlapping, so the equation holds; (iv) $x$ intersects with both $X$ and $Y$. The equation holds since for the left side it counts also the intersection of the AOI's with the elements in $Y$, that it would not count for the right side.
Finally, since we have proved that our greedy algorithm is submodular, non-negative and monotone we can assert that in the worst case we obtain an approximation of $(1-1/e)$.

\section{Result}

This section presents the description of the metrics and a selection of experimental results evaluating the key performances of the approach.

\subsection{Metrics}

To evaluate our approach we considered two different metrics. 
The first metric evaluates the coverage of peers AOI. We refer to this metric as \textit{AC}. \textit{AC} is a value in the interval $(0,1)$ and, given a peer at an arbitrary iteration, is defined as  the ratio between the AOI coverage obtained by the P's view and the best AOI coverage defined by considering all the peers in the virtual environment. 
The second metrics measures the difference between the local replica of the peer' state against the server state.
To measure this difference, we exploit a slightly modified version of the \textit{Jaccard similarity coefficient} \cite{lee1999measures}.
Let us consider $C$ as the local replica of a peer and $S$ as the remote replica of the server.
The original Jaccard coefficient is computed as $S \cap C / S \cup C$.
However, this formulation either does not take in account the difference of the positions of the entities, or considers entities with different positions as distinct.
In order to take into account at the same time the difference in position and the presence of the entities we exploit the following formula to compute the Jaccard coefficient (in short \textit{JC}):

\begin{equation}
JC = \frac{
\sum_{x_S, x_C \in S \cap C} 1 - |\frac{dist(x_S, x_C)}{d_{MAX}} | }
{S \cup C}
\end{equation}

\noindent
where $d_{MAX}$ is the diameter of the peer's AOI.
A peer with $JC=1$ has its local replica perfectly synchronized with the state of the server while
$JC=0$ implies that the replica is completely out-of-sync with that of the server. Any value in between 0 and 1 gives a quantitative evaluation on the quality of the synchronization.

While AC measures how good the heuristics performs in a dynamic environment, JC measures the quality of the approach in terms of the quality of the application. A direct correlation between AC and JC would be desirable. 
The experimental results supports the existence of this correlation.

\subsection{Behaviour over $T_s$}

\begin{figure}[tbh]
\centering
\includegraphics[width=0.8\textwidth]{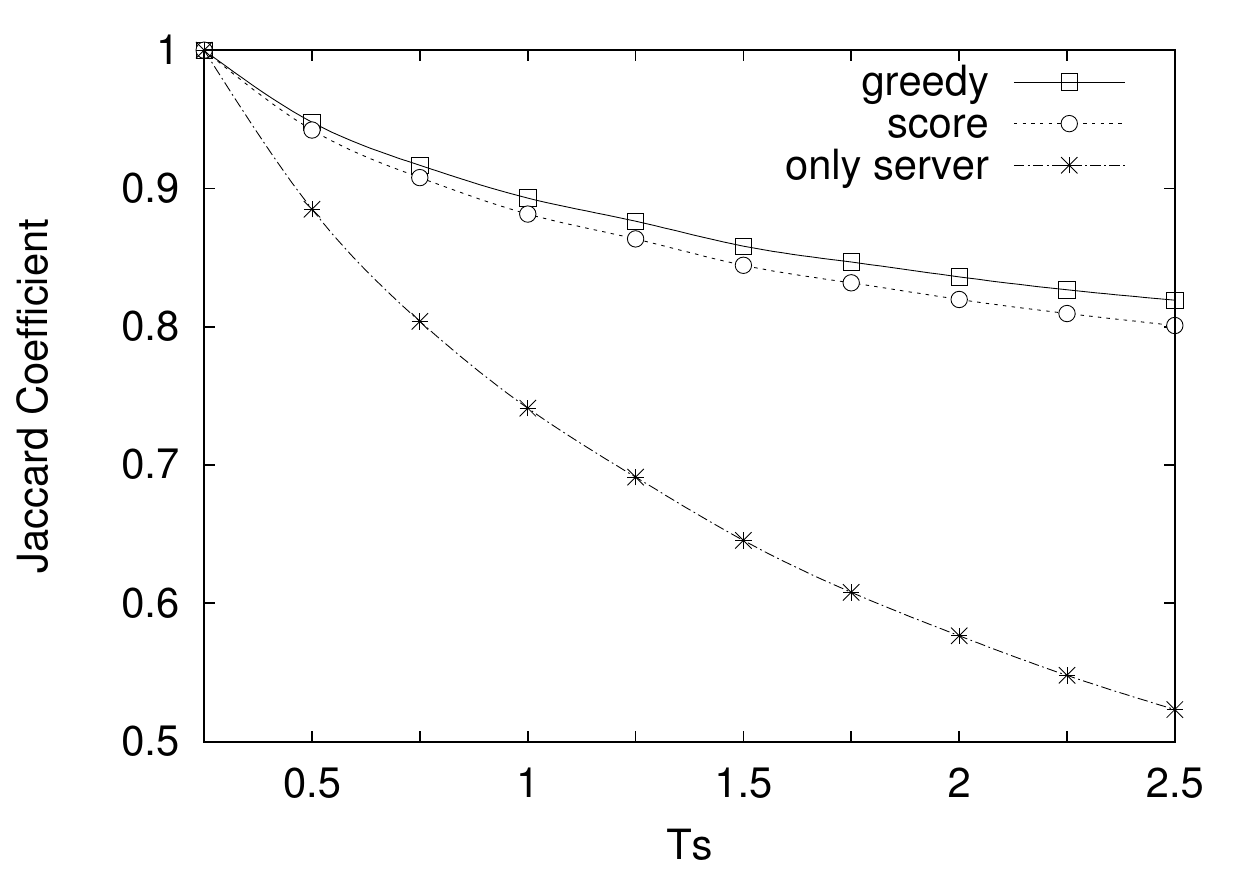}
\caption{Comparison between score-based heuristics and greedy-based heuristics}
\label{graph:heur}
\end{figure}

In this section we discuss the result of several simulation runs by varying the interval of time ($T_s$) between two consecutive communications to the central server. A gossip cycle and a query to the overlay are executed every 0.25 seconds. For instance, with $T_s = 1$ there is a server communication followed by three requests to overlay in row and then another server communication. 
%
%
Where it is not indicated differently, the simulations consider: 500 nodes with a cache of 10 elements each, 1000 objects, and an AOI approximation of 32x32 tiles. 
Figure \ref{graph:heur} shows the comparison of the JC between the greedy heuristic and the score heuristic algorithms.
In general, we can observe how the reduction in the JC is limited even with high values of $T_s$.
For instance, with $T_s = 1$ the average JC value for both the heuristic algorithms is around 0.9.
From an application point of view, this means that the mechanism is able to fully support IM.
As expected, further increments of $T_s$ imply a JC reduction. Note however, that even with the $T_s = 2.5$ and the support of the PAM-overlay, the JC is still around 0.8.
The effectiveness of the mechanism is further supported by the values of the JC when using only the server.
In other words, increasing $T_s$ would be problematic if not supported by the PAM-overlay. For example, with $T_s = 1.5$, the JC with the support of the overlay is around 0.9, whereas is 0.65 using only the server.

%
As regards the comparison between heuristic algorithms, the greedy slightly outperforms the score heuristics.
With these simulation parameters, the AC, which is independent from $T_s$, is 0.8 and 0.85 respectively for the greedy and the score heuristics. This suggests a correlation between AC and the JC.

\begin{figure}[tbh]
\centering
\includegraphics[width=0.8\textwidth]{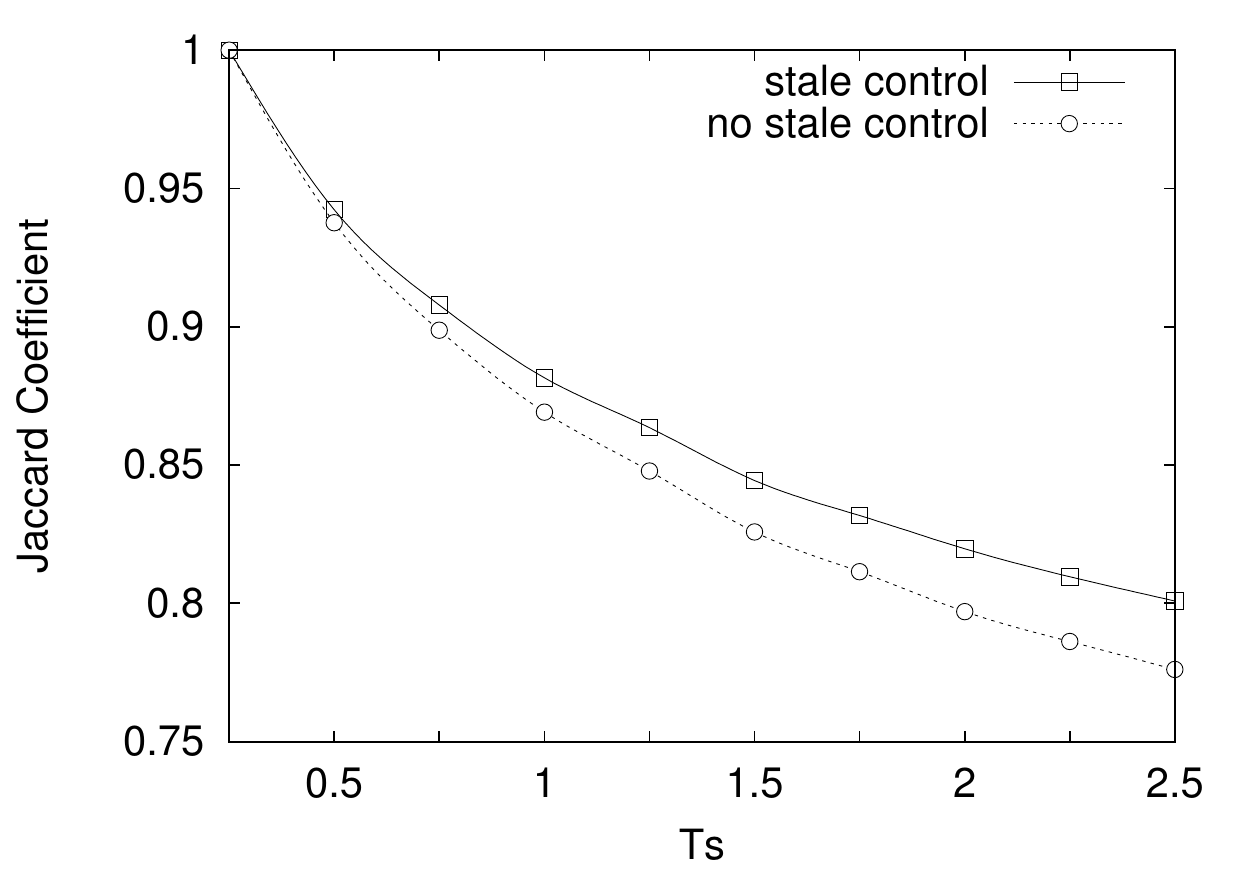}
\caption{Score heuristics, comparison between considering or not freshness of entries}
\label{graph:ts}
\end{figure}

Figure \ref{graph:ts} shows the JC when selecting the more fresh entries during a gossip iteration. The ranking algorithm considered is the score, but similar results have been obtained with the the greedy heuristics. The results are evident and not surprising: to prefer fresh entries gives a neat increment on the performance. 
As the previous, even this result indicates a  correlation between AC and JC as the score's $AC=0.80$ with stale control and $AC=0.70$ without.

\begin{figure}[tbh]
\centering
\includegraphics[width=0.8\textwidth]{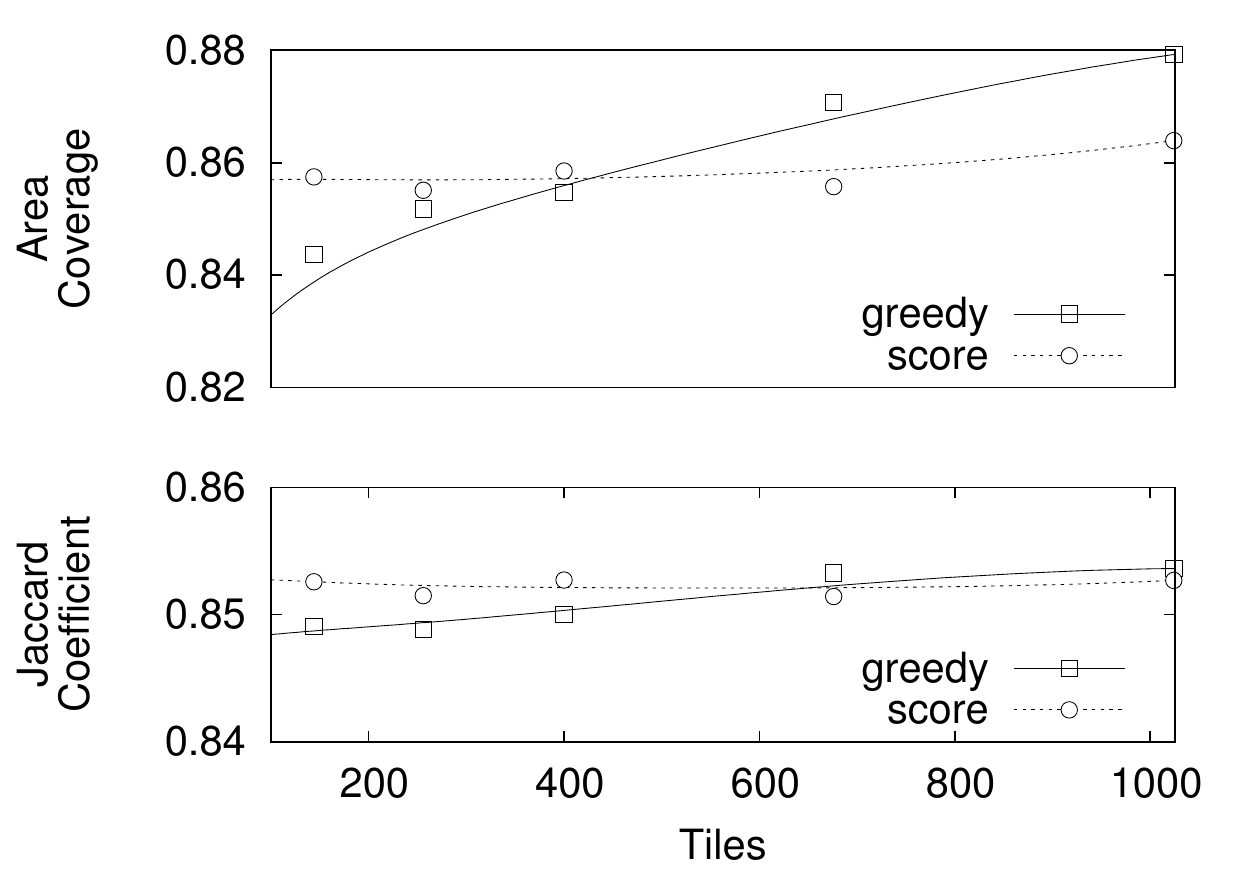}
\caption{AC and JC with different number of tiles}
\label{graph:tiles}
\end{figure}

\subsection{Tiles Variation}

Figure \ref{graph:tiles} shows the JC and the AC of the greedy and score heuristics with various degree of AOI approximation (from 16 to 1024 tiles). The data in the plot have been obtained by averaging the outcome of 20 independent simulation runs with a $T_s = 1$.
The results show that the score-based heuristics is basically agnostic to the approximation whereas the increment in the number of tiles implies an increment of the performance of the greedy-based heuristics. From the graph it is clear how with approximations larger than 400 tiles for AC and 600 for JC, the greedy-based heuristics outperforms the score-based.
The reason why it happens lies on the order used by the greedy heuristics for choosing the areas. Indeed, with higher approximation, the greedy has greater chance to choose a worse area. When the approximation is reduced, the greedy heuristics performance increases.

\begin{figure}[tbh]
\centering
\includegraphics[width=0.8\textwidth]{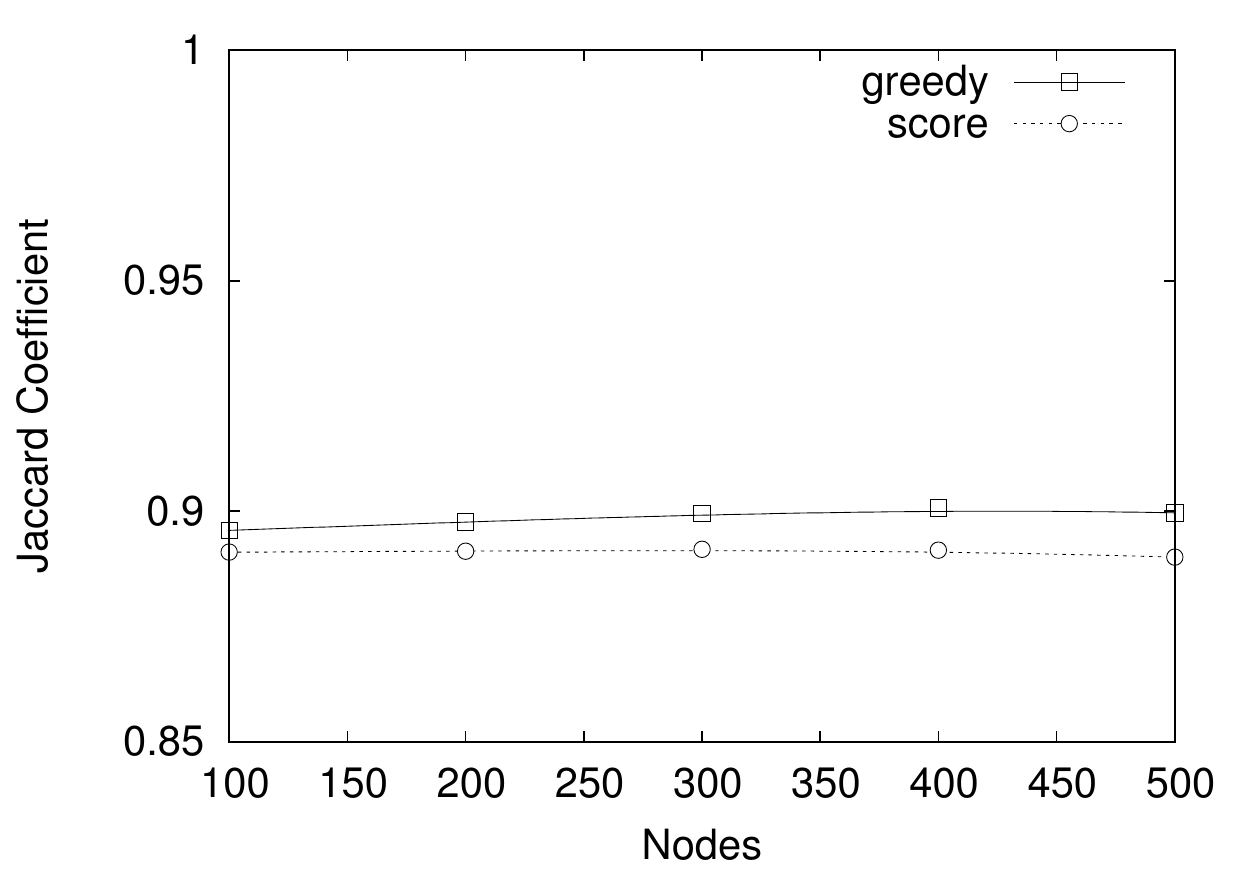}
\caption{JC with different values for network sizes}
\label{graph:scal}
\end{figure}

\begin{figure}[tbh]
\centering
\includegraphics[width=0.8\textwidth]{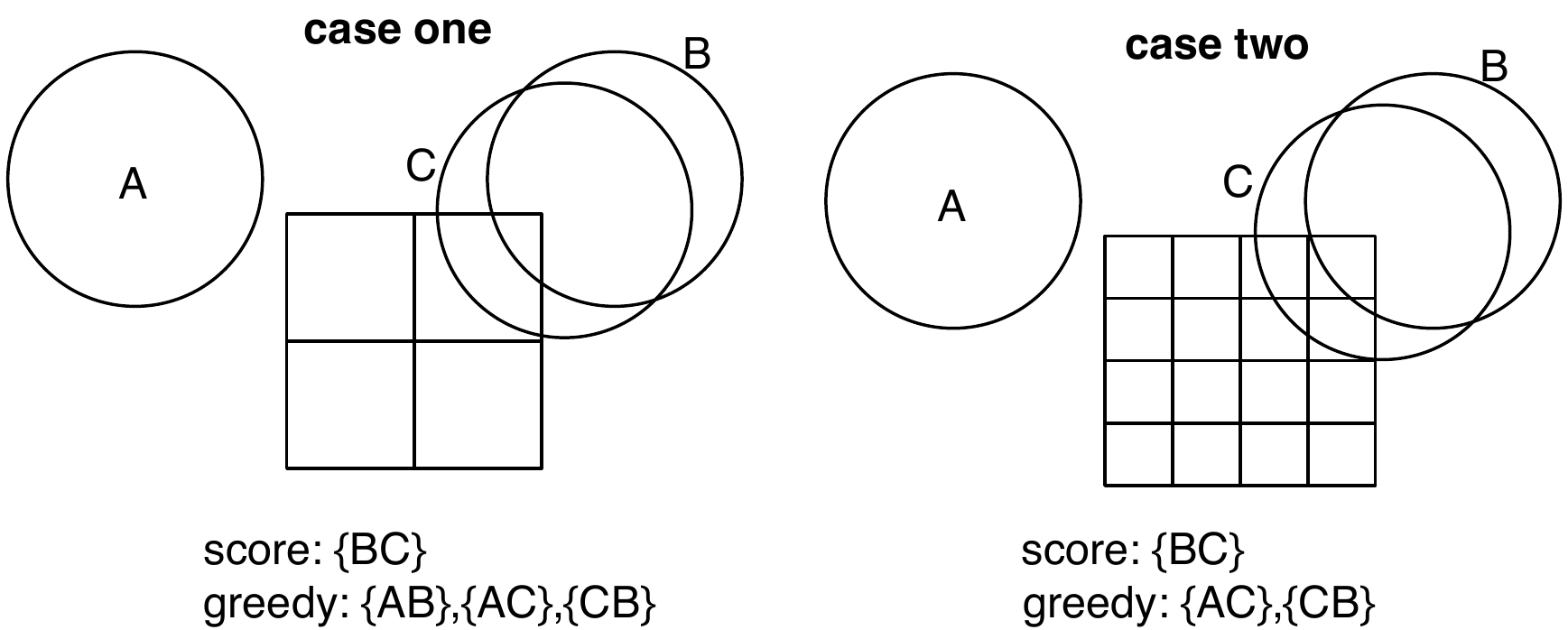}
\caption{Impact on the heuristics with different tiles approximation}\label{fig:approx}
\end{figure}

The reason of this can be explained clearly with a simple example, depicted in Figure \ref{fig:approx}.
Let us consider the same case with two different AOI approximation degrees, and also consider the maximize coverage problem with
$d = 2$ and $N = \{A,B,C\}$. In the first case (4 tiles) the greedy heuristic would choose either $AB$, $AC$ or $CB$. This is because at the first greedy step, either $C$ or $B$ would be selected, even if $C$ has a better coverage then $B$.
In the case with a better approximation (16 tiles), the greedy-heuristic would choose $C$, which leads to $CA$ or $CB$ that are both optimal solutions. When the approximation is further reduced, then the greedy-based heuristic works better than the score-base one, due to the better performance with respect to the optimum.

\subsection{Number of Peers}

Figure \ref{graph:scal} shows how the amount of peers affects the JC of greedy and score heuristics.
Each point in the plot is the average of the outcome of 20 independent runs. The simulations have been run with a 32x32 AOI approximation, a fixed cache size of 10 elements, and a $T_S = 1$.

As expected, the greedy overcomes the score, but they show a similar behaviour. 
Their performance are essentially independent from the number of nodes, even with a fixed-size cache.
In fact, there is a slight increment on the JC as the number of peers increases, due to the higher changes of crowded zones with more nodes.
This situation allows the node to exploit the knowledge of the neighbours more often.

\section{Conclusion}

In this chapter we described a gossip-based mechanism to build overlay for best-effort Neighbours Discovery in MMOGs.
Conversely to the other approach in the field, we trade some precision in the result to keep the mechanism fast, simple and lightweight.
The simulations have shown encouraging results: even when the delay between two consecutive communications with the server is very large the PAM overlay is able to obtain good results. Further, an increasing in the number of nodes increases the performance of the overlay.

Our proposal can be further extended and studied.
To this end, we plan to improve the precision of the result by considering additional information when ranking peers, such as movement forecasts and different neighbours selection functions. As to further validate our solution, we intend to test it with movement traces from different mobility models and to compare it with the non best-effort work presents in literature.
Part of these lines have been further studied, and they are explained in
Chapter \ref{chap:architecture}.
In any case, we expect that other solutions will be proposed following this line.
Indeed, techniques and mechanisms to combine Peer-to-Peer approaches (including best-effort ones) and centralized solutions (like on-demand computing) are one of the research topics for the next-generation MMOG infrastructures.

\chapter{Toward a Complete Architecture}
\label{chap:architecture} 

MMOG architectures must support a number of features whose requirements are often in contrast with each other.
A single MMOG can be accessed by multiple user concurrently, therefore the architecture must account for maintaining consistency.
To scale up to thousands of users, interest management and load balancing schemas are a necessity. Since the nodes of the network can potentially be untrustworthy and unreliable, security and fault tolerance mechanisms must be considered.
As we have described in detail these aspects in Chapter \ref{chap:background}, here we focus on the design of a full, concrete architecture for MMOGs.

The State Action Manager (SAM) and the Positional Action Manager (PAM) were presented through this work as two stand-alone components. The SAM (see Chapter \ref{chap:sam}) manages the state actions, by exploiting a Distributed Hash Table to distribute the effort on management of the entities to multiple resources, including user-provided and on-demand resources. The PAM (see Chapter \ref{chap:pam}) is the component devoted to the management of the positional actions. It employs a combination of a centralized server and gossip protocols to acquire the position of relevant entities in the proximity of the users.

This chapter presents a preliminary study on the combination of SAM and PAM in a concrete architecture.
This combination is described more in detail in the next section, where we consider a client-centric perspective.
A seamless integration of PAM and SAM also requires that they share the same mechanisms to recruit and release on-demand computing resources. 
The PAM-server presented in Chapter \ref{chap:pam} is described as a single server architecture.
Section \ref{sec:multiserver} presents a multi-server version of the PAM, with an insight of possible mechanisms for load distribution in PAM.

\section{Combining PAM and SAM}

The combination of PAM and SAM in a seamless architecture exposes two main issues.
First, it is mandatory the definition of a client capable of correctly exploiting SAM and PAM, in order to be able to participate to a virtual environment. Second, a common infrastructural platform that comprehend SAM and PAM must be defined. For example, the same instance of an on-demand resource may run the SAM or the PAM component in different moments. 
Before entering in the details of the this second issue, we provide the description of the PAM that includes the multi server support.

\subsection{Client's Perspective}

\begin{figure}[tbh]
\centering
\includegraphics[width=0.7\textwidth]{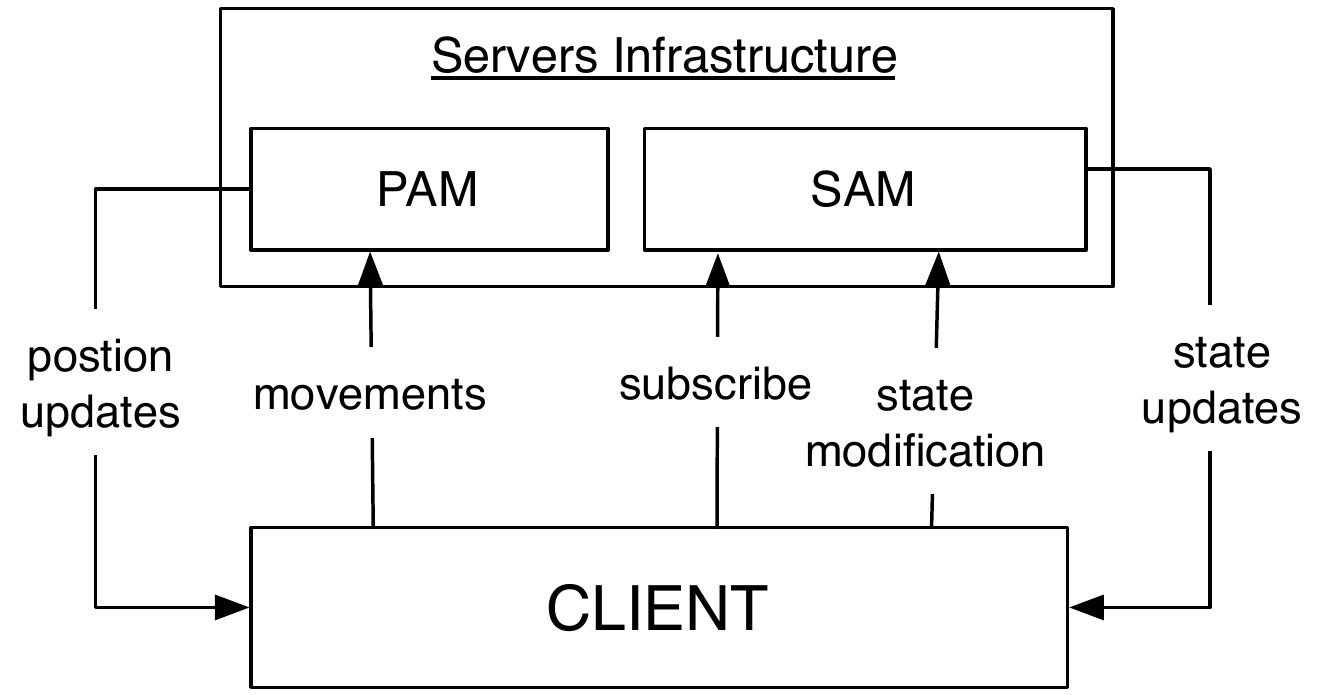}
\caption{Client-centric view of the proposed architecture}\label{fig:interest-management}
\end{figure}

From a client perspective, PAM and SAM should be used together to provide a MMOG.
In this section we describe the interaction that a client has with the two components to reach this goal.
The interactions are presented as messages, which are depicted in Figure \ref{fig:interest-management}.

\paragraph{Positional actions}
Clients periodically send its current position to the PAM server. The frequency rate of the positional action depends on the particular virtual environment. It is expectable that fast-paced MMOGs have an higher frequency with respect to slow-paced MMOGs. In any case, the PAM-server can employ an \textit{optimistic} approach to compute the position of the avatar between two positional actions. This kind of approaches, of which the most popular is Dead Reckoning \cite{pantel2002suitability}, are explained in Chapter \ref{chap:background}.

\paragraph{Positions updates}
The client receives from the PAM the positions of the entities in its AOI.
This information can be provided either by the PAM-server or by the PAM-overlay, as described in Chapter \ref{chap:pam}.
The frequency whereby the PAM server sends position updates to the clients is defined by the game operator.
Either way, the client considers a larger sized AOI with respect to the one actually visualized by the players.
This allows the client to perform a sort of pre-fetching for possible interesting entities for the player.
With this optimization, the client can visualize it immediately to the player, as soon as the entity enters in the player's real AOI.
However, the size of the pre-fetching AOI must be correctly chosen.
It must be defined a trade-off between the increment in performances and the overhead due to the maintaining of a larger AOI.

\paragraph{Subscription}
In order to receive state updates from the SAM, clients must subscribe to the entities contained in their AOI.
From a client's perspective, the subscription process consists of: (i) providing the client's IP to the servers that manages the entities, so it can receive state updates and (ii) store the IP of the server, to send possible modifications of the state of the entity.
To explain the subscription process, let us proceed with an example.
Let us assume that a generic client needs to subscribe for the entity $e$.
To do so, the client can contact an arbitrary node in SAM.
This is possible since internally the SAM is organized like a Distributed Hash Table (see Chapter \ref{chap:sam}). 
In a DHT overlay, each node of the DHT is able to find the node that manages any entity in $O(\log n)$ steps, where $n$ is the number of nodes. However, even few hops may result in a high latency delay.
Fortunately, the client already considers a larger pre-fetching AOI, which can also be used to mask (part of) this latency.
Another issue is the definition of the node to contact for subscribing.
It could be either a specific node, or a server currently serving the client.
Further, it can stay the same or change over time. All these matters must be verified with additional simulations.
When an entity leaves the (pre-fetching) AOI of a player, the client must unsubscribe from the server.
Compared with the subscription, this action is rather simple. The client sends the proper message to server, of which it knows the IP already, and the server removes the client from the list of the subscribers.

\paragraph{State modification}
The client sends \textit{state actions} (Chapter \ref{chap:background}) to the PAM-server. Upon the reception of the state actions, the SAM server resolve possible conflicts and computes the authoritative version of the state of the entities. 

\paragraph{State updates}
Periodically, the SAM server sends state updates to the clients. Note that the updates are sent for the entities in the larger pre-fetching AOI, so to have them available as soon as they are needed. Upon reception of this message, the client updates the internal state of the local entities.\\

\begin{figure}[tbh]
        \centering
        \begin{subfigure}[b]{0.45\textwidth}
                \centering
                \includegraphics[width=\textwidth]{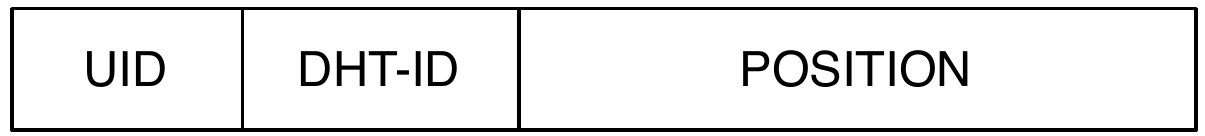}
                \caption{PAM descriptor}
                \label{fig:pam-descriptor}
        \end{subfigure}%
        \quad
        \begin{subfigure}[b]{0.45\textwidth}
                \centering
                \includegraphics[width=\textwidth]{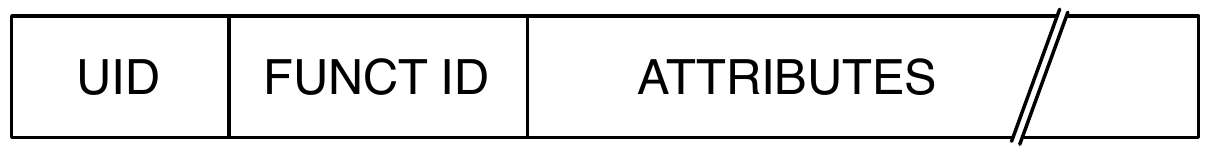}
                \caption{SAM descriptor}
                \label{fig:sam-descriptor}
        \end{subfigure}
        \caption{The entity descriptor is split between PAM and SAM}\label{fig:sam-pam-desc}
\end{figure}

The combination of PAM and SAM implies the entities descriptor (as defined in Chapter \ref{chap:background}) to be split between the two components. Figure \ref{fig:sam-pam-desc} graphically compares the two descriptors. In particular, the PAM descriptor considers the position of the entity, whereas the SAM descriptor stores the list of the attributes.

Another important step in the combination of PAM and SAM is to employ a common provisioner for the on-demand resources.
The idea is to extend the SAM's manager with the capability to provision also for the PAM.
The definition of a cost model for the PAM is a fundamental requisite for this task.
This model must represent and, if possible predict, the amount of bandwidth necessary to satisfy the players requests.
Also, by employing a prediction mechanism similar to the one in SAM, it would be possible to pro-actively provision on-demand resources.

\section{Multi-Server PAM}
\label{sec:multiserver}

In Chapter \ref{chap:pam} we described PAM by considering an infrastructure composed by a single server.
However, even if PAM's scalability is increased by the gossip network, a single server may be a bottleneck when the number of players overtakes certain limit. Here we propose the design of a multi-server architecture for PAM.
The management of the MMOG among multiple server requires a strategy to distribute the virtual environment.
Since PAM manages only the positional actions, it makes sense to employ a spatial distribution of the area. 
%

\begin{figure}[tbh]
        \centering
        \begin{subfigure}[b]{0.45\textwidth}
                \centering
                \includegraphics[width=\textwidth]{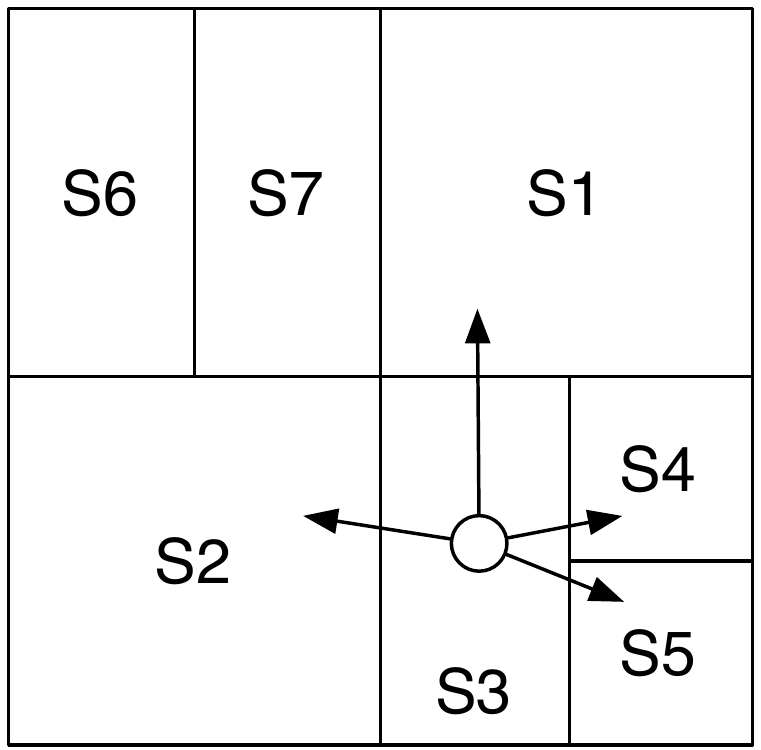}
                \caption{Original CAN. Servers can manage only a single region, and have link with side neighbours}
                \label{fig:can-original}
        \end{subfigure}%
        \quad
        \begin{subfigure}[b]{0.45\textwidth}
                \centering
                \includegraphics[width=\textwidth]{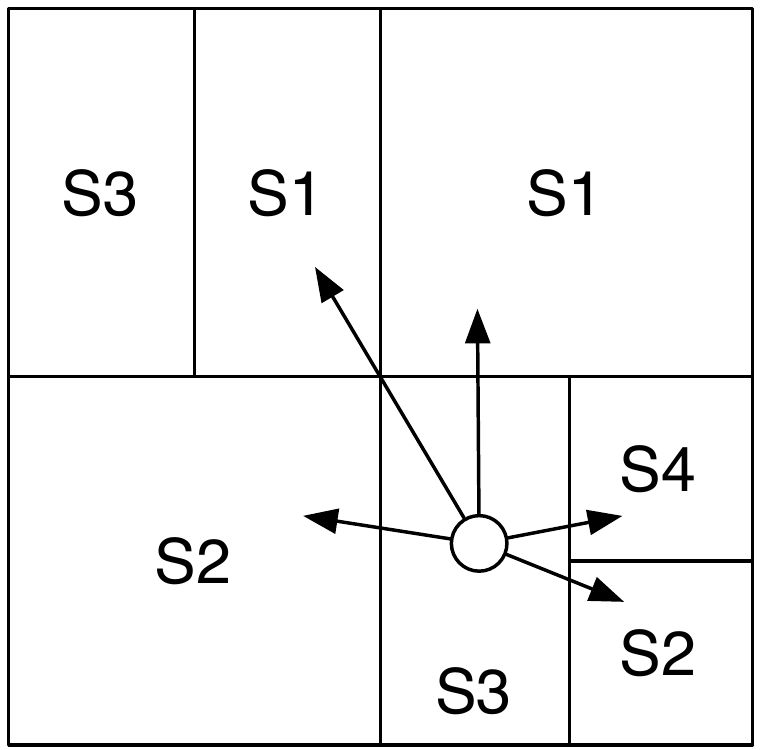}
                \caption{Enhanced CAN. Servers can manage more regions, and have link with side and diagonal neighbours}
                \label{fig:can-super}
        \end{subfigure}
        \caption{Original vs Enhanced CAN}\label{fig:can-vs}
\end{figure}

We enable this distribution by employing the CAN DHT \cite{ratnasamy2001scalable}.
CAN considers a bi-dimensional address space, which is divided into squared areas, called regions.
Each regions is associated to a node. Every data is represented by a 2-dimensional point.
Each node handles all the data whose point lies in the managed region.
CAN handles the joining and the leaving of a node by recursively merging (when joining) or splitting (when leaving) the regions.
In our context, we exploit CAN by creating a correspondence between the CAN and the virtual environment regions.
Any entity of the MMOG is managed according to its virtual position, as it was data in the original CAN.
Figure \ref{fig:can-original} shows an example of space in CAN.

In order to fully exploit the CAN DHT for our purposes, we extends some of its basic behaviour.
First, we have made possible for a node to manage more regions (Figure \ref{fig:can-super}).
There is no limit on the number of region a node can manage, and it can manage regions that are non-adjacent.
The mechanism is similar to the virtual server in SAM (Chapter \ref{chap:sam}), and so are the benefits.
However, here we apply the concept in a spatial address space, rather that the flat one considered in SAM.
The immediate advantage of this design is the possibility for the server to migrate regions between themselves.
This allows the infrastructure a mean to orchestrate the load, so to optimize the performances.

Second, we have provided CAN nodes with the ability of resolving \textit{spatial range queries}.
Conversely to a basic single item query, a spatial range query requires to find all items in an area.
Spatial range queries are periodically resolved to provide the with the entities in their AOIs.
In the single server PAM this operation is straightforward as the server has knowledge of all the positions of the entities.
However, in the multi-server PAM an avatar's AOI can overlap one or more servers.

In the original CAN overlay each node has a link with the servers that manage the adjacent neighbour regions. (Figure \ref{fig:can-original}). To support spatial range queries we have improved the base CAN overlay, by adding also links with the diagonal regions (Figure \ref{fig:can-super}).
This simple addition greatly improves the efficiencies of spatial queries. 
To clarify this point, let us consider a player whose avatar is in the region managed by S3 in Figure \ref{fig:can-vs}. 
Let us also consider that the player's AOI overlap with the server S7.
In the original CAN, S3 would have contacted S1, which in turn would have contacted S7.
Conversely, in the improved CAN S1 can directly contact S7.

To be sure that every spatial query is resolved within a single hop, we put a limit on the dimension of a region, so that it cannot be smaller than the AOI of a player. In this way, by imposing the minimum size of regions as the AOI size, and being the regions squared, an arbitrary AOI can overlap at maximum four regions.

\paragraph{Avatar Movement}

As soon as a client connects to the PAM, it contacts a bootstrap node (which can be any node arbitrary node of the CAN DHT) to know the server to connect. The client maintains a direct connection with this server for two purposes. The client pushes the position of the avatar, and the server informs the clients with the entity in the avatar's AOI. However, when an avatars move to another region, it has to switch the connection to the new server. This change is fully handled server-side.

When the server receives a position update, it checks if the avatar is still in the same region. 
If it is the case, then it just updates the new position of the client.
Otherwise, the server identifies the neighbour server that is the new server for the avatar.
If we assume that the player cannot move larger that the size of a region in a single step, all this information is enough to find the new server. 
Now the server sends an hand-off message to the new server, by communicating the new incoming avatar and its position.
Afterwards, it removes the avatar from its own list. The new server adds the client to its list, and notifies to the client the successful operation. 
It can be the case that the new zone belongs to the same server. In this case the operation is completely transparent to the client.

\subsection{PAM Load Distribution}

In this section we analyse a load balancing mechanism for the multi servers PAM.
How we have seen during the whole thesis, the load in a MMOG follows seasonal patters.
Hence, here we consider not only load balancing, but also the orchestration of the on-demand resources for the multi-server PAM.

As we have seen in Chapter \ref{chap:pam}, an average server can manage few thousands players concurrently.
Even if the gossip-based interest management of PAM help increasing this limit, it eventually comes the time when a server is overloaded.
Here we consider a single node, let us call it \textit{orchestrator} managing  the load balancing and the recruit/releases of the on-demand resources. 
Here, we do not specifically address how the node for the orchestrator is selected. Consider for example that the first server to participate to PAM takes the role of orchestrator.

The initial consideration is on the definition of load.
For a PAM server, the load mostly originates from the resolution of the range queries.
Hence we measure the load as the average number of range query resolved per period of time.
This number should also include the requests that come from neighbours regions.
Also, we argue that the orchestrator should control the load per region, rather than per server.
This would give the possibility to have a finer control on the load distribution.

By considering the load of the region, and the assignment regions to servers, the orchestrator performs the following actions:

\begin{itemize}
\item In case of unbalanced load, the orchestrator either migrates a region from a heavy loaded server to an unloaded server, or splits up the region and migrates only part of it.
\item In case the number of servers was not enough to support the load, the orchestrator would recruit a new on-demand resource. The new server can be assigned with a newly created region (by splitting a overloaded region) or by migrating an existing region from a heavy loaded server. In the situation where the capacity of the servers is too high (over-provisioning), the orchestrator can release a server by assigning its regions to under-loaded servers.
\end{itemize}

Whatever the decision of the orchestrator, an important factor is the migration of servers.
In particular, it is important to consider and mask the migration time as much as possible.
In this case, it comes in help the design of the PAM.
The servers inform the players periodically, with a period defined as $T_s$ in Chapter \ref{chap:pam}.
If we were able to keep the migration time shorter than $T_s$, than the migration would be completely hidden from the user.

\section{Cheating}

As we pointed out in Chapter \ref{chap:background}, cheating is defined in \cite{Neumann2007} as "an unauthorized interaction with the system aimed at offering an advantage to the cheater".
In this section we provide an overview on possible cheating exploitation in SAM and PAM, as long as possible solutions to mitigate the issue. 
As in general in security, the following ideas are based on the principle of making the cheating hard to exploit, rather than providing bulletproof data protection.

\subsection{SAM}

A malicious player can exploit two possibilities to cheat in SAM:
\begin{itemize}
\item By taking advantage of her role as a client, and exploiting well-known cheating techniques used for centralized MMOGs, such as the \textit{suppressed update} (see Chapter \ref{chap:background}). We will refer to this kind of cheating as \textit{client-cheating}.

\item By taking advantage of her role as the server, in case virtual servers are assigned to her machine. 
For instance, it would be possible to favour updates of an ally player against updates from an enemy player.
We will refer to this kind of cheating as \textit{server-cheating}
\end{itemize}

As discussed in \cite{Webb}, employing a referee-based schema provides a good level of protection against client-cheating.
In a referee-based mechanism, entity updates must pass through a special node called \textit{referee}, whose task is to validate those updates.
In case of invalid updates, these are discarded. 
This schema fits the SAM architecture, as every server can play the role of referee.
In case of cheating evidence, the cheater can be reported or disconnected.

Allowing user-provided server to play the referee role exposes the architecture to server-cheating. 
For instance, a malicious server can mark as invalid the updates of a player, just to force her disconnection or reporting.
In this case, further measures should be taken into consideration.
For instance, backup nodes might perform checks on a sample of the updates, in place of the user-provided servers.
However, information exposure on the state of the other players can still provide an unfair advantage to players running a server.
A possible solution in this case would be to encrypt the entity values, so that only the backup server and the clients can decrypt the content. 
Due to the rapid staleness of entities' information, the encryption/decryption algorithm can be relatively simple, provided that the encryption key is refreshed often.
Also, we would a user-provided server to not manage an entity if the respective client is currently accessing it.
These ideas can also be integrated with rewarding mechanisms for virtuous user-provided servers, so to diminishing the value of the cheating.

\subsection{PAM}

As in the SAM, a referee-based approach is useful to mitigate the client cheating, with the servers playing the role of the referees.
However, differently from the SAM, players' nodes of the PAM communicate directly with each other.
In this scenario, it would be important to provide the authenticity of the sender and the integrity of the information.
A possible solution would be to use signed certificates to guarantee the provenance of data.
However, even if this mechanism is robust, it may be too clumsy for fast-paced applications like MMOGs.

A more flexible approach would be to apply a distributed reputation-based mechanism.
In this case, a node has the ability to report a malicious node to the nodes in its proximity.
In order to mark an arbitrary node $P$ as malicious, nodes check $P$'s information against the authoritative version of information coming by the server. If the difference between the two is suspicious (e.g. the information provided by $P$ is not compatible with the information from the server) then a report about $P$ is spread in the network.
When nodes receive enough reports about $P$ they could ignore the information coming from $P$, and, possibly, remove $P$ from the peer sampling, in fact isolating it from the network.

\section{Conclusion}

In perspective, our independent analysis of PAM and SAM have obtained encouraging results.
In this Chapter we have provided ideas to combine PAM and SAM in a complete architecture for MMOGs.
However, to fully validate this approach several pieces are still absent.
First, a unifying cost model for PAM and SAM must be designed to properly orchestrate the resource between the two components. Further, an extensive evaluation must be performed to fully validate the solution.



\chapter{Related Work}
\label{chap:related}


This thesis considers the integration of pure distributed and on-demand computing models to effectively support MMOGs.
The application of these two aspects has requested a wide and detailed study of the related work on the field.
In this chapter we discuss and, when possible, compare the approaches that, to the best of our knowledge, are more relevant with respect to our work. 

Due to the inherent heterogeneity of the aspects discussed, this chapter has been divided in several sections to ease its fruition.
Section \ref{centralized} offers an overview on the design of centralized infrastructures for MMOGs. In particular, we discuss the emerging research issues of the last few years, that is the application of on-demand platform to MMOG infrastructures.
Section \ref{p2p} collects a summary of work tackling the problem of building MMOG infrastructures in a pure distributed fashion.
Section \ref{hybrid} concludes the part dedicated to the infrastructure design, by describing the approaches that employ a combination of centralized and distributed computing models for MMOGs. 
Section \ref{mobilitymodels} shows an overview on the mobility models for virtual environments.
Section \ref{casestudies} selects several interesting case studies, which enclose most of the problematic discussed in the Chapter.
Finally, Section \ref{related-conc} concludes the chapter.

\section{Centralized Infrastructures}
\label{centralized}

One of the main design choice for a MMOG is related to the distribution of the virtual environment.
On one hand, centralized infrastructures rely a server or a cluster of servers to manage the system.
On the other hand, user-assisted systems exploit the resource provided by the users, mostly using P2P technologies. In this section we provide an overview of the characteristics of centralized infrastructures MMOGs.


Centralized infrastructures rely on a cluster of servers, typically located in a single data center, to manage the virtual environment.
In such systems a set of client machines, paired with the users, share the game state by connecting directly to the cluster, which acts as a point of centralization. Whenever a client issues an action, this is sent to the server that updates the state of the world accordingly, and notifies the new state to the interested set of clients.

In a centralized infrastructure, clients merely work as interfaces to present the virtual world to the users. 
Differently from user-assisted infrastructures (which we discuss in the following) clients of a centralized infrastructure do not manage any part of the state that is not related directly with their user. Also, in centralized infrastructure there are no direct connections among the users.

The first proposals for centralized multi-user virtual environment go back to the middle nineties.
Most of such works, have focused on how to overcome the lack of scalability of a single server machine.
One of the principal method, which is still used nowadays, is to limit the communications between the server and user by sending only the necessary set of entities. For example, \cite{funkhouser1995ring} presents detailed algorithms to compute users visibility in the virtual environment, and avoids to broadcast updates to the user that are not in their proximity.

Centralized architectures exploit multiple servers, often organized in clusters, to manage the virtual environment. When considering multiple severs, the first issue is to properly distribute the virtual environment among the servers and to setup a pattern of communication among the servers. Multiple distribution schemas have been proposed in the last decade.
A common classification \cite{prodan2009prediction,glinka2007rtf}  considers these as the most frequently used models: \textit{Instancing}, \textit{Mirroring} and \textit{Zoning}.

\begin{figure}[tbh]
        \begin{subfigure}[b]{0.3\textwidth}
                \centering
                \includegraphics[width=\textwidth]{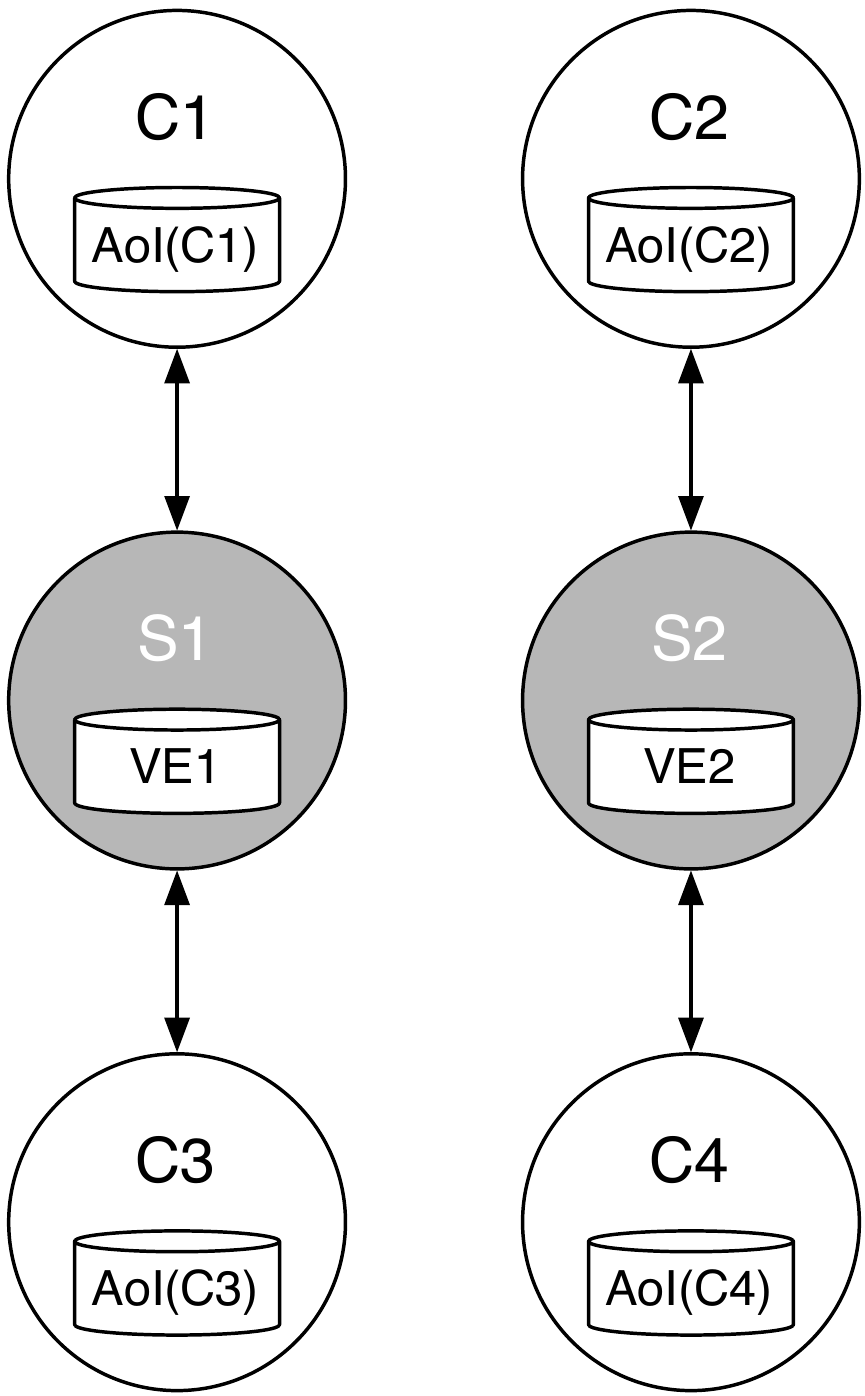}
                \caption{Instancing}
                \label{fig:instancing}
        \end{subfigure}
        \quad
        \begin{subfigure}[b]{0.3\textwidth}
                \centering
                \includegraphics[width=\textwidth]{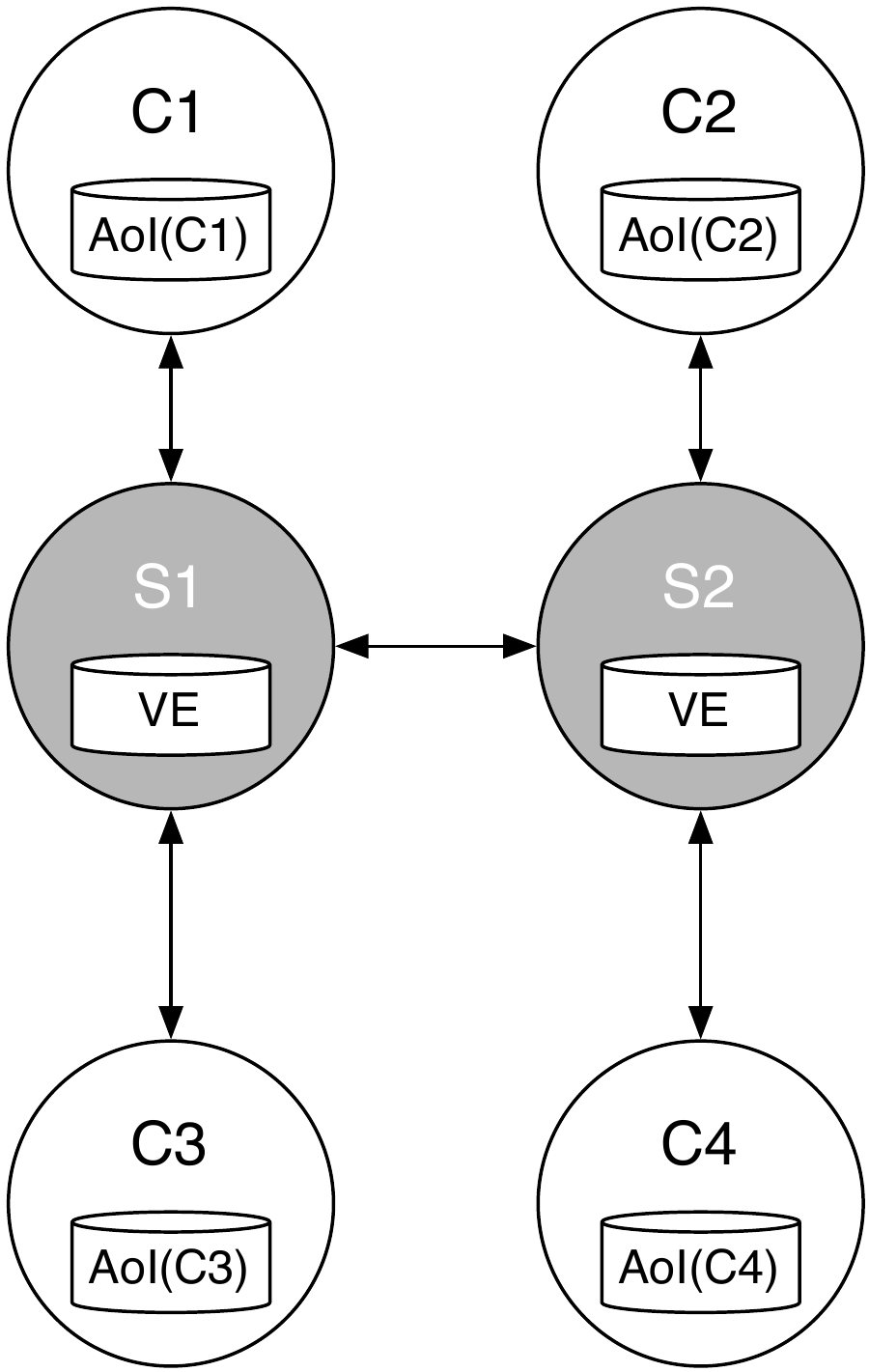}
                \caption{Mirroring}
                \label{fig:mirroring}
        \end{subfigure}
        \quad
        \begin{subfigure}[b]{0.3\textwidth}
                \centering
                \includegraphics[width=\textwidth]{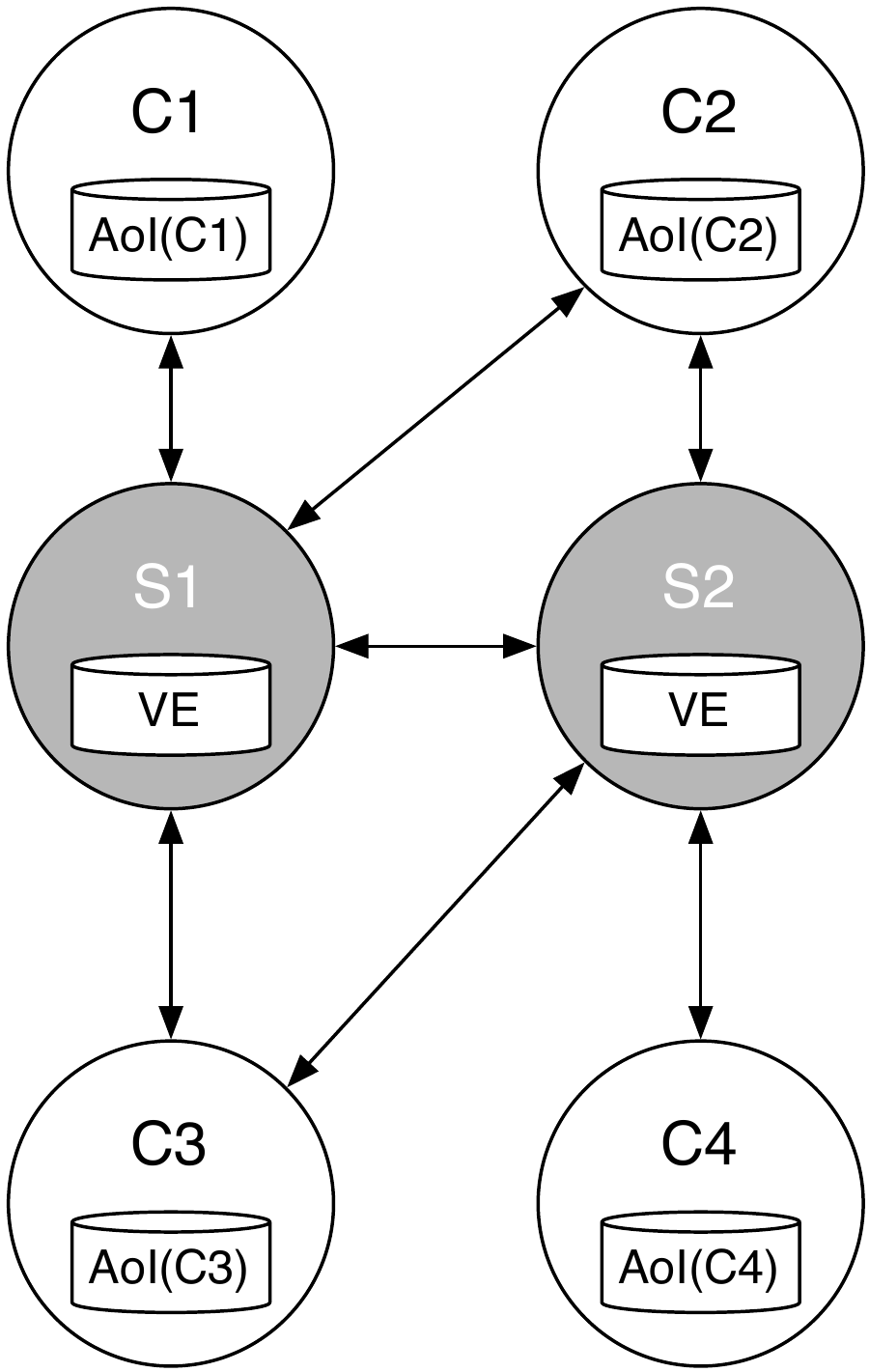}
                \caption{Zoning}
                \label{fig:zoning}
        \end{subfigure}
        \caption{Distribution strategies}\label{fig:dist-strategies}
\end{figure}

\paragraph{Instancing}
In this solution, the issues in managing the concurrency between multiple servers is avoided.
In practice, a portion of the virtual environment is replicated on multiple servers, each one maintaining an independent version of the state.
Each client is connected to a single server, and share the virtual environment only with the client connected to the same server.
For instance, in Figure \ref{fig:instancing} the clients $C_1$ and $C_3$ share a copy of the virtual environment on the server $S_1$, while $C_2$ and $C_4$ share another copy of the virtual environment on $S_2$.
In other words this solution basically replicate the centralized solution on more instances.
This solution is used in commercial applications as well as in several research work, such as in \cite{Barri2011}.

\paragraph{Mirroring}
In mirroring, as in instancing, the state of the VE is replicated in multiple servers.
However, unlike instancing, multiple servers can manage the same portion of the virtual environment.
Also, the servers are connected to each other in order to synchronize the state.
Each client connects to one server, but multiple clients in the same region can be connected to different servers.
Figure \ref{fig:mirroring}  shows an example of mirroring. 
In this case, all the clients share the same virtual environment using two servers.
Clients $C_1$ and $C_3$ are connected to the server $S_1$, while $C_2$ and $C_4$ are connected to $S_2$.
In case $C_1$ modifies the state of $S_1$, $S_1$ will synchronize with $S_2$ in order to share the same modification also with $C_3$ and $C_4$. This technique has been explored in different research works, such as \cite{cronin2004efficient,mauve2002generic}. 

\paragraph{Zoning}
In zoning, the virtual environment is divided into a set of regions, normally contiguous and non overlapping.
The shape of the regions is generally squared or hexagonal. 
One or more regions are assigned to a server, but the same region is assigned only to a single server.
Clients connect to one or more servers according to their position. In fact, if the area of interest of a user overlap multiple regions, the client must connect to the corresponding servers.
Zoning is depicted in Figure \ref{fig:zoning}, where $C_2$ and $C_3$ are connected to $S_1$ and $S_2$ at the same time.
This techniques has been studied in various research works, such as \cite{carlini2010integration,Kim2004,greenhalgh1995massive,lee2002atlas}.\\

These techniques are not exclusive to each other, rather a combination of them is possible. For instance, it is possible to divide the virtual environment into regions and for each regions apply mirroring. In fact, several proposal employ mixed approach to distribute the state of the virtual environment.

\subsection{On-demand Platforms}
\label{cloud}

Exploiting on-demand resources for MMOGs is a relatively young but very active line of resources.
Here we present two of the most relevant paper regarding this field.

To exploit the potential of on-demand provisioning, \cite{Marzolla} proposes a multi-tier cloud architecture. The first layer of the architecture contains a set of gateways, responsible for handling basic gaming protocol
checking and verification. The second level exploits Zone Partitioning by defining a set of cell servers each one  controlling a small area of the virtual environment.
Finally, the database servers manage the persistent game state information. Each layer of the architecture contains a set of parallel servers whose number is elastically defined at run time.
To this end, a monitor periodically collects several system statistics, and triggers the provisioner when the system response time deviates from a given threshold. A Queueing Network performance model is exploited where each server is modelled  with exponentially distributed inter-arrival times, exponentially distributed service times and FIFO service discipline. Finally, a greedy algorithm computes the number of servers in each level required for maintaining the response time under a given threshold and reserves the corresponding resources on the cloud.

In the same context, \cite{IosupTPDS} proposes an analytical load model
for MMOGs taking into account the main resources used by MMOGs: CPU, memory, and network. 
The model describes the machine load that has to consider the computation of the interaction
between pairs of entities, the reception of event messages from each client, and the update of entity states received from/sent to another machine.
\cite{IosupTPDS} also shows that, even if simple prediction algorithms are computationally inexpensive,
they exhibit a low predictive power. More elaborated prediction algorithms like autoregressive (AR), integrated (I), moving
average (MA) models, and combinations of these are time consuming and resource intensive, so that they are
not suitable for highly dynamic MMOGs. \cite{IosupTPDS} proposes an alternative approach, based on low complexity neural networks and shows that this approach enables precise resource provisioning.

\section{User-assisted Infrastructures}
\label{p2p}

In user-assisted infrastructures, clients actively participate at the management of the virtual environment. 
In other words, clients manage a part of the virtual environment, in fact by assuming the role of the server for that part.
User-assisted infrastructures are characterized by a pattern of communication between client nodes, which we generally refer to as \textit{overlay}. The nature and type of the overlay change according to the kind of infrastructure considered; in fact, user-assisted infrastructures can be classified in two ways, according to the clients that cover the role of server.

In \textit{hierarchical infrastructures} a set of (super) clients, also called \textit{Super Peers (SP)}, have enhanced knowledge with respect to regular clients. Normally SPs are connected by means of a dedicated overlay.
In \textit{flat} approaches, there is no neat distinction between super and regular clients.
These approaches typically employ an overlay that is common to all the clients in the network.


\subsection{Super-Peers Infrastructures}

\begin{table}[tbh]
\centering
\begin{tabular}{|c|p{1.3cm}|p{1.4 cm}|p{1.8cm}|p{1.7 cm}|}
\hline
Name & SP \newline overlay & Non-local \newline SPs & Event \newline notification & Space \newline Partitioning \\
\hline
\hline
HYMS \cite{Kim2004} & none & no & unicast SP & Square\\
\hline
VSM \cite{hu2008voronoi} & Voronoi & yes & unicast SP & dynamic \\
\hline
P2P-Arch \cite{Hampel2006} & DHT & yes & unicast SP & hexagonal\\
\hline
MOPAR \cite{Vuong2005} & DHT & yes & unicast P2P & hexagonal\\
\hline
\end{tabular}
\caption{Super Peer Approaches}\label{tab:sp}
\end{table}

In Super-Peer infrastructures, a selection of clients actively participate to the management of the virtual environment.
In the last years, several SP-based approaches have been proposed \cite{Kim2004,hu2008voronoi,Hampel2006,Vuong2005}.
These works perform a partitioning of the virtual world into regions.
Each region is managed by a super peer (sometimes called \textit{region controller} \cite{Hampel2006} or \textit{arbitrator} \cite{hu2008voronoi}) along with a set of backup SPs in order to increase robustness in case of failure of the main super peer.
To be considered as a super peer, a regular peer must satisfy particular requirements in term of hardware capability (bandwidth, CPU, RAM) and in term of stability. 
Regular peers receive state updates from the SP that manages their regions.

In \cite{Kim2004} the world is divided into square regions that, at the beginning, are assigned to a central server.
The first peer with enough computational and bandwidth capabilities to enter a region manages the region.
No overlay is provided between super peers, i.e. super peers do not have the possibility to communicate.
This represents a limit of the approach, as the view of a user is limited to a single region.
Also, this work performs local SP assignment, i.e. a peer can become a Super Peer for an arbitrary region $R$ only if the correspondent avatar is in $R$.  This can increase the probability of cheating.
Conversely, In \cite{Hampel2006} and \cite{Vuong2005} hexagonal regions are assigned to Super Peers in a random fashion, by exploiting a DHT.
Because of the random mapping, it is unlikely that a Super Peer manages a region where the correspondent user in playing.
This non-local SP assignment helps to reduce the possibility of cheating.

Voronoi State Management \cite{hu2008voronoi} partitions the virtual environment with a Voronoi tessellation.
Given a number $N$ of points on a two-dimensional plane, a Voronoi diagram partitions the space into $N$ non-overlapping regions and each region contains all the points closer to the region site. 
In a Voronoi overlay network each site corresponds to a peer in the network. 
Each peer $p$ maintains a Voronoi diagram of a subset of the space and connections with its Voronoi neighbours in the two-dimensional space.
The result is an unstructured overlay where each peer manages the space correspondent to its Voronoi region.
Each Super Peer manages a Voronoi cell, by receiving and communicating the updates to the clients in the cell.
A big advantage of the Voronoi partitioning is the possibility to resize the region managed by super peers, which helps to balance and distribute the load.

\subsection{Peer-to-Peer Infrastructures}

\begin{table}[tbh]
\centering
\begin{tabular}{|p{2.2cm}|p{1.3cm}|p{2.1cm}|p{2.2cm}|p{1.2cm}|}
\hline
Name & Space \newline Part. & P2P \newline Overlay & Event \newline Notification & Objects \newline Mngmt. \\
\hline
\hline
SOLIPSIS \cite{Frey2008} & Voronoi & Delauney & P2P unicast & yes \\ 
\hline
VON \cite{Hu2006} & none & unstructured & P2P unicast & no \\
\hline 
Colyseus \cite{Bharambea} & -- & DHT & pubsub over DHT & yes \\ 
\hline
SimMud \cite{Knutsson2004} & regions (static) & DHT & Positions: P2P multicast & yes \\ 
\hline
APOLO \cite{lee2006apolo} & none & unstructured (quadrant) & Controlled Flooding & no \\
\hline
Compass \cite{ricci2011aoi} & voronoi & delauney & P2P multicast trees & no \\
\hline
Peer Clustering \cite{Chen} & regions (static) & DHT & region manager unicast &  yes \\ 
\hline
\end{tabular}
\caption{Flat approaches. Other works cited in the section, like VON-Forwarding \cite{chen2007forwarding} and FiboCast \cite{Jiang2009} are optimization of VON.}\label{tab:flat}
\end{table}

In this kind of solutions, neither supernodes or servers are considered. 
All users participating to the virtual environment manage a portion of it.
These infrastructures are typically structured considering the position of the user in the virtual environment.
Flat approaches can be classified according to the degree of structuredness of the overlay between nodes.

Several works exploits Distributed Hash Tables (DHTs) as the main server overlay \cite{Chen,Bharambea,Knutsson2004}.
For instance, \cite{Chen} exploits the randomness of DHT objects placement in order to assign regions to their controller.
Each controller receives the notification of updates and forward them to the interested node.
Similarly, SimMud \cite{Knutsson2004} divides the virtual world into regions, and each region has assigned a coordinator.
The coordinator serves two tasks.
First, it permits the creation of a full connected overlay between nodes. 
This overlay is then used by the node to notify each other their movements.
Second, the coordinator works as the root of a multicast tree for the region. To create the multicast paths, SimMud exploits Scribe \cite{Castro2002}, a well know approach to build multicast infrastructures over structured P2P networks. Peers generate events and notify them to their region coordinator, which in turn forward them along the multicast tree.
Even if from a structural point of view Scribe is able to manage dynamic membership and large groups, a potential problem is the latency of messages. In fact, the number of hops and the length of the paths may dramatically increase the latency. 
Also, it has been pointed out in \cite{Bharambe2005} that application-level multicast may saturate the bandwidth of nodes in presence of heterogeneous bandwidth capability, which is the case in wide distributed MMOGs.
This strategy assures low values for messages latency, since each recipient is always one or two hops away from the source. However, as the number of recipient nodes grows, this method may oversaturate the bandwidth capability of the source. 

Another kind of solutions consider a dynamic partitioning of the virtual environments.
These solutions are based on the Voronoi tessellation.
They employ an event forwarding schema, in order to deliver events to other possible interested recipients.
These solutions have usually high scalability, since the necessary bandwidth to deliver events is split among a number of nodes. On the other hand, forwarding-based solutions may increase latency since event source and recipient may be separated by multiple hops.
Compared with DHT-based approach, these mechanisms yield two relevant advantages. First, they have no overhead for peer churn, since they work without any long term and synchronized structure. Second, only local peer information is exploited to forward messages.

One of the first solutions based on unstructured overlay is APOLO \cite{lee2006apolo}. Each peer divides its space of interest into quadrants and maintains a link to the closest neighbour in each quadrant. In order to notify an event, a peer sends the message to these four neighbours, which in turn recursively forward the message until it reaches all the possible interested peers. This solution strictly bounds the number of outgoing connections per peer, nevertheless, it may dramatically increase the number of hop and the bandwidth consumption in case of crowded situations.

A later approach, VON-forwarding \cite{chen2007forwarding}, divides the space according to a Voronoi diagram.
Each peer broadcasts a message to all its Voronoi neighbours in order to notify the peer in its AOI. This solution exploits that, on average, a peer in a Voronoi diagram has six neighbours. Compared with the direct link approach, VON-forwarding helps reduce the number of messages per event sent by peers. 
Compared with APOLO, the number of hops decreases due to the wider degree of the AOI-cast tree.
In spite of that, the bandwidth usage is not efficient due the elevated number of messages replication in the network.
This model has been subsequently refined with VoroCast and FiboCast \cite{Jiang2009}.
VoroCast builds a multicast spanning tree using the underlying Delauney network and sends the notifications of events along the edges of this tree. FiboCast is a further optimization of VoroCast. It models messages frequency rates using the Fibonacci sequence, in a way that farthest nodes from the source receive updates less frequently than nearby nodes.
%
The main disadvantage of these systems is the fact that they require non-local information to correctly forward messages. 
In particular they need to know the neighbours of the neighbour of a node, and since it depends on the position of the peers, this information has to be updated frequently. This may cause an increasing of bandwidth consumption, especially in crowded situation, where the Voronoi diagram change rapidly. Another aspect to consider is that VoroCast and FiboCast do not take into account the effects of the latency when considering the position of the peers (i.e. the positional drift). Due to this reasons, delivered messages may be duplicated and travel along path that are longer than necessary.

Ricci et al \cite{ricci2011aoi} propose a Delauney-based AOI-cast that copes with these two drawbacks. First, they employ a forwarding schema based on compass routing that exploits only information local to peers, i.e. theirs one-hop neighbours. This avoids the extra-usage of bandwidth for maintaining n-hop neighbours, which happens in approaches like VoroCast. Second, their solution takes into account the latency in information diffusion, by considering the possible positional drift occurred to the peer when computing AOI-cast paths. This reduces messages redundancy and decreases the probability of message losses.

\subsection{Anti-Cheating}

The mechanisms to contrast cheating are called Anti-Cheating (AC).
One of the first AC solutions, Lockstep \cite{Baughman2007}, divides the time into rounds and requires every player to submit its moves for that round before the next round is allowed to begin. 
Unfortunately this approach slows down the experience and it is not applicable for fast-paced virtual environment.
Asynchronous Synchronization \cite{Baughman2007} (AS) and Sliding Pipeline \cite{Cronin} (SP) strive to improve the performance of Lockstep. 
AS relaxes the constraints of Lockstep by requiring only players in a region to work as Lockstep. 
SP permits the updates to be pipelined and the use of dead reckoning in order to improve the smoothness of the simulation. However, both these approaches suffer of the same problem of Lockstep, i.e. they force a user to wait until the duration of a round before validate its state.
New Event Ordering (NEO, \cite{Gauthierdickey}) aims to reach a distributed consensus among a set of distributed clients. NEO explicitly bounds the round duration, and each round is divided into two halves. 
Clients must send the actions to half of a group within the half of a round duration, in order to consider the update committed.
In the other half of the round, players send their key for security checks. 
%
In \cite{Webb}, authors propose the Referee Anti-Cheat Schema (RACS), an anti-cheating schema suitable for centralized and hybrid approaches. RACS uses a central server, called \textit{referee}, to receive, simulate and validate the events. 

\section{Hybrid}
\label{hybrid}

Hybrid architectures aim for the combination of user-assisted and centralized infrastructures. A wide-used method divides the \textit{Virtual Environment} (VE) into \textit{regions} or \textit{cells}, whose dimension can be either fixed or variable. These regions are in turn assigned to a peer or a server, which becomes the \textit{manager} of the entities in that region.
%
Region assignment in hybrid architecture mostly follows two different approaches: (i) a region can be assigned to either a peer or
to a server without any restriction, or (ii) only a subset of cells can be assigned to peers.

The work proposed in \cite{Kim2004} belongs to the first category. The authors consider square cells, which are initially managed by a central server. The first peer with enough computational and bandwidth capabilities to enter a cell becomes the cell manager. Afterwards,
a fixed number of peers that enters the same cell act as backup managers in order to increase failure robustness. Similarly, \cite{Barri2010} proposes an hybrid system, including a central server and a pool of peers. The central server runs the MMOG and, as soon as it reaches the maximum of its capacity, delegates part of the load to the peers. 

The same authors of \cite{Barri2010} propose in \cite{Barri2011} an approach belonging to the second category. A central server executes the main game whereas the peer run \textit{auxiliary games} which are typical of certain games genres, such as Massively Multiplayer Online Role-Playing Games (commonly called MMORPGs). Auxiliary games are separated instance of the MMOG, shared only by a fixed (and usually not high) number of players. 
In a similar way, \cite{Chen} proposes a functional partition of the DVE tasks. Central servers operate user authentication, game persistence and manage regions characterized by high-density user interactions, whereas peer support only low-density interaction regions. 
%
Authors of \cite{Jardine2008} provide an interesting distinction between \textit{positional} and \textit{state-changing} actions. They propose an hybrid architecture where peers manage positional actions, which are more frequent and prone to be maintained locally. Central servers handle state-changing actions, that are not transitory and require a larger amount computational power. 

The idea of distinguishing positional and state-changing actions is in fact an interesting idea which we have exploited 
in the design of our architecture.
However, rather than assigning different actions to different type of nodes, we define two different and independent distributed structures that manage, respectively, positional and state-changing actions.
The management of the nodes can be assigned  to a peer or to a cloud node.

In other words, we exploit an intermediate approach. On one hand, some functionalities, like authentication, must be handled by centralized and full controllable servers. On the other hand, other functionalities may be mapped
to central servers or to peers. This requires a complete dynamic strategy allowing for more flexibility in load distribution, which requests a fine-grained management of the resources by the MMOG operator. Resources control is very important for our approach, since the seamless combination of Cloud and P2P requires to keep under control the cost and to effectively deal with the implicit uncertainty related to peers. 
Therefore, a basic issue for the exploitation of hybrid architectures is the definition
of effective load distribution mechanisms.

\section{Mobility Models in MMOGs}
\label{mobilitymodels}

The evaluation of the fist generation of  MMOG architectures was generally performed by exploiting mobility models originally designed to reproduce the movements of human beings, such as those exploited to evaluate ad-hoc wireless networks.

The {\em Random Way Point model} RWPM \cite{Hong1999} was one of the most widely exploited mobility model.
In RWPM a set of {\em way points} are placed uniformly at random locations in the virtual environment.
Each entity independently moves toward them.  As soon as an entity reaches a way point, it stops there for a time interval. 
Afterwards, it chooses another way point, and so on. 
RWPM has been adapted to describe different kind of scenarios in a MMOG by tuning the spatial distribution and the number of the waypoints, the speed of the entities and the criteria to select the waypoints at each step.
While most mobility models for ad-hoc wireless mobile networks focus on the  motion behaviour of each entity separately, mobility models taking into account the behaviour of {\em group of entities} have been proposed.
The {\em Reference Point Group Mobility model (RPGM)} \cite{Hong1999} has been proposed as an extension of the RWPM, by 
introducing in the model the concept of group. The model can be exploited to simulate the behaviour of teams of players.

Although very simple to generate, the mobility patterns created with RWPM-based mobility models are not precise enough 
to represent the characteristics of nowadays MMOGs.
Indeed, since players participating to large scale MMOGs usually have the possibility to move freely around the virtual world, the distribution of the players in MMOGs is usually not uniform \cite{La2008}.
Players tend to gather in well determined positions of the virtual environments, creating the so called hotspots.
Furthermore, players behaviour inside hotspots results to be highly non-uniform: players move slowly and chaotically within the hotspots, while the movement between hotspots is straight and fast \cite{Liang2008}.

These considerations lead to the design of mobility models specifically developed for MMOGs.
For example, \cite{Tan2005} provides a design and evaluation of a mobility model based on a popular on-line game\footnote{Quake II, http://www.idsoftware.com/games/quake/quake2}. The model is based on a RWPM whose parameters are evaluated by using model fitting techniques on traces. These traces have been used also in \cite{Bharambea} in order to evaluate their solution. They propose a model based on real traces where players tend to move between popular regions of the map and the popularity distribution of these regions follows a well specific trend. 
\textit{Blue Banana} \cite{Legtchenko2010} provides the design of a mobility model based on Second Life \cite{sl-site}.
This model characterizes the virtual environment between desert areas and hotspots.
The model assumes the movement of the avatars to be slow and chaotic in the hotspots, while fast and predictable in the desert areas of the MMOG. The model exploits an automaton defined by three states, the {\em halted state}, where the avatar does not move, the {\em exploring state} where the avatar moves within a hotspot and the {\em travelling state} where
the avatar moves from one hotspot toward another one.
The definition of hotspot as an invariant for a MMOG mobility model is also one of main finding of the work of  Miller and Crowcroft \cite{Miller2009}. They measure and analyse players movements in a World of Warcraft (WOW, \cite{wow-site}) scenario, which is representative of the team-oriented interaction that modern MMOGs encourage into the game. 
However, \cite{Miller2009} does not define any mobility model. The main findings of their work state that a way point-based model is not enough to describe complex movements of MMOGs, that the level of gathering of players in groups is less than expected and that hotspots based mobility is a realistic pattern of movement in MMOGs.

\section{Case Studies}
\label{casestudies}

In this section we describe several core proposal in the field of the distributed virtual environment.
We chose three different and heterogeneous approach that, in our opinion, best represent the issues in designing a distributed virtual environment.

\subsection{SimMud}
SimMud \cite{Knutsson2004} is a support for Massively Multiplayer Games built on top of the Pastry DHT \cite{rowstron2001pastry}. SimMud uses Scribe \cite{Castro2002}, an application-layer multicast built on top of Pastry, as the main communication pattern to disseminate game state. The design of SimMud is based upon the limited movement speed and sensing capabilities of the avatars, so that the locality of interest can be exploited. SimMud maps both the participating peers and the MMOG's objects onto uniformly distributed IDs in the circular 128-bit namespace of the DHT. Object insertions and lookup are done by exploiting the classical DHT primitives.
Objects are managed by nodes whose ID is numerically closest to the object ID. In SimMud, "closeness" it is related to the numerical ID and no geographical or topological optimizations are considered.

Scribe is a scalable multicast infrastructure that maps the information about multicast groups to the Pastry DHT. A multicast tree associated with the group is built by merging the Pastry routes from each group member to the group ID's root, which also acts as the root of the multicast tree. A multicast message from the root reaches the members by following the reverse paths of the multicast tree.


The world is statically partitioned into rectangular regions and the nodes in the same region form an interest group for that portion of the map. The region updates are sent within the group only. Whenever a player goes from a region to another, the group membership changes accordingly. 
A node whose ID is the closest to the region ID serves as the \emph{coordinator} for that region. The coordinator manages all the objects in its region and also acts as the root of the multicast tree. The load can be distributed by creating a different ID for each type of objects in the region, thus mapping them on to different peers.
     
SimMud defines different classes of game state and pairs different consistency maintenance strategies with each class.
\begin{itemize}
\item The \emph{player state} is accessed according to a single-writer multiple-reader pattern. Each player updates its own location as it moves around. Player-player interactions, such as fighting and trading, only affect the states (e.g. the life points) of the players involved. Since position change is the most common event in a game, the position of each player is disseminated through multicast messages at fixed intervals to all other players in the same region. The interval is determined during game design, according to the requirements of the game.

\item The \emph{object state} is managed by a coordinator-based mechanism to keep shared objects consistent. A certain degree of replication is provided. Each object is assigned to a coordinator that manages its updates. A replica is maintained by a node close to the coordinator in the DHT space. 
The coordinator both resolves conflicting updates, and stores the current object value. 
Successful updates are multicasted to the region to update each player's local copy.
\end{itemize}

SimMud exploits shared state replication to manage peer failures. 
The copies are kept consistent, in spite of node and network failures, through a lightweight primary-backup mechanism that tolerates failures of the network and nodes. These failures are detected by exploiting messages of regular game events (i.e. peer movements), without any additional network traffic.

\subsection{Colyseus}

The main focus of Colyseus \cite{Bharambea} is the management of the game state. The world is seen as a collection of objects, both mutable (e.g. items, characters) and immutable (e.g. map geometry, graphics). Colyseus manages the collection of mutable objects through a component called \textit{global object store}. Each mutable object is associated with a \textit{think} function that determines the behaviour of the object.
The architecture of a generic peer is composed by the following modules:
\begin{itemize}
\item a \textit{local object store} i.e. the collection of primary objects and replicas
\item a \textit{replica manager} that manages the synchronization of primary and replicas
\item a \textit{object placer} that decides where to place and migrate primary replicas
\item a \textit{object locator} that connects to a DHT overlay indexing all the objects in the game. 
\end{itemize}


Each object in the global object store has one \textit{primary} copy that resides onto one node. Updates to an object are sent to the primary node, which takes care of the ordering of updates.
A node executes only the think function associated with primary objects in its local store. The execution of such functions may require access to objects that a node is not the primary owner of. In order to facilitate the execution of this code, a node create a secondary \textit{replica}. The node periodically registers an interest with the node hosting the primary object. Replicas are weakly consistent and are synchronized with the primary copy. In detail, at each frame, whenever the primary object is modified, an update is sent to all the replicas. Similarly, whenever a secondary replica is modified, an update is shipped to the primary owner.

The replicas are fetched using a multi-attribute range query DHT. Colyseus proposes two approaches in order to guarantee low latency on lookup queries. First, it exploits spatial and temporal locality in object movements in order to obtain prediction of subscriptions. This allow the DHT to execute speculative pre-fetching of replicas. 
Second, it enables soft caching of both publications of the objects and subscriptions. By storing publications, a subscription can immediately match with a recent publication. 

\subsection{Voronoi Based Overlay Network}
\label{sec-von}

Voronoi based Overlay Network (VON) \cite{Hu2006} is a P2P overlay network based on Voronoi Diagrams which preserves high consistency of the overlay topology. 
The initial proposal of VON defines a direct connection model, where each node of the  virtual environment is directly connected to all the nodes located in its AOI. Due to the limited bandwidth of each node, this model may constrain the number of neighbours that may appear within the area of interest of a given node. VON defines different kinds of neighbours of a node. The enclosing neighbours of a node $n$ are the nodes whose regions immediately surround the Voronoi region defined by $n$.
The boundary neighbours are the nodes whose Voronoi regions intersect the border of AOI(n).
Finally, the AOI neighbours are all further nodes belonging to AOI(n).  Each node keeps a Voronoi Diagram including its enclosing, boundary and AOI neighbours. In VON, each node acts as a "watchman" for another one in discovering approaching neighbours. When an entity moves, it sends its new position to all the neighbours belonging to its Voronoi Diagram. If the receiver is a boundary neighbour, it performs an overlap-check, i.e. checks whether the Voronoi region of one of its enclosing neighbours overlaps the AOI of the mover. The receiver notifies the mover if a new overlap occurs, i.e. previously disjoint regions currently overlap. In this case the boundary neighbour explicitly notifies the mover about the new neighbours.
The moving entity becomes aware of neighbours outside its AOI with minimal network overhead, since position notification may be exploited to discover new neighbours.
Whenever a node leaves the virtual environment or fails, its neighbours update their Voronoi Diagram by removing that node.
The direct connection model may require a large amount of bandwidth, especially when crowding occurs. 
Event notifications are propagated to each AOI neighbour by forwarding notification through neighbour nodes.

\subsection{On-demand Provisioning}

The work in \cite{prodan2009prediction} has been one of the fist proposal for dynamic provisioning of on-demand resources for MMOGs applications.
Its architecture is composed by two core services: a \textit{load prediction service}, and a \textit{resource allocation service}.

The load prediction service has the task of predicting the future distribution of avatars in an area of the MMOG. 
The load prediction service exploits a \textit{neural network} to estimate the numbers of avatars in an arbitrary region.
The number of avatars is then used to compute the CPU time requirements for the region.
The neural network is trained with a series of traces that are implemented by the authors.
The traces consider different avatar's profile, such as aggressive (frequently interact with opponents), team player (mostly acts in groups), scout (mostly acts alone) and camper (hides and waits for opponents). Their result shows that the neural network over-performs other prediction mechanisms, such as Moving average and Exponential smoothing.

The resource allocation service (presented in \cite{Nae2008}) exploits the prediction results to drive the on-demand allocation of resources. The resource allocation service executes two main tasks.
First, it recruits more servers to accommodate more players during peak hours. 
The new servers are recruited considered the prediction on the CPU load made by the neural network.
Second, the resources allocation service releases under-utilized servers to optimize the resource utilization.

\section{Conclusion}
\label{related-conc}

This chapter has provided an analysis of the state of the art in several research fields on MMOGs.
A clear trend of the research related to MMOGs emerges from this analysis.
The first pioneering works in late '90 have considered centralized approaches.
At the time, issues on interest management and on consistency-interactivity tradeoff have emerged.
In the early 2000, MMOGs research turned on widely distributed infrastructure, together with the concurrent explosion of the P2P computing. P2P and widely distributed infrastructures have posed new challenged that had not been considered before.
Extensive load distribution, overlay maintenance and fault tolerance are few examples of the issues faced in that period.

Nowadays, the next-generation MMOG platforms definitely leads forward to on-demand computing models.
This trend is evident since most of the recent works on MMOGs architectures strive with this thematic. 
However, we believe that a decade of research in P2P-based MMOGS architecture have still its role to play. 
For this reason, this thesis treats the combination of P2P and on-demand computing models, and, as far as we know, we are the first in this direction.

\chapter{Conclusion}
\label{chap:conclusion}

This thesis has presented two different and independent components for a MMOGs architecture that integrates the illusion of infinite resources provided by the Cloud, with the few cost associated to the exploitation of user-provided resources. We designed the two components by allowing the MMOG operator to control the trade-off between performance and economical cost.

The idea of combining these two different computing models came from some work we carried out in the field of cloud computing \cite{carlini2012cloud} and P2P architecture for on-line gaming infrastructures \cite{ricci2011aoi}. We first proposed the general concepts in \cite{carlini2010integration}, along with several preliminary ideas.
In \cite{ricci2012tutorial} we presented a comprehensive state of the art on on-line games infrastructures, also considering emerging computing models as the Cloud.
However, at the best of our knowledge, we are among the first that have proposed the \textit{combination of Cloud and P2P} for large scale MMOGs infrastructures.

The Positional Action Manager (PAM) is based on a combination of a cloud server and a best-effort P2P overlay providing support for interest management in large scale online games. To build the P2P overlay, PAM employs a \textit{two-layer gossip-based} protocol. 
PAM is fully described in \cite{gossipim,netgames12}.
As far as we know, this is the first time a gossip protocol is used as an active mean to resolve interest management. Besides its originality, the PAM-overlay has provided encouraging results. 
Experimental results show that PAM is able to obtain performance comparable to a server solution, while reducing the expenses for the game operator. In accordance with our view, the operator can further tune the tradeoff between performances and economical cost, by trading some precision in the result for a more economic infrastructure.

The State Action Manager (SAM) exploits a Distributed Hash Table, equipped with Virtual Servers to distribute the effort on management of the entities to multiple resources, including user-provided ones. 
We presented an initial version of the SAM \cite{pos}, where we exploited the knowledge acquired with our prior work on the distributed hash tables \cite{carlini2010reducing,carlini2011probabilistic}. In the thesis, we have presented a refined and enhanced version of the SAM. 
In order to pro-actively distribute the entities of a MMOG among the nodes of the SAM, we employed a \textit{greedy heuristics} that minimizes the operational costs while keeping the availability and the fraction of non overloaded nodes above the given threshold. 

In order to test and evaluate the two components, we built a \textit{realistic bandwidth consumption workload}. 
We considered both the load from direct players interactions and the load from the interactions of the players with the objects of the virtual environment. 
This represents a difference with work in literature, that tends to considers these approaches separately.
The movement traces of the avatars were obtained by exploiting a Second Life mobility model.
We have observed in \cite{carlini2011evaluating} that this model assures a fair balancing in players movements.

Finally, we proposed some future work, having the common goal to unify the two aforementioned components in a full infrastructure for MMOGs.
The design of a smart client would allow the players to exploit the advantages of SAM and PAM at the same time. A multi server PAM would be able to scale up to ten thousands of players, while keeping the economical costs acceptable. 
We strongly believe that the combination of Cloud Computing and Peer-to-Peer is the next milestone for MMOGs architectures.

{\small
	\renewcommand{\bibname}{References} 
	\bibliographystyle{alpha}
	\bibliography{backmatter/bib/thesis}
}

\makecopyright

\end{document}